\documentclass{article}
\pdfoutput=1

\usepackage{amsmath, amssymb, amsthm, amscd, amsfonts}
\usepackage[margin=1in]{geometry}
\usepackage{tikz}
\usetikzlibrary{calc,shapes.multipart,chains,arrows}
\usepackage{graphicx}
\usepackage{caption}
\usepackage{subcaption}
\usepackage{enumitem}
\usetikzlibrary{shapes.symbols,trees,positioning,fit,arrows,decorations.pathreplacing,backgrounds}

\usepackage{comment}
\usepackage{thmtools, thm-restate}

\theoremstyle{plain} \numberwithin{equation}{section}
\newtheorem{theorem}{Theorem}[section]

\newtheorem{lemma}[theorem]{Lemma}

\theoremstyle{definition}

\newcommand{\true}{true}
\newcommand{\false}{false}

\newcommand{\ifcode}{{\bf if} \,}
\newcommand{\thencode}{{\bf then} \,}

\newcommand{\goto}{{\bf go to} \,}
\newcommand{\waittill}{{\bf wait till} \,}
\newcommand{\return}{{\bf return} \,}

\newcommand{\forcode}{{\bf for} \,}

\newcommand{\cas}{\mbox{CAS}}
\newcommand{\fas}{\mbox{FAS}}

\renewcommand{\cas}{\mbox{\bf CAS}}
\renewcommand{\fas}{\mbox{\bf FAS}}

\newcommand{\reset}{\mbox{\sc Reset}}

\newcommand{\tail}{\mbox{\sc Tail}}
\newcommand{\nodes}{\mbox{Node}}

\newcommand{\barrier}{\mbox{\sc Barrier}}

\renewcommand{\barrier}{\mbox{\sc CSowner}}

\newcommand{\csowner}{\barrier}
\newcommand{\seq}{\mbox{\sc Seq}}
\newcommand{\mlock}{\mbox{\sc Lock}}
\newcommand{\alock}{\mbox{\sc Lock}}

\newcommand{\stopwait}{\ensuremath{\mbox{\sc Stop}}}
\newcommand{\csownerdsm}{\mbox{\tt CSowner}}
\newcommand{\stopwaitdsm}{\mbox{\tt Stop}}

\newcommand{\set}{\mbox{\sc Set}}
\newcommand{\tryset}{\mbox{\sc TrySet}}
\newcommand{\csset}{\mbox{\sc CSSet}}
\newcommand{\exitset}{\mbox{\sc ExitSet}}

\newcommand{\crashedset}{\mbox{\sc CrashSet}}

\newcommand{\inuse}{\mbox{Active}}

\newcommand{\inusep}{\inuse_p}
\newcommand{\myseq}{\mbox{\sc S}}

\newcommand{\myseqp}{\myseq_p}
\newcommand{\isactive}{\inuse}
\newcommand{\activep}{\isactive_p}

\newcommand{\lock}{\mbox{\sc Lock}}

\newcommand{\pc}[1]{\mbox{\ensuremath{PC_{#1}}}}
\newcommand{\pcp}{\pc{p}}

\newcommand{\prev}[1]{\ensuremath{prev_{#1}}}
\newcommand{\prevp}{\ensuremath{\prev{p}}}

\newcommand{\initialize}[1]{\ensuremath{\texttt{reset}_{#1}}}
\newcommand{\initializep}{\ensuremath{\initialize{p}}}

\newcommand{\tryproc}[1]{\ensuremath{\texttt{try}_{#1}}}
\newcommand{\exitproc}[1]{\ensuremath{\texttt{exit}_{#1}}}
\newcommand{\recoverproc}[1]{\ensuremath{\texttt{recover}_{#1}}}

\newcommand{\tryprocp}{\ensuremath{\tryproc{p}}}
\newcommand{\exitprocp}{\ensuremath{\exitproc{p}}}
\newcommand{\recoverprocp}{\ensuremath{\recoverproc{p}}}

\newcommand{\mread}[2][p]{\ensuremath{\texttt{read}_{#1}(#2)}}
\newcommand{\mwrite}[2][p]{\ensuremath{\texttt{write}_{#1}(#2)}}
\newcommand{\mwait}[2][p]{\ensuremath{\texttt{wait}_{#1}(#2)}}
\newcommand{\mbwait}[2][p]{\ensuremath{\texttt{wait}_{#1}(#2)}}
\newcommand{\mrelease}[2][p]{\ensuremath{\texttt{release}_{#1}(#2)}}
\newcommand{\mcapture}[2][p]{\ensuremath{\texttt{capture}_{#1}(#2)}}
\newcommand{\mset}[2][p]{\ensuremath{\texttt{set}_{#1}(#2)}}
\newcommand{\mbset}[2][p]{\ensuremath{\texttt{set}_{#1}(#2)}}
\newcommand{\mreset}[2][p]{\ensuremath{\texttt{reset}_{#1}(#2)}}
\newcommand{\mabandon}[2][p]{\ensuremath{\texttt{exit}_{#1}(#2)}}

\newcommand{\limplies}{\Rightarrow}

\newcommand{\none}{\ensuremath{\perp}}
\newcommand{\isequal}{=}

\newcounter{linecounter}
\newcommand*{\procline}[1]{{\bf \refstepcounter{linecounter}\thelinecounter\label{ln:#1}.}}
\newcommand*{\refln}[1]{{\bf \ref{ln:#1}}}
\def\final{True}
\newcommand{\cond}[1]{%
	\label{inv:fullcrash:cond#1}
	\ifnum\pdfstrcmp{\final}{True}=0 \unskip\else #1 \fi\ignorespaces}

\newcommand{\lockobj}{\ensuremath{\ell}}

\renewcommand{\L}{{\mathcal{L}}}

\newcommand{\cso}{\csowner}
\newcommand{\lck}{\mlock}
\renewcommand{\sp}{\myseq_p}

\newcommand{\gotocs}{\ensuremath{\mbox{IN\_CS}}}
\newcommand{\gotorem}{\ensuremath{\mbox{IN\_REM}}}

\newcommand{\status}{\mbox{\em status}}
\newcommand{\good}{\mbox{\em good}}

\newcommand{\rectry}{\mbox{\em recover-from-try}}
\newcommand{\reccs}{\mbox{\em recover-from-cs}}
\newcommand{\recexit}{\mbox{\em recover-from-exit}}
\newcommand{\recrem}{\mbox{\em recover-from-rem}}
\newcommand{\procset}{\ensuremath{{\mathcal{P}}}}

\newcommand{\upto}{\texttt{-}}
\newcommand{\lockstate}{\mbox{\sc State}}

\newcommand{\face}[1]{\ensuremath{\text{Face}_{#1}}}
\newcommand{\facep}{\ensuremath{\face{p}}}

\newcommand{\tok}[0]{\texttt{token}}

\newcommand{\X}[0]{X}
\newcommand{\W}[0]{W}

\newcommand{\Go}[1][p]{\text{\sc Go}_{#1}}

\renewcommand{\b}[1][p]{b_{#1}}
\renewcommand{\i}[1][p]{i_{#1}}
\newcommand{\x}[1][p]{x_{#1}}
\newcommand{\ptr}[1][p]{ptr_{#1}}

\newcommand{\fflag}[0]{flag}
\newcommand{\fseq}[0]{seq}

\newcommand{\object}[1]{\mathcal{#1}}
\newcommand{\C}[0]{\object{C}}
\newcommand{\Sig}[0]{\object{S}}

\AtBeginDocument{%
  \providecommand\BibTeX{{%
    \normalfont B\kern-0.5em{\scshape i\kern-0.25em b}\kern-0.8em\TeX}}}

\title{Constant RMR Recoverable Mutex under System-wide Crashes}

\author{
Prasad Jayanti\footnote{Dartmouth College; prasad.jayanti@dartmouth.edu} 
\and 
Siddhartha V. Jayanti\footnote{Google Research and MIT; sjayanti@google.com, siddhartha@csail.mit.edu} 
\and 
Anup Joshi\footnote{Yahoo; anup.s.joshi.gr@dartmouth.edu} 
}

\date{January 17, 2023}

\begin{document}

\maketitle

\begin{abstract}
We design two Recoverable Mutual Exclusion (RME) locks for the system-wide crash model.
Our first algorithm requires only $O(1)$ space per process, and achieves $O(1)$ worst-case RMR complexity in the CC model.
Our second algorithm enhances the first algorithm to achieve (the same) $O(1)$ space per process and $O(1)$ worst-case RMR complexity in {\em both} the CC and DSM models.
Furthermore, both algorithms allow dynamically created threads of arbitrary names to join the protocol and access the locks. 
To our knowledge, these are the only RME locks to achieve worst-case $O(1)$ RMR complexity assuming nothing more than standard hardware support.
In light of Chan and Woelfel's $\Omega(\log n / \log\log n)$ worst-case RMR lower bound for RME in the individual crash model, our results show a separation between the system-wide crash and individual crash models in worst-case RMR complexity in both the CC and DSM models.
\end{abstract}

\section{Introduction}

The emergence and widespread commercial availability of {\em non-volatile memory (NVM)}, which retains its state despite {\em system-wide crashes} caused by power outages \cite{nvmm:pcm, nvmm:memristor, nvmm:mram} has ushered in the era of {\em recoverable} (a.k.a. {\em durable}) algorithms.
Such algorithms allow processes that {\em crash}, and lose the contents of their process registers due to a system-wide power outage, to smoothly {\em recover} upon restart and resume computing from the time of the crash  \cite{IzraelevitzMS16, attiya:rlin, FriedmanHMP18, FriedmanPetrankRamalhete, Golab:rmutex, Golab:rmutex2, Golab:rmutex3, Golab:rmutex, jayanti:fasasmutex, jayanti:rmesublog, jayanti:fcfsmutex, jayanti:rmeabort, Chan, ChanWoelfelLB}.
Due to the centrality of {\em mutual exclusion (mutex)} locks in concurrent computing \cite{Dijkstra:mutex}, the {\em recoverable mutual exclusion} (RME) lock has received a lot of attention  \cite{Golab:rmutex, Golab:rmutex2, Golab:rmutex3, Golab:rmutex, jayanti:fasasmutex, jayanti:rmesublog, jayanti:fcfsmutex, jayanti:rmeabort, Chan, ChanWoelfelLB}.
The principal goal of RME lock research has been to design algorithms that have low ``RMR'' complexity for both ``cache-coherent'' (CC) and ``distributed shared memory'' (DSM) systems, in the face of crashes.
In this paper, we design RME locks that tolerate system-wide crashes for both CC and DSM systems.
Our locks have just constant worst-case RMR complexity, allow access to dynamically created threads, and require only constant space per accessing thread.
Our algorithm uses only the read, write, compare-and-swap (CAS) and fetch-and-store (FAS) instructions, which are commonly supported on modern machines.

In the rest of this section, we review the background, state our contributions, and compare them to the state of the art.

\subsection{Failure models}

In practice, crashes are generally caused by power-failures, which cause all processes in the system to crash simultaneously \cite{IzraelevitzMS16}.
Recoverable algorithms for such {\em system-wide crashes}, have been widely studied in both the systems and theory communities \cite{IzraelevitzMS16, Golab:rmutex3, FriedmanHMP18, FriedmanPetrankRamalhete, LiGolab, ConditNightingaleFrostIpek, CoburnEtAlConference, CoburnEtAlJournal, KolliPelleySaidiChenWenisch}.
A lot of theory literature, particularly related to RME, has also studied the {\em individual crash model} \cite{Golab:rmutex2, Golab:rmutex, jayanti:fasasmutex, jayanti:rmesublog, jayanti:fcfsmutex, jayanti:rmeabort, Chan, ChanWoelfelLB, BenDavid, Attiya} in which individual processes can crash and restart even as other processes run unaffected.
Algorithms designed for this individual crash model work for system-wide crashes, because a sequence of individual crashes can simulate a system-wide crash.
Algorithms designed directly for system-wide crashes however, can exploit the structure in the simultaneous failure to potentially achieve better efficiency guarantees.
The main result of this paper---a constant worst-case RMR algorithm for RME tolerating system-wide crashes---displays such an efficiency boost in light of Chan and Woelfel's $\Omega(\log n / \log\log n)$ worst-case RMR lower bound for RME in the individual crash model; here $n$ is the number of processes participating in the algorithm.

\subsection{RME Lock: Problem Statement}\label{stmt}

A standard mutex lock $\ell$ consists of two methods ---$\ell.\mbox{\em try}_p()$ and 
$\ell.\mbox{\em exit}_p()$ for each process $p$---and
a {\em recoverable} mutex (RME) lock consists of one additional method, called $\ell.\mbox{\em recover}_p()$.
Initially each process is in the {\em remainder section} and in a {\em correct} state.
From the remainder section, when in a correct state, a process $p$ may invoke and execute $\ell.\mbox{\em try}_p()$.
When $\ell.\mbox{\em try}_p()$ returns, $p$ is said to be in the {\em critical section (CS)}.
From the CS, $p$ next invokes and executes $\ell.\mbox{\em exit}_p()$.
When this method completes, $p$ is back in the remainder section and in a correct state.

At any time, a process may execute a {\em normal step} or a system-wide crash may occur.
In a normal step of a process $p$, $p$ executes the instruction pointed to by its program counter $\pcp$.
When a system-wide crash occurs, for each process $p$,  $p$ enters a {\em crashed state}, $\pcp$ is set to $p$'s remainder section, and all $p$'s other registers and local variables that are not in the NVM are set to arbitrary values.
When $p$ subsequently restarts, it is required to invoke $\ell.\mbox{\em recover}_p()$.
When this method completes, it returns either $\gotorem$ or $\gotocs$ and $p$ is said to move back from the {\em crashed} to the {\em correct} state.
A return value of $\gotorem$ by $\ell.\mbox{\em recover}_p()$ puts $p$ in the remainder section and a return value of $\gotocs$ puts $p$ in the CS.

Two fundamental properties of an RME lock are:

\begin{itemize}
    \item
	\underline{Mutual Exclusion}: At most one process is in the CS at any time.
	\item
	\underline{Critical Section Reentry} (CSR) \cite{Golab:rmutex}: If a process $p$ crashes while in the CS, no other process may enter the CS before $p$ reenters the CS. In other words, if $p$ crashes in the CS, the subsequent execution of $\ell.\mbox{\em recover}_p()$ that returns, must put $p$ back in the CS.
\end{itemize}

The CSR property is required because $p$ could be manipulating some shared data structure in the CS at the time of the crash, in which case the crash could leave the data structure in an inconsistent state.
The CSR property ensures that $p$ gets the opportunity to enter the CS before anyone else and clean up its act.
(We assume, as is standard in the RME literature, that the CS is idempotent.)

An RME lock must also ensure progress in {\em fair} runs---runs where no process stays in the CS forever and no process permanently stops taking steps, unless it is in the Remainder in a correct state.
Our algorithm satisfies the following liveness property:

\begin{itemize}
    \item
	\underline{Starvation Freedom}: In every fair run, if a process $p$ executes $\ell.\mbox{\em try}_p()$ and no crash occurs during this execution of the try method, then the method will eventually complete, leading to $p$ entering the CS.
\end{itemize}



Two other important properties are \cite{jayanti:fcfsmutex}:

\begin{itemize}
	\item
	\underline{Bounded Recovery}:	
	If a process $p$ executes $\ell.\mbox{\em recover}_p()$ without crashing, $p$ returns from the method in a constant number of its own steps.
	\item
	\underline{Bounded Exit}:	
	If a process $p$ executes $\ell.\mbox{\em exit}_p()$ without crashing, $p$ returns from the method in a constant number of its own steps.
\end{itemize}

Bounded Recovery ensures that if $p$ crashes while in the CS, once it restarts, it will be able to walk back into the CS without being obstructed by others \cite{jayanti:fcfsmutex}. 
Bounded Exit stipulates that there should be no hindrances when all that a process wants to do is to give up the lock.

\subsection{Dynamic Joining and Adaptive Space}

Previous RME locks were designed for a fixed set of $N$ processes labeled $1,\ldots,N$, and have required pre-allocated shared-arrays of length $O(N)$ or $O(N^2)$ for communication.
We design RME locks that allow {\em dynamic joining}, i.e., processes that are created on-the-fly  (a.k.a. {\em threads}) of arbitrary names can access our locks.
Thus, there is no pre-defined limit $N$ on the number of processes that may access our locks.
To achieve this property, we design all communication structures in our protocol to be pointer-based, so we can eliminate the need for pre-allocated fixed-length arrays.

Our space usage is {\em adaptive}, i.e., the space grows with the actual number of processes $n$ that use our lock
(in contrast to space being a function of a pre-defined maximum number of processes $N$ for which the lock is designed).

\subsection{RMR Complexity}

The complexity of mutual exclusion algorithms is commonly studied in two multiprocessor models:
{\em cache-coherent} (CC) and {\em distributed shared memory} (DSM).

In the CC model, each process has a cache.
A read operation by a process $p$ on a shared variable $X$ fetches a copy of $X$ from 
shared memory to $p$'s cache, if a copy is not already present in the cache.
Any non-read operation on $X$ by any process invalidates copies of $X$ at all caches.
An operation on $X$ by $p$ counts as a
{\em remote memory reference} (RMR) if either the operation is not a read
or $X$'s copy is not already present in $p$'s cache.
When a process crashes, its cache contents are lost.

In the DSM model, shared memory is partitioned between the processes.
Each shared variable $X$ resides in exactly one of the parts.
{\em Any} operation on $X$ by a process $p$ counts as an RMR if and only if $X$ is not in $p$'s part of shared memory.

A {\em passage} of a process $p$ in a run starts from the time it leaves the remainder (in the correct or the crashed state) to the earliest subsequent time when $p$ returns to the remainder
(in the correct or the crashed state).


The {\em passage complexity} or the {\em RMR complexity}
of an RME algorithm is the worst-case number of RMRs that a process incurs in a passage. 

\subsection{Our Contribution}

We design two RME algorithms that tolerate system-wide crashes.
Our first algorithm has $O(1)$ worst-case RMR complexity for CC systems.
Our second algorithm enhances the first, and has $O(1)$ worst-case RMR complexity for {\em both} CC and DSM systems.
Both algorithms require only $O(1)$ space per process, allow dynamic joining, and satisfy all of the properties stated in Section~\ref{stmt}---Mutual Exclusion, CSR, Starvation Freedom, Bounded Recovery, and Bounded Exit.
Both algorithms use just the standard {\em fetch-and-store (FAS)} and {\em compare-and-swap (CAS)} instructions for synchronization.\footnote{The operation $r \leftarrow \mbox{\bf FAS}(X, v)$ has the following semantics: if $X$'s value is $u$ immediately before the instruction, it writes $u$ in the CPU register $r$ and updates $X$'s value to $v$. The operation $r \gets \mbox{\bf CAS}(X, u, v)$ has the following semantics: 
if $X$'s value is $u$, the operation changes $X$'s value to $v$ and writes {\em true} in the CPU register $r$;
otherwise, $X$ remains unchanged and the operation writes {\em false} in $r$.}

\subsection{Comparison to the state of the art}

Four previous works explore the possibility of constant RMR solutions for RME locks \cite{Golab:rmutex2, jayanti:fasasmutex, Golab:rmutex3, Chan}.
Two of these works require hardware support for instructions that atomically manipulate two unrelated words of shared-memory---the {\em fetch-and-store-and-store} (FASAS) instruction and the {\em double-word-compare-and-swap} (DCAS) instruction---to achieve $O(1)$ worst-case RMR complexity in the individual process crash model \cite{Golab:rmutex2, jayanti:fasasmutex}.
To our knowledge, no current systems support these instructions.

Chan and Woelfel designed an RME algorithm for the independent crash model, which achieves {\em amortized} $O(1)$ RMR complexity and requires unbounded space \cite{Chan}.
A lower bound by the same authors shows that $O(1)$ {\em worst-case} RMR complexity cannot be achieved in the individual crash model \cite{ChanWoelfelLB}, and confirms that the worst-case $\Theta(\log n/\log\log n)$ RMR algorithms of Golab and Hendler \cite{Golab:rmutex2} and Jayanti et al. \cite{jayanti:rmesublog} are the best possible for CC and DSM in that model of computation, where $n$ is number of processes accessing the lock.

Golab and Hendler (GH) designed a pair of RME algorithms for the system-wide crash model, which assume that ``processes receive additional information from the environment regarding the occurrence of the failure,'' to achieve $O(1)$ worst-case RMR \cite{Golab:rmutex3}.
Specifically, the GH algorithms assume that processes have access to a shared failure-counter, which an out-of-band environmental mechanism must increase after each crash event.
Their CC algorithm can be implemented to allow dynamic joining and uses only constant space per process, like ours.  
Their DSM algorithm however, does not allow dynamic joining, requires a knowledge of the maximum number of processes $N$ that might access the lock, and uses $O(N)$ space per process, i.e., a total of $O(N^2)$ space.
Both our CC and DSM algorithms allow dynamic joining and use only $O(1)$ space per process, i.e., a total of $O(n)$ space, where $n$ is the actual number of processes that access the lock in the run (hence, $n$ can be much smaller than $N$).

The GH algorithms do not satisfy the Bounded Recovery and Bounded Exit properties, but satisfy a stronger version of Starvation Freedom, which states that 
in an infinite run with infinitely many super passages, every process that executes the try section eventually reaches the critical section.

To our knowledge, our algorithms are the only RME locks to achieve worst-case $O(1)$ RMR complexity assuming nothing more than standard hardware support. 
They are also the first to achieve the Bounded Recovery and Bounded Exit properties, alongside $O(1)$ RMR complexity, and the first to achieve $O(1)$ space per process in the DSM model.
Furthermore, in light of Chan and Woelfel's $\Omega(\log n / \log\log n)$ worst-case RMR lower bound for RME in the individual crash model, our results show a separation between the system-wide crash and individual crash models in worst-case RMR complexity in both the CC and DSM models (without assuming a failure detector).

\subsection{Related research on RME}

Golab and Ramaraju formalized the RME problem and designed several algorithms by adapting traditional mutual exclusion algorithms \cite{Golab:rmutex}.
Ramaraju \cite{ramaraju:rglock}, Jayanti and Joshi \cite{jayanti:fcfsmutex}, and Jayanti, Jayanti, and Joshi \cite{jayanti:fasasmutex} 
designed RME algorithms that support the First-Come-First-Served property \cite{Lamport:fcfsmutex}.
Golab and Hendler \cite{Golab:rmutex2} and Jayanti, Jayanti, and Joshi \cite{jayanti:rmesublog} presented RME algorithms that have sub-logarithmic RMR complexity.
Dhoked and Mittal gave a lock with sub-logarithmic RMR complexity that is additionally adaptive \cite{Mittal}.
A recent lower bound by Chan and Woelfel \cite{ChanWoelfelLB} matches the upper bounds of Jayanti, Jayanti, and Joshi \cite{jayanti:rmesublog} and Golab and Hendler \cite{Golab:rmutex2} to pin down the worst-case RMR complexity of RME in the individual crash model as $\Theta(\log n/\log\log n)$ for both the CC and DSM models.
Interestingly, Chan and Woelfel's previous work shows that, given unbounded space, $O(1)$ RMR complexity can be achieved in the {\em amortized} sense.
RME locks that are abortable have also been designed:
the first by Jayanti and Joshi uses CAS and has logarithmic RMR complexity \cite{jayanti:rmeabort}, and a subsequent one by Katzan and Morrison achieves sublogarithmic RMR complexity using CAS and Fetch\&Add \cite{KatzanMorrison2020}.
All these results apply to the individual crash model of failure.
As discussed above, Golab and Hendler \cite{Golab:rmutex3} presented an algorithm that has $O(1)$ RMR complexity in the system-wide crash model, assuming a failure detector.

\subsection{Organization}

In Section~\ref{sec:mainalg}, we present an algorithm for an RME lock that has $O(1)$ RMR complexity for only the CC model.
In Section~\ref{sec:dsmalg}, we adapt this algorithm to achieve $O(1)$ RMR complexity for both CC and DSM.
We conclude in Section~\ref{sec:conclusion}.


\section{RME lock for CC} 
\label{sec:mainalg}

In this section, we present an algorithm for implementing an RME lock $\L$ for the system-wide crash model.
This algorithm has $O(1)$ RMR complexity on CC machines, and is displayed in Figure~\ref{algo:fullcrash}.
We explain the ideas underlying its design in Section~\ref{sec:highlevel}, and provide a line-by-line commentary in Section~\ref{sec:commentary}.
We then provide an invariant based proof of correctness in Section~\ref{sec:correc}.


\subsection{High level ideas} \label{sec:highlevel}

The algorithm uses the following persistent (NVM) variables.
Some are shared and the others, subscripted with $p$, are local to process $p$.

\begin{itemize}
	\item
	\underline{$\seq$ and $\sp$}: $\seq$ stores a sequence number that grows monotonically,
	and the local variable $\sp$ holds what process $p$ believes to be the current sequence number.
	If $p$ crashes and subsequently restarts, it increases $\seq$ to $\sp +1$.
	
	\item
	\underline{$\lck[0]$, $\lck[1]$, $\lck[2]$}: 
	These are three instances of a standard (non-recoverable) mutex lock. Since our RME lock's properties will depend on the properties of these underlying ``base locks'', we instantiate these to be Lee's queue locks \cite{lee:twonodeme}, which we have presented in Figure~\ref{algo:qlock} in Appendix A, along with its properties.
	
	To compete with other processes in order to obtain the ownership of the CS, $p$ uses the lock numbered $\sp \% 3$.
	
	\item
	\underline{$\stopwait[0]$, $\stopwait[1]$, $\stopwait[2]$}:
	These are boolean variables and the variable $\stopwait[\sp \% 3]$ denotes whether $\lck[\sp \% 3]$ is in good condition or not.
	That is to say that, if a crash occurred while a process was actively using the lock $\lck[\sp \% 3]$ and the lock was not subsequently reset, 
	then the lock is not in good condition and hence should not be used to gain access to the CS.
	
	\item
	\underline{$\cso$}: 
	This variable stores the name of the process in the CS of the implemented recoverable lock $\L$ (i.e., the real CS and not that of any of the three base lock).
	Its value is $\bot$ if no process is in the CS of $\L$.
	
	\item
	\underline{$\inuse_p$}: This boolean local variable holds {\em true} while $p$ executes the algorithm.
	Therefore, if $p$ jumps to the Remainder section because it crashed while executing the algorithm,
	$\inuse_p$ has $\true$.
	On the other hand, if $p$ completes the algorithm normally (e.g., without crashing),
	$\inuse_p$ has $\false$ when $p$ reaches the Remainder section.
	
%
%
\end{itemize}

The idea is that, in order to compete for the recoverable lock $\L$,
each process $p$ reads into $\sp$ the sequence in $\seq$ and attempts to acquire 
the base lock $\lck[\sp \% 3]$ (by executing $\lck[\sp \% 3].\tryprocp()$).
For example, if processes $p$ and $q$ compete for $\L$ when the sequence number is 10,
they try to acquire $\lck[10 \% 3]$ (or $\lck[1]$).
If there are no crashes, then each process acquires $\lck[1]$, enters the CS of $\L$,
and then releases $\lck[1]$.
Thus, $p$ and $q$ enter the CS of $\L$ one after the other and return to the Remainder section.

For a more complex scenario, suppose that the system crashes while $p$ and $q$ are competing for $\lck[1]$.
The crash wipes out these process' CPU registers, thereby rendering $\lck[1]$ unusable in the future.
When one of these processes, say $p$, subsequently restarts and enters $\recoverprocp()$ of $\L$,
it infers from the value of $\true$ in $\inuse_p$ that it must have crashed while executing the algorithm.
So, $p$ advances $\seq$ by writing 11 and goes on to acquire $\lck[11 \% 3]$, or $\lck[2]$, if necessary.
However, suppose that a third process $r$ executes the algorithm before $p$ changes $\seq$ to 11.
Since $r$ has no previous context, when it reads 10 in $\seq$, it simply writes 10 in $\myseq_r$
and competes for $\lck[10 \% 3]$ (i.e., in $\lck[1]$), which could be stuck, but then $r$ doesn't know that.
If $p$ now takes steps, it writes 11 in $\seq$ and proceeds to compete for $\lck[11 \% 3]$ (i.e., $\lck[2]$).
Thus, at this point, there are processes waiting at two different locks, namely, $\lck[1]$ and $\lck[2]$.
To prevent $r$ from being stuck forever at $\lck[10 \% 3]$, 
we require that while waiting in the $\tryproc{r}()$ procedure of $\lck[10 \% 3]$, $r$ parallely also checks whether $\stopwait[10 \% 3]$ changes to $\true$.
On noticing a change in $\stopwait[10 \% 3]$, $r$ can promptly switch to $\lck[11 \% 3]$.

Could there be a scenario where processes could be waiting at three locks?
The answer is no.
To see this, suppose that a system-wide crash occurs when $\seq$ has some value $v$.
Processes that have not witnessed the value $v$ in $\seq$ in the past (such as process $r$ in the previous scenario)
as well as processes that saw $v$ in $\seq$ but exited the algorithm normally
will proceed to $\lck[v \% 3]$, while those that had seen $v$ in $\seq$ and experienced the crash
write $v+1$ in $\seq$ and wait at  $\lck[(v+1) \% 3]$.
Thus, there is no possibility of any process flocking to a third lock.
In particular, of the three base locks employed in the algorithm, we can be certain that
no process waits at $\lck[(v-1) \% 3]$.
Since this lock could be stuck from prior crashes, now is the ripe time to reset and keep it ready for future use;
so, $p$ resets $\lck[(v-1) \% 3]$ as it changes $\seq$ from $v$ to $v+1$.
This observation that three base locks suffice
and that $\lck[(v-1) \% 3]$ should be reset when changing $\seq$ from $v$ to $v+1$
is a crucial insight in our algorithm.


\setcounter{linecounter}{0}
\begin{figure}[!ht]
	\begin{footnotesize}
		\hrule
		\vspace{-3mm}
		\begin{tabbing}
			\hspace{0in} \=  \hspace{0.2in} \= \hspace{0.1in} \=  \hspace{0.2in} \= \hspace{0.2in} \= \hspace{0.2in} \=\\
			{\bf Shared variables (stored in NVM)} \\ 
			\hspace{0in} \=  \hspace{0.1in} \= $\seq \in \mathbb{N}$, initially $1$. \\
			\> \> $\mlock[0 \cdots 2]$ is an array of base mutual exclusion locks, as implemented in Figure~\ref{algo:qlock}. \\
			\> \> $\stopwait[0 \cdots 2]$ is an array of booleans, each initially {\em false}. \\
			\> \> $\barrier$ stores a process identifier or $\none$, initially $\none$.  \\
			\> {\bf Persistent variables local to process $p$ (stored in NVM)} \\
			\> \> $\inusep$ is a boolean, initially {\em false}. \\
			\> \> $\myseqp \in \mathbb{N}$, initially $1$. 
		\end{tabbing}
		
		\begin{minipage}[t]{.95\linewidth}
			\begin{tabbing}
				\hspace{0in} \= \hspace{0.25in} \= \hspace{0.2in} \=  \hspace{0.2in} \= \hspace{0.2in} \=\\
				\> \procline{fullcrash:rem:1} \> {\texttt{Remainder Section}} \\
				\\
				\> \> \underline{\texttt{procedure $\tryprocp()$}}\\
				\> \procline{fullcrash:try:1}\> $\inusep \gets \true$ \\
				\> \procline{fullcrash:try:2} \> $\myseqp \gets \seq$ \\
				\> \procline{fullcrash:try:3} \> $\mlock[\myseqp \% 3].\tryprocp()$ \hspace{.2in} $\parallel$ \hspace{0.1in} \waittill $\stopwait[\myseqp \% 3] = \true$  \\
				\> \procline{fullcrash:try:4} \> \ifcode $\seq \neq \myseqp$: \goto Line~\refln{fullcrash:try:8} \\
				\> \procline{fullcrash:try:5} \> \waittill $\csowner \isequal \perp$ \hspace{.1in} $\parallel$ \hspace{.1in} \waittill $\stopwait[\myseqp \% 3] = \true$ \\
				\> \hspace{1.55in} \goto Line~\refln{fullcrash:try:8} \\
				\> \procline{fullcrash:try:6} \> \ifcode $\cas(\csowner, \perp, p)$: \return \gotocs \\
				\> \procline{fullcrash:try:7} \> \ifcode $\seq \neq \myseqp$: \\
				\> \procline{fullcrash:try:8} \> \> $\myseqp \gets \myseqp + 1$ \\
				\> \procline{fullcrash:try:9} \> \> $\mlock[\myseqp \% 3].\tryprocp()$ \\
				\> \procline{fullcrash:try:10} \> \> \waittill $\csowner \isequal \perp$ \\
				\> \procline{fullcrash:try:11} \> \> \ifcode $\cas(\csowner, \perp, p)$: \return \gotocs \\
				\> \procline{fullcrash:try:12} \> \waittill $\csowner \isequal \perp$ \\
				\> \procline{fullcrash:try:13} \> $\csowner \gets p$ \\
				\> \procline{fullcrash:try:14} \> \return \gotocs \\
				\\
				\> \procline{fullcrash:cs:1} \> {\texttt{Critical Section}} \\
				\\
				\> \> \underline{\texttt{procedure $\exitprocp()$}}\\
				\> \procline{fullcrash:exit:1} \> \ifcode $\myseqp \isequal \seq$: \\
				\> \procline{fullcrash:exit:2} \> \> $\mlock[\myseqp \% 3].\exitprocp()$ \\
				\> \procline{fullcrash:exit:3} \> $\csowner \gets \none$ \\
				\> \procline{fullcrash:exit:4} \> $\inusep \gets \mbox{\it false}$ \\
				\\
				\> \> \underline{\texttt{procedure $\recoverprocp()$}}\\
				\> \procline{fullcrash:rec:1} \> \ifcode $\inusep \wedge \seq \isequal \myseqp$: \\
				\> \procline{fullcrash:rec:2} \> \> $\mlock[(\myseqp - 1) \% 3].\initializep()$ \\ 
				\> \procline{fullcrash:rec:3} \> \> $\stopwait[(\myseqp - 1) \% 3] \gets \mbox{\it false}$ \\
				\> \procline{fullcrash:rec:4} \> \> $\seq \gets \myseqp + 1$ \\
				\> \procline{fullcrash:rec:5} \> \> $\stopwait[\myseqp \% 3] \gets \true$ \\
				\> \procline{fullcrash:rec:6} \> \ifcode $\csowner \isequal p$: \return \gotocs \\
				\> \procline{fullcrash:rec:7} \> $\inusep \gets \mbox{\it false}$ \\
				\> \procline{fullcrash:rec:8} \> \return \gotorem
			\end{tabbing}  
		\end{minipage}
		\vspace*{-2mm}
		\captionsetup{labelfont=bf}
		\caption{Algorithm for an RME lock $\L$ for an arbitrary number of processes of arbitrary names, for CC machines. Code shown for a process $p$.} 
		\label{algo:fullcrash}
		\hrule
	\end{footnotesize}
\end{figure}

\subsection{Line-by-line commentary} \label{sec:commentary}
In this section we informally describe the working of our algorithm presented in Figure~\ref{algo:fullcrash}.
We first describe how a process $p$ would execute the algorithm in the absence of a crash, 
and then proceed to explain the working of the algorithm in presence of a crash.

\subsubsection*{Crash-free attempt.}
A process $p$ starts an attempt to enter the CS from the Remainder Section by invoking the $\tryprocp()$ procedure.
At Line~\refln{fullcrash:try:1}, $p$ sets $\inusep$ to $\true$ making a note to itself that it has started an attempt.
Subsequently, $p$ reads the current sequence number $\seq$ into its local variable $\myseqp$ at Line~\refln{fullcrash:try:2}, and proceeds by trying to acquire the corresponding lock $\mlock[\myseqp \% 3]$ at Line~\refln{fullcrash:try:3}.
While executing $\lck[\myseq_p \% 3].\tryprocp()$, 
$p$ simultaneously monitors the value of the corresponding stop flag, $\stopwait[\myseqp\% 3]$, by interleaving its steps between the try method and the \waittill at Line~\refln{fullcrash:try:3}.
If $\myseqp\%3$ is indeed the current lock, then $p$'s try attempt will eventually succeed.
Otherwise, if another process $q$ that previously crashed while active in the lock with sequence number $\myseqp$ restarts, it will increment $\seq$ and set the flag  $\stopwait[\myseqp\% 3]$ to {\em true}.
When $p$ finishes Line 4, it could have been because it successfully obtained the lock or because the stop flag was raised.
Thus, it checks whether the sequence number $\seq$ was incremented at Line 5, if so it abandons the $\lck[\myseqp\%3]$ and proceeds to update its sequence number at Line 9 and proceeds to try to obtain the newly installed lock (as we explain later).
Otherwise, if $\myseqp$ is still current (i.e. it still equals $\seq$), then $p$ waits for the CS to become freed by spinning on $\csowner$ at Line 6.
Once again, $p$ is aware that the sequence number could be updated, so it simultaneously monitors $\stopwait[\myseqp\%3]$ and (just as before) proceeds to Line 9 if the stop flag gets raised.
If $p$ eventually finds the CS vacated (i.e. $\csowner = \perp$ at Line~\refln{fullcrash:try:5}), it tries to establish ownership of the CS by CASing its name into the $\csowner$ field at Line~\refln{fullcrash:try:6}.
If $p$ succeeds in the \cas, then it moves to the CS.
Otherwise, as we explain in the following, there could be two reasons that $p$ failed in the \cas\ after having read that the CS is empty at Line~\refln{fullcrash:try:5}.
In the first case, the lock used by $p$ must have been an older lock that was part of a crash before $p$ enqueued itself into it,
and subsequently a new lock was installed that allowed another process to enter the CS (all of this not noticed by $p$).
In the second case $p$ itself was enqueued in the latest lock, however, it lost to another process $q$ that had enqueued itself into a older lock (just like $p$ in the first case above).
To find out which of the above two cases caused this \cas\ failure, $p$ compares $\seq$ with $\myseqp$ at Line~\refln{fullcrash:try:7}.
If the two are the same, $p$ knows it is in the right lock, hence, it goes to Line~\refln{fullcrash:try:12} to wait for its turn to occupy the CS.
Otherwise, $p$ tries for the latest active base lock next, hence, it goes to Line~\refln{fullcrash:try:8}.

From the above description, we see that $p$ could reach Line~\refln{fullcrash:try:8} in three different ways: from Line~5, Line~6, or Line~8. 
In all the cases, $p$ knows that the sequence number increased after its execution of Line 3.
Since, $p$ knows that the sequence number could go up only once between two crashes (an invariant maintained by our algorithm), it increments $\myseqp$ by 1 (Line~\refln{fullcrash:try:8}) to the current value of $\seq$.
This time, when $p$ tries for the lock $\lck[\myseqp\%3]$ at Line~10, it is guaranteed to eventually succeed (if there is no system-wide crash), and thus does not need to simultaneously monitor a stop flag.
At Line~\refln{fullcrash:try:10} $p$ waits to ensure that the CS is empty.
It needs to ensure this for one of two reasons: either a process acquired the CS before the crash and still continues to do so,
or a process succeeded in the \cas\ at Line~\refln{fullcrash:try:6} by coming through a lock before it was replaced by the current lock that $p$ came from.
Once $p$ is past Line~\refln{fullcrash:try:10}, it attempts to \cas\ its own name into $\csowner$ if the CS is still empty.
If $p$ succeeds in the \cas, then it moves to the CS.
Otherwise $p$ failed only because some other process must have succeeded in the \cas\ at Line~\refln{fullcrash:try:6} as described above.
$p$ waits one more time at Line~\refln{fullcrash:try:12} to ensure that the CS is empty.
Once $p$ reaches Line~\refln{fullcrash:try:13} it is sure that it reached the line by enqueuing into the latest active lock and there could not be any more competition to it.
Hence it writes its own name into $\csowner$ and moves to CS (Line~\refln{fullcrash:try:14}).

In the $\exitprocp()$ procedure, $p$ first checks if the base lock that it queued up and entered the CS from is still the active lock (Line~\refln{fullcrash:exit:1}).
If it is indeed still the active lock, $p$ executes the $\exitprocp()$ procedure of that lock at Line~\refln{fullcrash:exit:2}.
Otherwise, $p$ knows that it is no longer in the queue of the lock it used and hence doesn't execute the $\exitprocp()$ method of that lock.
At Line~\refln{fullcrash:exit:3} $p$ informs other processes that it is giving up the CS by writing $\perp$ to $\csowner$.
Finally, at Line~\refln{fullcrash:exit:4} $p$ sets $\inusep$ to $\false$ to complete its attempt.

\subsubsection*{Recovery from a crash.}
We now discuss what happens when $p$ crashes while executing an attempt.
The idea behind our algorithm is to keep an active instance of the base lock and the number in $\seq$ determines which of the three instances is active.
Therefore, in any configuration, $\mlock[\seq \% 3]$ is the active base lock.
This idea guides a process on whether after a crash it needs to setup a new lock or not.
Thus, at Line~\refln{fullcrash:rec:1}, $p$ determines if it took any meaningful steps prior to the crash, and if so, whether it might have attempted to queue up in the currently active base lock.
If $\inusep = \false$, $p$ is sure that it didn't take any meaningful steps before the crash and hence continues to Line~\refln{fullcrash:rec:6} of the procedure
(from there it is sure to find $\csowner \neq p$ and thus will return to the Remainder completing the attempt).
If $\inusep = \true$ but $\myseqp \neq \seq$, then $p$ knows that it no longer needs to setup a new lock because either a new lock was already setup
or the currently active lock is not the one that $p$ queued itself into.
Thus, $p$ would continue to Line~\refln{fullcrash:rec:6}.
In the case where $\inusep = \true$ and $\myseqp = \seq$, $p$ might have enqueued itself into $\mlock[\seq \% 3]$ at either of Lines~\refln{fullcrash:try:3} or \refln{fullcrash:try:9}.
Hence, assuming that $p$ itself has broken the queue structure of $\mlock[\seq \% 3]$, it moves to setup a new lock at Lines~\refln{fullcrash:rec:2}\upto\refln{fullcrash:rec:5}.
Our algorithm maintains the invariant that $\mlock[(\seq + 1) \% 3]$ is always initialized, i.e., the next instance of $\mlock$ is always kept ready to use.
Therefore, before installing the next lock in sequence, $p$ readies the lock after the next lock in sequence to maintain the invariant.
To this purpose, at Line~\refln{fullcrash:rec:2}, $p$ first resets the lock $\mlock[(\myseqp - 1) \% 3]$, 
which is the lock coming up after the next lock in sequence (i.e., lock numbered $(\seq + 2) \% 3$ or $(\seq - 1) \% 3$ is the lock coming up after the next lock in sequence).
$p$ then resets the $\stopwait$ flag associated with lock $\mlock[(\myseqp - 1) \% 3]$ at Line~\refln{fullcrash:rec:3}.
Next $p$ moves to signal that $\lck[\myseqp \% 3]$ should no longer be used, therefore, it increments $\seq$ by $1$ at Line~\refln{fullcrash:rec:4}.
It then sets $\stopwait[\myseqp \% 3]$ to $\true$ (at Line~\refln{fullcrash:rec:5}) so that any processes that queued up at lock $\lck[\myseqp \% 3]$ in the time between the previous crash and now know that they should move to the next lock in sequence, i.e., $\lck[\seq \% 3]$.
An invariant our algorithm maintains is that between two consecutive crashes, the value of $\seq$ is incremented at most once.
To see how that is achieved consider the following example. 
Suppose $p$, $q$, and $r$ committed to the value 10 by reading $\seq$ at Line~\refln{fullcrash:try:2} and noting into their respective variables $\myseqp$, $\myseq_q$, and $\myseq_r$.
Assume that a crash occurs after that and $p$ and $q$ restart immediately actively taking steps.
$p$ and $q$ both read $\seq = \myseq_p$ and $\seq = \myseq_q$ and thus execute Lines~\refln{fullcrash:rec:2}-\refln{fullcrash:rec:3} one after another.
After that $q$ executes Line~\refln{fullcrash:rec:4} first to set $\seq$ to 11 for the first time.
Following this, $p$ will execute Line~\refln{fullcrash:rec:4} at most once before the next crash, 
and that step wouldn't change the value of $\seq$ because $p$ would attempt to set it to 11, a value $\seq$ already holds. 
Also, following the step by $q$, no matter when $r$ starts taking steps, it will always find $\seq = 11$ which is not the same as what $\myseq_r$ held prior to the crash.
Hence, $r$ would never change the value of $\seq$.
Thus, with the execution of Line~\refln{fullcrash:rec:5}, $p$ completes the repair and moves to execute Line~\refln{fullcrash:rec:6}.

At Line~\refln{fullcrash:rec:6} $p$ checks if it acquired access to the CS prior to the crash by checking if $\csowner = p$.
If so, $p$ moves to the CS  by returning $\gotocs$ from $\recoverprocp()$.
Otherwise, $p$ is sure that it has repaired any issues that might have caused due to the crash and it can't go in the CS any more in the current attempt.
Thus it sets the $\inusep$ flag to $\false$ (Line~\refln{fullcrash:rec:7}) to signal the end of the attempt and returns $\gotorem$ (Line~\refln{fullcrash:rec:8}) to go back to the Remainder.

In the above, Line~\refln{fullcrash:rec:6} may put $p$ into the CS, hence satisfying the CSR property.
Thus, in the $\exitprocp()$ procedure $p$ checks if it entered the CS normally or due to the CSR property as mentioned above.
If $p$ finds that $\seq = \myseq_p$ at Line~\refln{fullcrash:exit:1}, it infers that it entered the CS normally, 
and hence executes the $\exitprocp()$ procedure of $\mlock[\myseq_p \% 3]$ at Line~\refln{fullcrash:exit:2}.
Otherwise, $p$ knows that it is no longer in the queue of the lock it used and hence doesn't execute $\exitprocp()$.
At Line~\refln{fullcrash:exit:3} $p$ informs other processes that it is giving up the CS by writing $\perp$ to $\csowner$.
Finally, at Line~\refln{fullcrash:exit:4} $p$ sets $\inuse_p$ to $\false$ to complete its attempt.


\section{Correctness of the algorithm} 
\label{sec:correc}

The proof of correctness is based on the inductive invariant of the algorithm, which we present in Figure~\ref{inv:fullcrash}.
For analysis purposes, we introduce an abstract variable $status_p$ for each process $p$ (only for the proof).
$status_p$ reflects what section of the protocol $p$ last crashed in (if at all).
Specifically, if $p$ crashes in the try section, CS, or exit section, $status_p$ is set to {\em recover-from-try, recover-from-CS, or recover-from-exit} respectively.
When $p$ subsequently executes the recover method to completion, it attains $good$ status again.
When in $good$ status, if $p$ calls the recover method, $status_p$ is set to {\em recover-from-rem}.

In Appendix~\ref{app:correc}, we prove this invariant by induction and use it to prove the the main result for CC machines.
Appendix~\ref{app:correc} is composed of several subsections:
\begin{itemize}
\item
In Section B.2, we prove the lemmas that establish that the three base locks are used correctly, i.e., respecting the use pattern.
\item
In Section B.3, we use the invariant to prove the properties of the algorithm, i.e., Mutual Exclusion, Starvation Freedom, CSR, Bounded Recovery, and Bounded Exit.
\item
In Section B.4, we argue that the RMR complexity of the algorithm is $O(1)$.
\item
In Appendix C, we provide the full inductive proof of the invariant.
\end{itemize}

To give a flavor of how we prove the properties using the invariant, we reproduce the proofs of Mutual Exclusion and Starvation Freedom below.
These proofs of course rely on other lemmas in the appendix, which are appropriately referenced.
Finally, we end the section with the statement of the main theorem that summarizes the results about the CC algorithm.\\

\noindent
{\bf Lemma 15 (Mutual Exclusion).} {\em
At most one process is in the CS in any configuration of any run.
}
\begin{proof}
	Assume for a contradiction that there are two processes $p$ and $q$ in the CS in the same configuration.
	Therefore, $\pcp = \refln{fullcrash:cs:1}$ and $\pc{q} = \refln{fullcrash:cs:1}$ in the same configuration.
	By Condition~\ref{inv:fullcrash:cond9} we have $\csowner = p$ as well as $\csowner = q$ in the same configuration.
	Thus $\csowner$ has two different values in the same configuration, a contradiction.
\end{proof}

\noindent
{\bf Lemma 17 (Starvation Freedom).} {\em
At most one process is in the CS in any configuration of any run.
In every fair run, if a process $p$ executes the try method and no crash occurs during this execution of the try method, then the method will eventually complete, leading to $p$ entering the CS.
}
\begin{proof}
Since there are only finitely many crash steps in the run,
for the purpose of the argument take a run and pick the earliest time $\tau$ in the run such that all the crashes have occurred by $\tau$.
We need to show that if a process $p$ invokes $\tryprocp()$ at a time $t > \tau$,
it is in the CS at some time $t' > t$.
From an inspection of the algorithm we note that when the method $\tryprocp()$ returns, 
it puts $p$ in the CS because every return statement returns the value $\gotocs$.
Thus, we need to show that process $p$ doesn't forever get stuck at Lines~\refln{fullcrash:try:3}, \refln{fullcrash:try:4}\upto\refln{fullcrash:try:5},
\refln{fullcrash:try:9}, \refln{fullcrash:try:10}, or \refln{fullcrash:try:12}, which will ensure that $p$ does return from $\tryprocp()$.
By Lemmas~\ref{lem:fullcrash:nowait1}, \ref{lem:fullcrash:nowait2}, and \ref{lem:fullcrash:nowait3} we know that $p$ eventually gets past the 
Lines~\refln{fullcrash:try:4}\upto\refln{fullcrash:try:5}, \refln{fullcrash:try:10}, and \refln{fullcrash:try:12} respectively.
When $\pcp = \refln{fullcrash:try:3}$, by Condition~\ref{inv:fullcrash:cond11}, $\myseq_p \in \{ \seq - 1, \seq \}$.
If $\myseq_p = \seq - 1$, by Condition~\ref{inv:fullcrash:cond15}, $\stopwait[\myseq_p \% 3] = \true \vee (\exists q, \pc{q} = \refln{fullcrash:rec:5} \wedge \myseq_{q} = \myseq_p)$.
In either case, $\stopwait[\myseq_p \% 3] = \true$ eventually, which $p$ notices at the wait loop of Line~\refln{fullcrash:try:3} and goes past the line.
Hence, we assume that whenever $p$ executes Lines~\refln{fullcrash:try:3} or \refln{fullcrash:try:9}, $\myseq_p = \seq$, 
because that is the only other possibility by Condition~\ref{inv:fullcrash:cond11}.
Therefore, we will argue next that the starvation freedom property of $\lock[\seq \% 3]$ is satisfied,
which will imply that $p$ gets past Lines~\refln{fullcrash:try:3} or \refln{fullcrash:try:9}.
We know from an inspection of the algorithm that any process that were to execute the $\initialize{}()$ at Line~\refln{fullcrash:rec:2},
would do so for $\mlock[\seq - 1]$ or $\mlock[\seq - 2]$, because by Condition~\ref{inv:fullcrash:cond11} $\myseq_p \in \{ \seq - 1, \seq \}$ for any process $p$ with $\pcp = \refln{fullcrash:rec:2}$.
It follows that from the last crash onwards, no process will execute $\mlock[\seq \% 3].\initialize{}()$, meeting the first condition for starvation freedom on the base lock.
It is straightforward that the second condition for starvation freedom on the base lock is met.
Lemmas~\ref{lem:fullcrash:nowait1}, \ref{lem:fullcrash:nowait2}, and \ref{lem:fullcrash:nowait3} ensure that the last condition for starvation freedom on the base lock is also met.
Hence we know that $\lock[\seq \% 3]$ satisfies starvation freedom.
Therefore, a process $p$ gets past Lines~\refln{fullcrash:try:3} or \refln{fullcrash:try:9} by the starvation freedom property of the base lock.
It follows that the claim holds.
\end{proof}

\makeatletter
\newcommand{\invlab}[2]{{\bf #2.}\def\@currentlabel{{\bf #2}}\label{inv:cond#1}}
\makeatother

\begin{figure}[!ht]
	\begin{footnotesize}
		\hrule
		\vspace{-3mm}
		\begin{tabbing}
			\hspace{0in} \=  \hspace{0.05in} \= \hspace{0.1in} \=  \hspace{0.2in} \= \hspace{0.2in} \= \hspace{0.2in} \=\\
			{\bf Definitions :} \\
			\>$\bullet$\> $\forall i \in [0, 2], \mlock[i].\lockstate = (\mlock[i].\tryset, \mlock[i].\csset, \mlock[i].\exitset)$. \\
			\>$\bullet$\> $\forall i \in [0, 2], \mlock[i].\set= \mlock[i].\tryset \cup \mlock[i].\csset \cup \mlock[i].\exitset$. \\
			{\bf Conditions :} \\
		\end{tabbing}
		\vspace{-4mm}
		\begin{enumerate}
			\item \cond{1} $1 \leq \myseqp \leq \seq$ 
			
			\item  \cond{2} $\mlock[(\seq + 1) \% 3].\lockstate = (\phi, \phi, \phi)$ $\wedge$ $(\pcp \in \{ \refln{fullcrash:rec:3}, \refln{fullcrash:rec:4} \}$ $\limplies$ 
			$\mlock[(\myseq_p - 1) \% 3].\lockstate = (\phi, \phi, \phi))$
			
			\item \cond{3} $\stopwait[\seq \% 3] = \false$ $\wedge$ $\stopwait[(\seq + 1) \% 3] = \false$ $\wedge$ $(\pcp = \refln{fullcrash:rec:4} \limplies \stopwait[(\myseq_p - 1) \% 3] = \false)$ 
			
			\item  \cond{4} $((\inuse_p = \false \vee \myseqp < \seq \vee \pcp \in \{ \refln{fullcrash:try:2},  \refln{fullcrash:exit:3}, \refln{fullcrash:exit:4}, \refln{fullcrash:rec:6}, \refln{fullcrash:rec:7} \})$ $\limplies$ $p \notin \mlock[\seq \% 3].\set)$ \\
			\hspace*{3mm} $\wedge$ $(\myseqp < \seq - 1 \limplies p \notin \mlock[\myseqp \% 3].\set)$
			
			\item  \cond{5} $\forall i \in [0,2]$, $p$ is in at most one of $\mlock[i].\tryset$, $\mlock[i].\csset$, or $\mlock[i].\exitset$. 
			
			\item \cond{6} $\pcp \in \{ \refln{fullcrash:try:3}, \refln{fullcrash:try:9} \}$ $\limplies$  $p \in \mlock[\myseq_p \% 3].\tryset$ 
			
			\item  \cond{7} $((\pcp = \refln{fullcrash:try:4} \wedge \stopwait[\myseq_p \% 3] = \false) \vee \pcp \in \{ \refln{fullcrash:try:5} \upto \refln{fullcrash:try:7}, \refln{fullcrash:try:10} \upto \refln{fullcrash:try:14} \}$ 
			$\vee$ $(\pcp \in \{ \refln{fullcrash:cs:1}, \refln{fullcrash:exit:1} \} \wedge \myseq_p = \seq))$  \\
			\hspace*{15mm} $\limplies$  $p \in \mlock[\myseqp \% 3].\csset$ 			
			
			\item \cond{8} $\pcp = \refln{fullcrash:exit:2}$ $\limplies$ $p \in \mlock[\myseq_p \% 3].\exitset$
			
			\item  \cond{9} $((\inuse_p = \false \vee \pcp \in \{ \refln{fullcrash:try:1} \upto \refln{fullcrash:try:12}, \refln{fullcrash:exit:4}, \refln{fullcrash:rec:7} \}) \limplies \csowner \neq p)$  
			$\wedge$ $(\pcp = \refln{fullcrash:try:13} \limplies \csowner = \perp)$ \\
			\hspace*{3mm} $\wedge$ $((\pcp \in \{ \refln{fullcrash:try:14} \upto \refln{fullcrash:exit:3} \} \vee \status_p = \reccs) \limplies \csowner = p)$
			
			\item \cond{10} $(\pcp \in \{ \refln{fullcrash:try:2} \upto \refln{fullcrash:exit:4}, \refln{fullcrash:rec:2} \upto \refln{fullcrash:rec:5} \} \limplies \inuse_p = \true)$ \\
			\hspace*{3mm} $\wedge$ $(((\pcp \in \{ \refln{fullcrash:rem:1}, \refln{fullcrash:rec:1} \upto \refln{fullcrash:rec:7} \} \wedge \status_p \in \{ \good, \recrem \} ) \vee \pcp \in \{ \refln{fullcrash:try:1}, \refln{fullcrash:rec:8} \} ) \limplies \inuse_p = \false)$
			
			\item  \cond{11} $(\pcp \in \{ \refln{fullcrash:try:3} \upto \refln{fullcrash:try:7}, \refln{fullcrash:exit:2}, \refln{fullcrash:rec:2} \upto \refln{fullcrash:rec:4} \} \limplies \myseqp \in \{ \seq - 1, \seq \})$ 
			$\wedge$ $(\pcp \in \{ \refln{fullcrash:try:8}, \refln{fullcrash:rec:5} \} \limplies \myseqp = \seq - 1)$ \\
			\hspace*{3mm} $\wedge$ $(\pcp \in [\refln{fullcrash:try:9}, \refln{fullcrash:try:14}] \limplies \myseqp = \seq)$ 
			$\wedge$ $( (\pcp = \refln{fullcrash:rec:6} \wedge \csowner = p) \limplies \myseq_p < \seq )$ 

			\item \cond{15} This condition says that if $\seq$ is already incremented, then $\stopwait[(\seq - 1) \% 3]$ is already $\true$ or some process is poised to set it to $\true$. \\
			$(\pcp \in \{ \refln{fullcrash:try:3} \upto \refln{fullcrash:try:5} \} \wedge \myseq_p = \seq -1)$ $\limplies$ $(\stopwait[\myseq_p \% 3] = \true \vee (\exists q, \pc{q} = \refln{fullcrash:rec:5} \wedge \myseq_{q} = \myseq_p))$ 

			\item  \cond{12} This condition is useful in proving that once $\seq$ is incremented, it will not be incremented again prior to the next crash. 
			The argument follows from the fact that once a process is in one of certain states in the run that can only occur after $\seq$ is incremented,
			there will be no other process that will eventually increment $\seq$ once more prior to the next crash. 
			
			$((\pcp \in \{ \refln{fullcrash:try:3} \upto \refln{fullcrash:try:6}, \refln{fullcrash:exit:2}, \refln{fullcrash:rec:2}\upto\refln{fullcrash:rec:5} \} \wedge \myseqp = \seq - 1)$ 
			$\vee$ $(\pcp = \refln{fullcrash:try:6} \wedge \csowner \neq \perp)$ 
			$\vee$ $\pcp \in \{ \refln{fullcrash:try:7} \upto \refln{fullcrash:try:14} \} )$ \\
			\hspace*{15mm} $\limplies$ $\forall q, (\neg ( \pc{q} \in \{ \refln{fullcrash:rem:1}, \refln{fullcrash:rec:1} \}$ $\wedge$ $\inuse_{q} = \true$ $\wedge$ $\myseq_{q} = \seq) \wedge (\pc{q} \in \{ \refln{fullcrash:rec:2} \upto \refln{fullcrash:rec:5} \} \limplies \myseq_{q} = \seq - 1))$

			\item \cond{14} This condition essentially means that there can be at most two working queues coming out of the three base locks. 
			Moreover, once a process $q$ from an older queue goes past Line~\refln{fullcrash:try:6}, that queue freezes for the remaining run upto the next crash.
			
			$(\forall q, (\pcp \in \{\refln{fullcrash:try:6}, \refln{fullcrash:try:7}, \refln{fullcrash:try:12} \} \wedge \myseq_p = \seq \wedge \csowner = q)$ \\
			\hspace*{15mm} $\limplies$ $(\myseq_{q} = \seq - 1 \wedge q \in \lock[\myseq_{q} \% 3].\csset$ \\
			\hspace*{25mm} $\wedge$ $\forall r, (q \neq r \wedge \myseq_{r} = \myseq_{q})$ \\
			\hspace*{35mm} $\limplies$ $(\pc{r} \in \{ \refln{fullcrash:rem:1} \upto \refln{fullcrash:try:3}, \refln{fullcrash:try:8}, \refln{fullcrash:exit:4}, \refln{fullcrash:rec:1} \upto \refln{fullcrash:rec:8} \}$
			$\vee$ $( \pc{r} = \refln{fullcrash:try:4} \limplies \stopwait[\myseq_r  \% 3] = \true)))$ \\
			\hspace*{3mm} $\wedge$ $(((\pcp \in \{ \refln{fullcrash:try:7}, \refln{fullcrash:try:12} \} \wedge \myseq_p = \seq \wedge \csowner = \perp) \vee \pcp \in \{ \refln{fullcrash:try:13}, \refln{fullcrash:try:14} \} )$ \\
			\hspace*{15mm} $\limplies$ $(\forall r, (\myseq_r \neq \seq - 1 \vee \pc{r} \in \{ \refln{fullcrash:rem:1} \upto \refln{fullcrash:try:3}, \refln{fullcrash:try:8}, \refln{fullcrash:exit:4}, \refln{fullcrash:rec:1} \upto \refln{fullcrash:rec:8} \}$
			$\vee$ $( \pc{r} = \refln{fullcrash:try:4} \limplies \stopwait[\myseq_r  \% 3] = \true))$ \\
			\hspace*{21mm} $\wedge$ $\exists q, q \in \lock[(\seq - 1) \% 3].\csset))$ 
		\end{enumerate}	
	
	\captionsetup{labelfont=bf}
	\caption{Invariant for the RME algorithm presented in Figure~\ref{algo:fullcrash}. The algorithm satisfies the conjunction of all the conditions given above in every configuration of every run.}
	\label{inv:fullcrash}
		\hrule
	\end{footnotesize}
\end{figure}

\begin{theorem}
	The algorithm in Figure~\ref{algo:fullcrash} solves the recoverable mutual exclusion problem on CC machines for an arbitrary number of processes of arbitrary names.
	It satisfies mutual exclusion, starvation freedom, critical section re-entry, bounded recovery, and bounded exit.
	The space complexity of the algorithm is $O(1)$ per process, and
	the RMR complexity of the algorithm on CC machines is $O(1)$ per passage.
\end{theorem}

\section{Enhancing the algorithm for DSM}\label{sec:dsmalg}

The CC algorithm just presented has an unbounded RMR complexity in the DSM model.
In this section, we enhance it to obtain a new algorithm that has $O(1)$ RMR complexity in both the DSM and CC models.
The enhanced algorithm so closely mirrors the CC algorithm that its correctness follows from that of the CC algorithm.
Below, we describe the main ideas behind how we transform the CC algorithm into an efficient DSM algorithm (Section~\ref{DSM-overview}), and then present the technical details (Sections~\ref{iss2} and \ref{iss3}).

\setcounter{linecounter}{0}
\begin{figure}[!ht]
	\begin{footnotesize}
		\hrule
		\vspace{-3mm}
		\begin{tabbing}
			\hspace{0in} \=  \hspace{0.2in} \= \hspace{0.1in} \=  \hspace{0.2in} \= \hspace{0.2in} \= \hspace{0.2in} \=\\
			{\bf Shared variables (stored in NVM)} \\ 
			\hspace{0in} \=  \hspace{0.1in} \= $\seq \in \mathbb{N}$, initially $1$. \\
			\> \> $\alock[0 \cdots 2]$ is an array of base mutual exclusion locks, as implemented in Figure~\ref{algo:dsmqlock}.  \\
			\> \> $\stopwait[0 \cdots 2]$ is an array of boolean signal objects, as implemented in Figure~\ref{algo:dsmsignal}, each initially {\em false}. \\
			\> \> $\barrier$ is a capturable object, as implemented in Figure~\ref{algo:capturable}, initially $\none$. \\
			\\
			\> {\bf Persistent variables local to process $p$ (stored in NVM)} \\
			\> \> $\inusep$ is a boolean, initially {\em false}. \\
			\> \> $\myseqp \in \mathbb{N}$, initially $1$. 
		\end{tabbing}
		
		\begin{minipage}[t]{.95\linewidth}
			\begin{tabbing}
				\hspace{0in} \= \hspace{0.25in} \= \hspace{0.2in} \=  \hspace{0.2in} \= \hspace{0.2in} \=\\
				\> \procline{dsmfullcrash:rem:1} \> {\texttt{Remainder Section}} \\
				\\
				\> \> \underline{\texttt{procedure $\tryprocp()$}}\\
				\> \procline{dsmfullcrash:try:1}\> $\inusep \gets \true$ \\
				\> \procline{dsmfullcrash:try:2} \> $\myseqp \gets \seq$ \\
				\> \procline{dsmfullcrash:try:3} \> $\alock[\myseqp \% 3].\tryprocp()$ \hspace{.1in}  \\
				\> \procline{dsmfullcrash:try:4} \> \ifcode $\seq \neq \myseqp$: \thencode \goto Line~\refln{dsmfullcrash:try:8} \\
				\> \procline{dsmfullcrash:try:5} \> $\csowner.\mwait{\myseqp \% 3}$ \hspace{.1in} $\parallel$ \hspace{.1in} $\stopwait[\myseqp \% 3].\mwait{}$ \\
				\> \hphantom{{\bf 6.}} \> \hphantom{$\csowner.\mwait{\myseqp \% 3}$ \hspace{.1in} $\parallel$ \hspace{.1in}} \goto Line~\refln{dsmfullcrash:try:8} \\
				\> \procline{dsmfullcrash:try:6} \> \ifcode $\csowner.\mcapture{}$ \thencode \return \gotocs \\
				\> \procline{dsmfullcrash:try:7} \> \ifcode $\seq \neq \myseqp$ \thencode \\
				\> \procline{dsmfullcrash:try:8} \> \> $\myseqp \gets \myseqp + 1$ \\
				\> \refln{dsmfullcrash:try:8}{\bf .1.} \> \> $\alock[(\myseqp - 1) \% 3].\mabandon{}$ \\
				\> \procline{dsmfullcrash:try:9} \> \> $\alock[\myseqp \% 3].\tryprocp()$ \\
				\> \procline{dsmfullcrash:try:10} \> \> $\csowner.\mwait{\myseqp \% 3}$ \\
				\> \procline{dsmfullcrash:try:11} \> \> \ifcode $\csowner.\mcapture{}$ \thencode \return \gotocs \\
				\> \procline{dsmfullcrash:try:12} \> $\csowner.\mwait{\myseqp \% 3}$ \\
				\> \procline{dsmfullcrash:try:13} \> $\csowner.\mwrite{}$ \\
				\> \procline{dsmfullcrash:try:14} \> \return \gotocs \\
				\\
				\> \procline{dsmfullcrash:cs:1} \> {\texttt{Critical Section}} \\
				\\
				\> \> \underline{\texttt{procedure $\exitprocp()$}}\\
				\> \procline{dsmfullcrash:exit:1} \> $x_p \gets \seq$ \\
				\> \procline{dsmfullcrash:exit:2} \> \ifcode $\myseqp \isequal x_p$ \thencode $\alock[\myseqp \% 3].\exitprocp()$ \\
				\> \refln{dsmfullcrash:exit:2}{\bf .1.} \> \ifcode $\myseqp \isequal x_p - 1$ \thencode $\alock[\myseqp \% 3].\mabandon{}$\\
				\> \procline{dsmfullcrash:exit:3} \> $\csowner.\mrelease{}$ \\
				\> \procline{dsmfullcrash:exit:4} \> $\inusep \gets \mbox{\it false}$ \\
				\\
				\> \> \underline{\texttt{procedure $\recoverprocp()$}}\\
				\> \procline{dsmfullcrash:rec:1} \> \ifcode $\inusep \wedge \seq \isequal \myseqp$: \\
				\> \procline{dsmfullcrash:rec:2} \> \> $\alock[(\myseqp - 1) \% 3].\initializep()$ \\ 
				\> \procline{dsmfullcrash:rec:3} \> \> $\stopwait[(\myseqp - 1) \% 3].\mreset{}$ \\
				\> \procline{dsmfullcrash:rec:4} \> \> $\seq \gets \myseqp + 1$ \\
				\> \procline{dsmfullcrash:rec:5} \> \> $\stopwait[\myseqp \% 3].\mset{}$ \\
				\> \refln{dsmfullcrash:rec:5}{\bf .1.} \> \ifcode $\myseqp = \seq - 1$: \thencode $\alock[\myseqp \% 3].\mabandon{}$ \\
				\> \procline{dsmfullcrash:rec:6} \> \ifcode $\csowner.\mread{} \isequal p$ \thencode \return \gotocs \\
				\> \procline{dsmfullcrash:rec:7} \> $\inusep \gets \mbox{\it false}$ \\
				\> \procline{dsmfullcrash:rec:8} \> \return \gotorem
			\end{tabbing}  
		\end{minipage}

		\captionsetup{labelfont=bf}
		\caption{
		Algorithm for an RME lock $\L$ for an arbitrary number of processes of arbitrary names, for DSM machines. Code shown for a process $p$.} 
		\label{algo:dsmfullcrash}
		\hrule
	\end{footnotesize}
\end{figure}

\subsection{Ideas underlying the DSM algorithm}\label{DSM-overview}

To achieve $O(1)$ RMR complexity for our RME algorithm in the DSM model, we need base mutex locks---$\mlock[i]$ for $i \in [0,2]$---that have $O(1)$ RMR complexity in the DSM model.
We obtain these by using Lee's second algorithm in Figure~\ref{algo:dsmqlock}, instead of his first algorithm.
Beyond this simple change, we must address the issue that an unbounded number of RMRs are incurred at each line of the algorithm where a process busy-waits.
In the following, we explain that there are three sources of busy-wait in the algorithm, and for each source, explain how we propose to bring the RMRs down to $O(1)$.

\begin{itemize}
\item
The busy-waiting on $\csowner$ at Lines 6, 11, and 13 of the earlier algorithm incurs unbounded RMRs on a DSM machine since the variable $\csowner$ must reside in a memory partition that is remote to all but one process.
To overcome this difficulty, we introduce a new object, which we call a {\em capturable object}, that supports a method for each operation the algorithm performs on $\csowner$, and implement each method so that it incurs only $O(1)$ RMRs. More specifically, the shared variable $\csowner$ of the CC algorithm is replaced by a capturable object $\csownerdsm$ in our DSM algorithm and the steps ``\waittill $\csowner \isequal \perp$'' (at Lines 6, 11, and 13), ``$\csowner \gets \none$'' (at Line 19), ``read $\csowner$'' (at Line 26), ``$\csowner \gets p$'' (at Line 14), and ``CAS$(\csowner, \bot, p)$'' (at Lines 7 and 12) of the earlier algorithm are replaced, respectively, by the constant-RMR methods $\csownerdsm.\mwait$, $\csownerdsm.\mrelease$, $\csownerdsm.\mread$, $\csownerdsm.\mwrite$, and $\csownerdsm.\mcapture$, as presented in Figure~\ref{algo:dsmfullcrash}.

\item
The statement ``\waittill $\stopwait[\myseqp\%3] \isequal \true$'' which appears at both Line 4 and Line 6, is a source of unbounded RMRs in the DSM model.
Although the wait statement at Line 4 is identical to the wait statement at Line 6, there is a significant difference between the two.
At Line 6, there can be at most one process busy waiting on $\stopwait[\myseqp\%3]$---the winner of $\mlock[\myseqp\%3]$.
In contrast, many processes can be busy waiting at Line 4.
We devise different solutions to reduce the 
the unbounded RMRs incurred at these two lines to $O(1)$.

To handle Line~6, we implement a new object, which we call a {\em boolean signal object}, that supports a method for each operation the algorithm performs on $\stopwait[i]$, and implement each method so that it incurs only $O(1)$ RMRs. More specifically, the shared variable $\stopwait[i]$ of the CC algorithm is replaced by a boolean signal object object $\stopwaitdsm[i]$ in our DSM algorithm and the steps ``\waittill $\stopwait[i] \isequal \true$'' (Line 6), ``$\stopwait[i] \leftarrow \true$'' (Line 25), and ``$\stopwait[i] \gets \mbox{\it false}$ (Line 23) of the earlier algorithm are replaced, respectively, by the constant-RMR methods $\stopwaitdsm[i].\mwait$, $\stopwaitdsm[i].\mset$, and $\stopwaitdsm[i].\mreset$, as can be seen by comparing Lines 6, 23, and 25 of Figures~\ref{algo:fullcrash} and \ref{algo:dsmfullcrash}.

\item
At Line 4 of the CC algorithm, $p$ repeatedly reads the $\stopwait[S_p \% 3]$ variable while performing in parallel the Try method on $\mlock[S_p \% 3]$.
This repeated reading is done because, if a process $q$ installs $\mlock[(S_p+1) \% 3]$ while $p$ is executing the Try method of $\mlock[S_p \% 3]$, the algorithm needs $p$ to detect this development and migrate to the newly installed lock.
Such a scenario manifests if events occur in the following order: a crash occurs while $q$ is in the try method of $\mlock[S_p \% 3]$; when processes subsequently restart after this crash, $p$ starts executing the algorithm, reads $S_p$ from $\seq$, and (clueless about the prior crash) executes the try method of the base lock $\mlock[S_p \% 3]$ (Line~4);
$q$ restarts, executes the recover method, and increments $\seq$ to $S_p+1$.

In the above scenario, since the newer base lock, namely $\mlock[(S_p+1) \% 3]$, has been installed, $p$ and any more such processes that are stuck at the try section of $\mlock[S_p \% 3]$ should release themselves and migrate to the try section of $\mlock[(S_p+1) \% 3]$. 
To meet this need while expending only $O(1)$ RMRs, we exploit the ``Release Property'' of Lee's lock, which guarantees that no process ever gets stuck in $\mlock[S_p \% 3]$ so long as all processes---including those that crash while using the $\mlock[S_p \% 3]$---eventually execute the exit section of that lock (Theorem~\ref{release} in Appendix A states the Release property and the paragraph preceding the theorem informally describes the property). 
This insight eliminates the need for the ``wait until'' statement at Line 4 (compare Line 4 of Figure~\ref{algo:fullcrash} with Line 4 of Figure~\ref{algo:dsmfullcrash}), and adds three lines (Lines 9.1, 18.1, and 25.1), where a process executes the exit method on an old lock as it installs or migrates to a new lock.

\end{itemize}

The above ideas give rise to the DSM algorithm in Figure~\ref{algo:dsmfullcrash}, which has three additional lines where $\exitprocp$ is called, as just explained, and every other line corresponds to the same numbered line of the earlier CC algorithm.

In the next two subsections, we present the specification and the implementation of the capturable and boolean signal objects, which completes the description of our DSM algorithm.

\subsection{Capturable object: Spec and Implementation}\label{iss2}

A {\em capturable object} $\C$ is specified as follows:
$\C.\mbox{\em state}$ is a pid or $\bot$, and
$\C$ supports the following operations.

\begin{itemize}
    \item $\C.\mread$: returns $\C.\mbox{\em state}$.
    \item $\C.\mwrite$: changes $\C.\mbox{\em state}$ to $p$.
    \item $\C.\mrelease$: changes $\C.\mbox{\em state}$ to $\bot$.
    \item $\C.\mcapture$: If $\C.\mbox{\em state} = \bot$, it changes $\C.\mbox{\em state}$ to $p$ and returns {\em true}; otherwise, it returns {\em false}, and leaves $\C.\mbox{\em state}$ unchanged.
    \item $\C.\mwait{i \in \{0,1,2\}}$: this method returns only if $\C.\mbox{\em state} = \bot$ at some point during the execution of the method. Thus, it is equivalent to the statement ``wait till $\C.\mbox{\em state} = \bot$''
    
    (The argument $i$ to the wait operation is irrelevant to the semantics of the operation, but we have introduced it for implementational convenience, as explained below.)
\end{itemize}

In the algorithm in Figure~\ref{algo:dsmfullcrash}, there are three lines where a process waits on $\csowner$ (Lines 6, 11, 13). In all of these lines together, for any $i \in \{0, 1, 2\}$, at most one process $p$ executes $\csowner.\mwait{i}$ at any given time.
So, for our purpose, it suffices to implement a capturable object that supports one waiting process at each $i$ (and an arbitrary number of processes that might execute the other operations).

The simple algorithm in Figure~\ref{algo:capturable} efficiently implements such an object, as summarized by the following theorem.

\setcounter{linecounter}{0}
\begin{figure}[!ht]
	\begin{footnotesize}
		\hrule
		\vspace{-3mm}
		\begin{tabbing}
			\hspace{0in} \=  \hspace{0.2in} \= \hspace{0.1in} \=  \hspace{0.2in} \= \hspace{0.2in} \= \hspace{0.2in} \=\\
			{\bf Shared variables (stored in NVM):} \\ 
			\hspace{0in} \=  \hspace{0.1in} \= $\X$ holds a process id or $\bot$; initialized to $\bot$. \\
			\> \> $\Go$ holds a pair $(\Go.\fseq, \Go.\fflag)$, where $\Go.\fseq$ is an integer and \\ \>\>$\Go.\fflag$ is a boolean; $\Go$ is initialized arbitrarily. \\
			\> \> $\Go$ resides in process $p$'s partition of shared memory. \\
			\> \> $\W[0\cdots2]$ is an array of pointers. Each pointer is initially $\bot$.
		\end{tabbing}
		
		\begin{minipage}[t]{.95\linewidth}
			\begin{tabbing}
				\hspace{0in} \= \hspace{0.25in} \= \hspace{0.15in} \=  \hspace{0.15in} \= \hspace{0.15in} \=\\
				\> \> \underline{\texttt{procedure $\C.\mread{}$}}\\
				\> \procline{capturable:read:1}\> \return $\X$ \\
                \\
				\> \> \underline{\texttt{procedure $\C.\mwrite{}$}}\\
				\> \procline{capturable:write:1}\> $\X \gets p$ \\
                \\
				\> \> \underline{\texttt{procedure $\C.\mcapture{}$}}\\
				\> \procline{capturable:capture:1}\> \return $\cas(\X, \bot, p)$ \\
                \\
				\> \> \underline{\texttt{procedure $\C.\mwait{\i \in \{0,1,2\}}$}}\\
				\> \procline{capturable:wait:1}\> $\x \gets \Go.\fseq$ \\
				\> \procline{capturable:wait:2}\> $\Go \gets (\x + 1, \mbox{\it false})$ \\
				\> \procline{capturable:wait:3}\> $\W[\i] \gets \&\Go$ \\
				\> \procline{capturable:wait:4}\> \ifcode $\X = \bot$ \return \\
				\> \procline{capturable:wait:5}\> \waittill $\Go.\fflag$ \\
				\\
				\> \> \underline{\texttt{procedure $\C.\mrelease{}$}}\\
				\> \procline{capturable:release:1}\> $\X \gets \bot$ \\
				\> \procline{capturable:release:2}\> \forcode $\i \in \{0,1,2\}$ \\
				\> \procline{capturable:release:3}\> \> $\ptr \gets \W[\i]$ \\
				\> \procline{capturable:release:4}\> \> \ifcode $\ptr \ne \bot$ \thencode \\
				\> \procline{capturable:release:5}\> \> \> $(\x, \b) \gets *\ptr$ \\
				\> \procline{capturable:release:6}\> \> \> \ifcode $\neg\b \wedge (\X = \bot)$ \thencode \\
				\> \procline{capturable:release:7}\> \> \> \> $\cas(*\ptr, (\x, \mbox{\it false}), (\x,\true))$ \\
			\end{tabbing}  
		\end{minipage}

		\captionsetup{labelfont=bf}
		\caption{
		Strictly-Linearizable implementation of a capturable object $\C$ that supports a single process waiting at each $i \in \{0,1,2\}$. 
		Code for process $p$.
		} 
		\label{algo:capturable}
		\hrule
	\end{footnotesize}
\end{figure}

\begin{theorem}{\em
For the algorithm in Figure~\ref{algo:capturable}, if for each $i \in \{0, 1, 2\}$, no two processes execute $\C.\texttt{wait}(i)$ concurrently, then:

\begin{itemize}
\item 
$\C$ is strictly linearizable \cite{StrictLinearizability}, i.e., (1) $\C$ is linearizable, and (2) if a crash occurs before the operation completes, then either the operation takes effect before the crash or it will never take effect.
\item
The operations $\mread$, $\mwrite$, $\mrelease$, and $\mcapture$ are wait-free.
\item
Suppose that no crash occurs since the time a process $p$ invokes $\C.\mwait{i_p}$, and suppose that $\C.\mbox{\em state} = \bot$ forever, starting from some arbitrary point in time.
Then, in a fair run, $p$ eventually returns from $\C.\mwait{i_p}$.
\item
A process incurs $O(1)$ RMRs to invoke and complete any of the five operations in both DSM and CC machines.
\end{itemize}
}\end{theorem}

\noindent{\em Proof sketch}: The algorithm maintains the key invariant that $\C.\mbox{\em state} = X$, and linearizes the read, write, capture, and release operations at Lines 1, 2, 3, and 9, respectively.
We linearize a wait operation only if it returns from the wait method; we observe that if a wait operation returns, then there is a point in time during the execution of the wait operation when $X = \bot$, and we linearize the operation at any such time.
Hence, we have Part (1) of the lemma. 
The other parts follow easily from an inspection of the code.
\hfill $\blacksquare$

\subsection{Signal Object: Spec and Implementation}\label{iss3}

A {\em boolean signal object} $\Sig$ is specified as follows:
$\Sig.\mbox{\em state}$ is Boolean, and
$\Sig$ supports the following three operations.

\begin{itemize}
    \item $\Sig.\mbset$: changes $\Sig.\mbox{\em state}$ to {\em true}.

    \item $\Sig.\mreset$: changes $\Sig.\mbox{\em state}$ to {\em false}

    \item $\Sig.\mbwait$: this method returns only if $\Sig.\mbox{\em state} = \mbox{\em true}$ at some point during the execution of the method. Thus, it is identical to the statement ``wait till $\Sig.\mbox{\em state} = \mbox{\em true}$''
\end{itemize}

In the algorithm in Figure~\ref{algo:dsmfullcrash}, for any $i \in \{0, 1, 2\}$, at most one process $p$ waits on $\stopwait[i]$ (at Line 6) at any given time.
So, for our purpose, it suffices to implement a boolean signal object that supports only one waiting process (and an arbitrary number of setting and resetting processes).
The simple algorithm in Figure~\ref{algo:dsmsignal} efficiently implements such an object, as summarized below:

\setcounter{linecounter}{0}
\begin{figure}[!ht]
	\begin{footnotesize}
		\hrule
		\vspace{-3mm}
		\begin{tabbing}
			\hspace{0in} \=  \hspace{0.2in} \= \hspace{0.1in} \=  \hspace{0.2in} \= \hspace{0.2in} \= \hspace{0.2in} \=\\
			{\bf Shared variables (stored in NVM):} \\ 
			\hspace{0in} \=  \hspace{0.1in} \= $\X$ is a boolean; initialized to {\em false}. \\
			\> \> $\Go$ holds a boolean; initialized arbitrarily. $\Go$ resides in $p$'s partition of memory. \\
			\> \> $\W$ is a pointer; initialized to $\bot$.
		\end{tabbing}
		
		\begin{minipage}[t]{.95\linewidth}
			\begin{tabbing}
				\hspace{0in} \= \hspace{0.25in} \= \hspace{0.15in} \=  \hspace{0.15in} \= \hspace{0.15in} \=\\
				\> \> \underline{\texttt{procedure $\Sig.\mwait{}$}}\\
				\> \procline{dsmsignal:wait:1}\> $\Go \gets \mbox{\it false}$ \\
				\> \procline{dsmsignal:wait:2}\> $\W \gets \&\Go$ \\
				\> \procline{dsmsignal:wait:3}\> \ifcode $\X$ \thencode \return \\
				\> \procline{dsmsignal:wait:4}\> \waittill $\Go$ \\
                \\
				\> \> \underline{\texttt{procedure $\Sig.\mset{}$}}\\
				\> \procline{dsmsignal:set:1}\> $\X \gets \true$ \\
				\> \procline{dsmsignal:set:2}\> $\ptr \gets \W$ \\
				\> \procline{dsmsignal:set:3}\> \ifcode $\ptr \ne \bot$ \thencode \\
				\> \procline{dsmsignal:set:4}\> \> $*\ptr \gets \true$ \\
				\\
				\> \> \underline{\texttt{procedure $\Sig.\mreset{}$}}\\
				\> \procline{dsmsignal:reset:1}\> $\X \gets \mbox{\it false}$ \\
			\end{tabbing}  
		\end{minipage}

		\captionsetup{labelfont=bf}
		\caption{
		Strictly-Linearizable implementation of a boolean signal object $\Sig$ that supports a single waiting process.
		Code for process $p$.
		} 
		\label{algo:dsmsignal}
		\hrule
	\end{footnotesize}
\end{figure}

\begin{theorem}{\em
For the algorithm in Figure~\ref{algo:dsmsignal}, if no two processes execute $\Sig.\texttt{wait}()$ concurrently, then:
\begin{itemize}
    \item 
    $\Sig$ is strictly linearizable.
    \item 
    The operations $\mbset$ and $\mreset{}$ are wait-free.
    \item
    Suppose that no crash occurs since the time a process $p$ invokes $\Sig.\mbwait$, and suppose that $\Sig.\mbox{\em state} = \true$ forever, starting from some arbitrary point in time.
    Then, in a fair run, $p$ eventually returns from $\Sig.\mbwait$.
    \item 
    A process incurs $O(1)$ RMRs to invoke and complete any of the three operations in both DSM and CC machines.
\end{itemize}

\noindent{\em Proof sketch}: The algorithm maintains the key invariant that $\Sig.\mbox{\em state} = X$, and linearizes the set and reset operations at Lines 5 and 9, respectively.
We linearize a wait operation only if it returns from the wait method; we observe that if a wait operation returns, then there is a point in time during the execution of the wait operation when $X = \true$, and we linearize the operation at any such time.
Hence, we have Part (1) of the lemma. 
The other parts follow easily from an inspection of the code.
\hfill $\blacksquare$

}\end{theorem}

\subsection{The final result}

The theorem below summarizes the results of this paper.

\begin{theorem}{\em
	The algorithm in Figure~\ref{algo:dsmfullcrash} solves the recoverable mutual exclusion problem on DSM and CC machines for an arbitrary number of processes of arbitrary names.
	It satisfies mutual exclusion, bounded exit, bounded critical section reentry, and starvation freedom.
	The space complexity of the algorithm is $O(1)$ per process, and
	the RMR complexity of the algorithm on DSM and CC machines is $O(1)$ per passage.
}\end{theorem}

\noindent
{\em Proof}: 
Since the algorithm in Figure~\ref{algo:dsmfullcrash} has the same invariant as that in Figure~\ref{algo:fullcrash}, the proofs of the properties are accordingly the same.
\hfill $\square$

\section{Conclusion}\label{sec:conclusion}

For the system-wide crashes, we have designed the first RME algorithm with $O(1)$ worst-case RMR complexity for both the CC and DSM models.
Our algorithm allows access to threads created dynamically, on-the-fly, and requires only $O(1)$ space per thread.

For the RME problem, the worst-case RMR complexity in the {\em individual crash model} for both CC and DSM machines (using realistic primitives) was settled to be $\Theta(\log n/\log\log n)$ by the upper bounds due to Jayanti et al. and Golab and Hendler \cite{jayanti:rmesublog, Golab:rmutex2} and the matching lower bounds due to Chan and Woelfel \cite{ChanWoelfelLB}.
This paper resolves the worst-case RMR complexity in the {\em system-wide crash model} to be $\Theta(1)$, thereby formalizing our intuition that individual crashes can be more expensive to cope with than system-wide crashes.

\noindent
We close with two open problems:

\begin{enumerate}
    \item 
    The algorithms in this paper use an unbounded sequence number.
    This is not a practical concern on modern 64-bit architectures, since it would take 500,000 years for the counter to overflow even at a rate of a million system-wide crashes per second.
    Nevertheless, from a theoretical stand point, it would be interesting to investigate whether we can do away with such variables.
    
    \item 
    Our algorithms do not satisfy the first-come-first-served (FCFS) property, but they satisfy the following weaker property: if an infinite run has only a finite number of crashes, then FCFS will hold for an infinite suffix of the run.
    Designing an FCFS algorithm without sacrificing on the other metrics is a tantalizing open problem.
\end{enumerate}

\bibliographystyle{acm}

\appendix
\section{Lee's Standard (Non-Recoverable) Mutual Exclusion Locks with Atomic Reset}\label{mutexspec}

We present two well known, standard (non-recoverable) queue locks due to Lee \cite{lee:twonodeme} with an atomic reset method.
The first lock, presented in Figure~\ref{algo:qlock}, has $O(1)$ RMR complexity in the CC model, and is used for the three base locks of our RME lock of Section~\ref{sec:mainalg}.
The second one, in Figure~\ref{algo:dsmqlock}, has $O(1)$ RMR complexity in both CC and DSM models, and is used for the three base locks of our RME lock of Section~\ref{sec:dsmalg}.

We capture the {\em status} of a lock $\ell$ by four sets---$\ell.\tryset$, $\ell.\csset$, $\ell.\exitset$, and $\ell.\crashedset$.
The first three are the set of processes in $\ell$'s try section, CS, and exit section, respectively.
When a system-wide crash occurs, the processes in these sets are moved to  $\ell.\crashedset$ (and the other three sets become empty).
Thus, $\ell.\crashedset$ consists of those processes that were using the lock when a crash occurred.
When atomic reset is performed on $\ell$, all four sets are set to $\emptyset$, reflecting that $\ell$ is once more ready for use as a fresh mutex lock; furthermore, reset can be performed at any time.

The {\em use-pattern} states the obvious: a process $p$ may enter $\ell.\tryset$ only when it is not in any of the four sets;
when $p$ leaves $\ell.\tryset$ (upon completing the try section), it gets into $\ell.\csset$; 
when $p$ leaves $\ell.\csset$, it gets into $\ell.\exitset$; and
when $p$ leaves $\ell.\exitset$, it is out of all sets.
Furthermore, $p$ may execute a step of the try, CS, or exit section only if it is in $\ell.\tryset$, $\ell.\csset$, or $\ell.\exitset$, respectively.


The properties of Lee's locks are summarized as follows.

\begin{theorem}{\em 
Lee's locks in Figures~\ref{algo:qlock} and \ref{algo:dsmqlock} satisfy the following properties, assuming that processes respect the use-pattern stated above: 

\begin{itemize}
    \item \underline{Mutual Exclusion}: $|\ell.\csset| \le 1$
    
    \item \underline{Starvation Freedom}: In every fair run, if a process $p$ executes $\ell.\mbox{\em try}_p()$ and $\ell.\crashedset$ is empty during the entire interval of this execution of the try method, then the method will eventually complete, leading to $p$ entering the CS.
    
    \item \underline{Bounded Exit}: If a process $p$ executes $\ell.\mbox{\em exit}_p()$ without crashing, $p$ returns from the method in a constant number of its own steps..
    
    \item \underline{Atomic Reset}: $\ell.\reset_p()$ is atomic and returns $\ell$ to its initial state.
    
    \item \underline{RMR and space complexity}:
    The lock in Figure~\ref{algo:qlock} has $O(1)$ RMR complexity in the CC model, the lock in Figure~\ref{algo:dsmqlock} has $O(1)$ RMR complexity in both CC and DSM models, and both locks have $O(1)$ space complexity per process.
\end{itemize}
}
\end{theorem}

Lee's algorithms have an additional property \cite{JJAbort}, which is explained as follows.
Suppose that a set $A$ of processes were using a lock $\ell$ when a system-wide crash occurred.
When processes subsequently restart, suppose that a set $B$ of processes (with $B$ disjoint from $A$) invoke the try method. 
Because processes in $B$ queue up behind those in $A$ that lost their state (due to the earlier crash), processes in $B$ could get stuck in the try section forever.
However, if {\em all} processes in $A$, when they restart, execute the exit section, some process in $B$ (informally, the first in the queue among the ones in $B$) will complete and return from its try section; furthermore, if each process in $B$, upon returning from the try section, executes the exit section, then {\em all} processes in $A$, one after the other, return from the try section.
The crashed processes in $A$, by executing the exit section while in $\ell.\crashedset$, violate the use-pattern.
Hence, Mutual Exclusion can be potentially violated, but the feature that none in $B$ gets stuck is useful when designing our RME algorithm for the DSM model.
The following theorem summarizes the property explained so far.

\begin{theorem}[Release Property]\label{release}{\em 
Let $\ell$ be a Lee's lock from Figure~\ref{algo:qlock} or \ref{algo:dsmqlock}.
Suppose that $\ell.\crashedset = A \neq \emptyset$ and $\ell.\tryset \neq \emptyset$.
If each process in $A$ executes the exit section and each process in $\ell.\tryset$, if and when it leaves $\ell.\tryset$, executes the exit section, then every process in $\ell.\tryset$ eventually leaves $\ell.\tryset$.
}
\end{theorem}

    


\begin{figure}[h!]
	\begin{footnotesize}
		\hrule
		\vspace{-3mm}
		\begin{tabbing}
			\hspace{0in} \=  \hspace{0.2in} \= \hspace{0.1in} \=  \hspace{0.2in} \= \hspace{0.2in} \= \hspace{0.2in} \=\\
			\\
			{\bf Shared variable (stored in NVM):} \\ 
			\hspace{0in} \=  \hspace{0.1in} \= $\tail$ is a pointer to a location that contains a boolean or $\bot$; $\tail$ is initialized to $\bot$.\\ 
			\> {\bf Persistent Variables local to process $p$ (stored in NVM):} \\
			\> \> $\nodes_p[0], \nodes_p[1]$ are pointers to booleans which are initialized arbitrarily. \\
			\> \> $\facep \in \{ 0, 1 \}$, initialized arbitrarily.
		\end{tabbing}
		\begin{minipage}[t]{.5\linewidth}
			\begin{tabbing}
				\hspace{0in} \= \hspace{0.25in} \= \hspace{0.2in} \=  \hspace{0.2in} \= \hspace{0.2in} \= \hspace{0.2in} \=\\
				\> \> \underline{\texttt{procedure $\lockobj.\tryprocp()$}}\\
				\> \procline{qlock:ltry:1} \> $\facep = 1 - \facep$ \\
				\> \procline{qlock:ltry:2} \> $* \nodes_p[\facep] \gets \mbox{\it false}$ \\
				\> \procline{qlock:ltry:3} \> $\prevp \gets \fas(\tail, \nodes_p[\facep])$ \\
				\> \procline{qlock:ltry:4} \> \ifcode $\prevp \neq \bot$ \thencode \waittill $* \prevp \isequal \true$ \\
				\\
				\> \> \underline{\texttt{procedure $\lockobj.\exitprocp()$}}\\
				\> \procline{qlock:lexit:1} \>$*\nodes_p[\facep] \gets \true$ \\
				\\
				\> \> \underline{\texttt{procedure $\lockobj.\initializep()$}}\\
				\> \procline{qlock:init:1} \> $\tail \gets \bot$
			\end{tabbing}  
		\end{minipage}
		\captionsetup{labelfont=bf}
		\caption{
		Lee's lock for CC machines \cite{lee:twonodeme}, with a line added for Reset. 
		Code shown for a process $p$.
		} 
		\label{algo:qlock}
		\hrule
	\end{footnotesize}
\end{figure}

\setcounter{linecounter}{0}
\begin{figure}[h!]
	\begin{footnotesize}
		\hrule
		\vspace{-3mm}
		\begin{tabbing}
			\hspace{0in} \=  \hspace{0.2in} \= \hspace{0.1in} \=  \hspace{0.2in} \= \hspace{0.2in} \= \hspace{0.2in} \=\\
			{\bf Shared variable (stored in NVM):} \\ 
			\hspace{0in} \=  \hspace{0.1in} \= $\tail$ is a pointer to a location that contains a pointer or $\bot$; $\tail$ is initialized to $\bot$.\\ 
			\> {\bf Persistent Variables local to process $p$ (stored in NVM):} \\
			\> \> $\nodes_p[0], \nodes_p[1]$ are pointers to memory-words which are initialized arbitrarily. \\
			\> \> $\facep \in \{ 0, 1 \}$, initialized arbitrarily.
		\end{tabbing}
		\begin{minipage}[t]{.5\linewidth}
			\begin{tabbing}
				\hspace{0in} \= \hspace{0.25in} \= \hspace{0.2in} \=  \hspace{0.2in} \= \hspace{0.2in} \= \hspace{0.2in} \=\\
				\> \> \underline{\texttt{procedure $\lockobj.\tryprocp()$}}\\
				\> \procline{dsmqlock:ltry:1} \> $\facep = 1 - \facep$ \\
				\> \procline{dsmqlock:ltry:2} \> $* \nodes_p[\facep] \gets \bot$ \\
				\> \procline{dsmqlock:ltry:3} \> $\prevp \gets \fas(\tail, \nodes_p[\facep])$ \\
				\> \procline{dsmqlock:ltry:4} \> \ifcode $\prevp \neq \bot$ \thencode \\
				\> \procline{dsmqlock:ltry:5} \> \> $\Go \gets \mbox{\it false}$ \\
				\> \procline{dsmqlock:ltry:6} \> \> \ifcode $\fas(*\prevp, \&\Go) = \tok$ \thencode \return \\
				\> \procline{dsmqlock:ltry:7} \> \> \waittill $\Go$ \\
				\\

				\> \> \underline{\texttt{procedure $\lockobj.\exitprocp()$}}\\
				\> \procline{dsmqlock:lexit:1} \>$\ptr \gets \fas(*\nodes_p[\facep], \tok)$ \\
				\> \procline{dsmqlock:lexit:2} \>\ifcode $\ptr \ne \bot$ \thencode $*\ptr \gets \true$ \\
				\\

				\> \> \underline{\texttt{procedure $\lockobj.\initializep()$}}\\
				\> \procline{dsmqlock:init:1} \> $\tail \gets \bot$ \\
				

			\end{tabbing}  
		\end{minipage}
		\captionsetup{labelfont=bf}
		\caption{
		Lee's queue lock for CC and DSM machines \cite{lee:twonodeme}, with a line added for Reset. Code shown for a process $p$.
		} 
		\label{algo:dsmqlock}
		\hrule
	\end{footnotesize}
\end{figure}


\section{Correctness of the algorithm} 
\label{app:correc}

We prove the correctness of the algorithm by first presenting the invariant satisfied by the algorithm in Section~\ref{sec:correcinv} and then use the invariant to prove correctness the correctness of the algorithm in Section~\ref{sec:correcproof}.

\subsection{Invariant of the algorithm} \label{sec:correcinv}
The invariant of the algorithm appears in Figure~\ref{inv:fullcrash}.

\subsection{Proofs required by the base lock}
In this section we prove certain lemmas that prove that our RME algorithm invokes the base lock as it is required by the lock.
This ensures that the base locks behave in the way they are expected to. 
Lemmas~\ref{lem:fullcrash:trans1}-\ref{lem:fullcrash:trans2} argue that the main algorithm follows the rules of transition as expected by the base lock.
Lemmas~\ref{lem:fullcrash:nowait1}-\ref{lem:fullcrash:nowait3} are important to argue that the pre-condition for starvation freedom on the base lock holds.
That is, no process stays in the CS forever: if any process $q$ is in ${\ell}.\csset$ at any time $t' >  t$, $q$ is in ${\ell}.\exitset$ after $t'$.
To see that, we observe from the algorithm that a process could loop only at certain places throughout the run.
When a process $p$ is at Line~\refln{fullcrash:try:3} or \refln{fullcrash:try:9}, we know that $p \in \mlock[\myseq_p \% 3].\tryset$.
However, it could be the case that $p \in \mlock[\myseq_p \% 3].\csset$ when $\pcp \in [\refln{fullcrash:try:4}, \refln{fullcrash:try:5}]$, or
certainly that $p \in \mlock[\myseq_p \% 3].\csset$ when $\pcp \in \{ \refln{fullcrash:try:10}, \refln{fullcrash:try:12} \}$.
While a process is resetting a lock $\ell$, no other process performs a step of the $\ell.\tryproc{}()$ or $\ell.\exitproc{}()$ methods
of that lock.
More precisely, for each lock $\ell$ and each process $q$, 
if, for a process $p$, $\pcp$ points to a step of ${\ell}.\initializep()$, 
then $\pc{q}$ does not point to a step in ${\ell}.\tryproc{q}()$ or ${\ell}.\exitproc{q}()$.
This is exactly what Lemma~\ref{lem:fullcrash:4rcathm1} argues about.
Lemma~\ref{lem:fullcrash:4rcathm1} below essentially implies that once a process $p$ finishes the execution of $\ell.\initializep()$, for any base lock $\ell$,
any process $q$, whether $q=p$ or not, goes back to the pattern of executing the $\ell.\tryproc{q}()$, CS, and $\ell.\exitproc{q}()$ methods, {\em beginning with} the invocation of $\ell.\tryproc{q}()$.
This follows from the fact that $\ell.\initializep()$ will reset the lock changing $\ell.\tryset$, $\ell.\csset$, and $\ell.\exitset$ to empty sets,
and By Lemmas~\ref{lem:fullcrash:trans1}-\ref{lem:fullcrash:trans2} we know that the main algorithm honors the transitions required by the base lock.

\begin{lemma}\label{lem:fullcrash:trans1}
A process $p$ may invoke $\mlock[\seq \% 3].\tryprocp()$ only if $p$ is not in any of the three sets.
\end{lemma}
\begin{proof}
	A process $p$ may invoke $\mlock[\seq \% 3].\tryprocp()$ either from Line~\refln{fullcrash:try:3} or \refln{fullcrash:try:9}.
	By Condition~\ref{inv:fullcrash:cond4}, when $\pcp = \refln{fullcrash:try:2}$, $p \notin \mlock[\seq \% 3].\set$, and we have the claim.
	When $\pcp = \refln{fullcrash:try:8}$, by Condition~\ref{inv:fullcrash:cond11}, $\myseq_p = \seq - 1$, and thus, by Condition~\ref{inv:fullcrash:cond4} again we have the claim.
\end{proof}

\begin{lemma}\label{lem:fullcrash:trans2}
A process $p$ may invoke $\mlock[\seq \% 3].\exitprocp()$ only if $p$ is in $\mlock[\seq \% 3].\csset$.
\end{lemma}
\begin{proof}
A process $p$ may invoke $\mlock[\seq \%3].\exitprocp()$ at Line~\refln{fullcrash:exit:2}.
This could happen only if $\myseq_p = \seq$ when $p$ takes a step at Line~\refln{fullcrash:exit:1}.
In that case, by Condition~\ref{inv:fullcrash:cond7}, $p \in \mlock[\myseq_p \%3].\csset$, which implies that the claim holds when $p$ invokes $\mlock[\seq \% 3].\exitprocp()$.
\end{proof}

\begin{lemma} \label{lem:fullcrash:wfcsr}
	There is a constant $c$ such that,
	if a crash occurs while a process $p$ is in the CS (i.e., when $\pcp = \refln{fullcrash:cs:1}$), then $p$ reenters the CS 
	before it executes $c$ of its steps that have no intervening crash steps.
\end{lemma}
\begin{proof}
	If a crash occurs while a process $p$ is in the CS,
	in the configuration immediately after the crash, $\status_p = \reccs$ and $\pcp = \refln{fullcrash:rem:1}$.
	It also holds that $\csowner = p$ in the configuration after the crash since it had that value prior to the crash by Condition~\ref{inv:fullcrash:cond9} and the crash doesn't change the value of $\csowner$.
	We know from the description of the control flow that $\pcp = \refln{fullcrash:rec:1}$ eventually.
	We note that once $\pcp = \refln{fullcrash:rec:1}$, $\pcp$ changes to $\refln{fullcrash:rec:6}$ in a constant number of $p$'s own normal steps,
	this follows from an inspection of the algorithm and since there are no loops between the execution path from Line~\refln{fullcrash:rec:1} to Line~\refln{fullcrash:rec:6}.
	
	We now argue by contradiction that $\csowner$ retains the value $p$ from the configuration immediately after $p$'s crash in CS upto the configuration prior to $p$ executing Line~\refln{fullcrash:rec:6}.
	Thus, assume for a contradiction that $\csowner$ is changed from $p$ to some other value after $p$'s crash in the CS but before $p$ ever executes Line~\refln{fullcrash:rec:6}.
	Let process $r$ be the earliest such process in the run to change the value of $\csowner$ from $p$ to some other value.
	The value of $\csowner$ is changed only in the $\tryproc{}()$ and $\exitproc{}()$ method and since $p$ is either at Line~\refln{fullcrash:rem:1} or the $\recoverprocp()$ method,
	$p \neq r$.
	We note that if $r$ changed it due to the CAS operation at Lines~\refln{fullcrash:try:6} or \refln{fullcrash:try:11},
	$\csowner = \perp$ in the configuration prior to $r$'s step, and hence $r$ is not the earliest process to change the value of $\csowner$ from $p$ to some other value.
	Similarly, if $r$ were to execute Line~\refln{fullcrash:try:13}, by Condition~\ref{inv:fullcrash:cond9}, $\csowner = \perp$ when $\pc{r} = \refln{fullcrash:try:13}$,
	and hence $r$ is not the earliest process to change the value of $\csowner$ from $p$ to some other value.
	Thus, by an inspection of the algorithm we see that $\csowner$ could change only to $\perp$ due to execution of Line~\refln{fullcrash:exit:3} by $r$.
	Let $C$ be the configuration when $\pc{r} = \refln{fullcrash:exit:3}$.
	We know that in $C$ $\csowner = p$.
	Since $\pc{r} = \refln{fullcrash:exit:3}$ in $C$, it follows from Condition~\ref{inv:fullcrash:cond9} that $\csowner = r$ in $C$.
	Therefore, $\csowner$ has two different values in $C$, a contradiction.
	It follows from the above that $\csowner$ retains the value $p$ from the configuration immediately after $p$'s crash in CS upto the configuration prior to $p$ executing Line~\refln{fullcrash:rec:6}.

	From the above it follows that if a crash occurs while a process $p$ is in the CS (i.e., when $\pcp = \refln{fullcrash:cs:1}$), 
	then $p$ reenters the CS before it executes $c$ of its steps that have no intervening crash steps
\end{proof}

\begin{lemma}\label{lem:fullcrash:csoperp}
	Assume a fair run in which $\csowner = p$ at time $t$, 
	$\csowner = \perp$ at time $t' > t$ in that run.
\end{lemma}
\begin{proof}
	Suppose we have a fair run in which $\csowner = p$ at time $t$.
	By Condition~\ref{inv:fullcrash:cond9}, $\activep = \true$ and $\pcp \in \{ \refln{fullcrash:try:14} \upto \refln{fullcrash:exit:3}, \refln{fullcrash:rec:1} \upto \refln{fullcrash:rec:6} \}$
	or, together with Condition~\ref{inv:fullcrash:cond9} and \ref{inv:fullcrash:cond10}, $\pcp = \refln{fullcrash:rem:1}$ and $\status_p \notin \{ \good, \recrem \}$.
	If  $\pcp = \refln{fullcrash:rem:1}$ and $\status_p \notin \{ \good, \recrem \}$, we know from the description of the control flow that $\pcp = \refln{fullcrash:rec:1}$ eventually.
	From an inspection of the algorithm, it follows that $p$ eventually executes Line~\refln{fullcrash:rec:6} in either of the cases when 
	$\activep = \true$ and $\pcp \in \{\refln{fullcrash:rec:1} \upto \refln{fullcrash:rec:6} \}$, or
	$\pcp = \refln{fullcrash:rem:1}$ and $\status_p \notin \{ \good, \recrem \}$.
	By the argument similar to one given for Lemma~\ref{lem:fullcrash:wfcsr}, 
	it follows that $\csowner$ retains the value $p$ upto the point where it executes Line~\refln{fullcrash:rec:6} and re-enters the CS and $\pcp = \refln{fullcrash:cs:1}$ eventually.
	Once $\pcp \in \{ \refln{fullcrash:try:14} \upto \refln{fullcrash:exit:3} \}$ it is straightforward to see that $p$ sets $\csowner$ to $\perp$ at Line~\refln{fullcrash:exit:3} at a later point in time,
	satisfying the claim.
\end{proof}

\begin{lemma} \label{lem:fullcrash:nowait1}
	If $p$ is at Lines~\refln{fullcrash:try:4}\upto\refln{fullcrash:try:5} at time $t$, 
	$p$ is at a different line at some time $t' > t$. 
\end{lemma}
\begin{proof}
	Suppose $p$ is at Lines~\refln{fullcrash:try:4}\upto\refln{fullcrash:try:5}.
	By Condition~\ref{inv:fullcrash:cond11}, $\myseq_p \in \{ \seq - 1, \seq \}$.
	We argue each of the cases as follows:
	\begin{itemize}
		\item[Case 1]
		Suppose $\myseq_p = \seq - 1$.
		By Condition~\ref{inv:fullcrash:cond15},
		$\stopwait[\myseq_p \% 3] = \true \vee (\exists q, \pc{q} = \refln{fullcrash:rec:5} \wedge \myseq_{q} = \myseq_p)$
		If $\stopwait[\myseq_p \% 3] = \true$, $p$ notices that the next time it executes Line~\refln{fullcrash:try:4} and comes out of the loop.
		Otherwise, we have $\exists q, \pc{q} = \refln{fullcrash:rec:5} \wedge \myseq_{q} = \myseq_p$.
		At Line~\refln{fullcrash:rec:5} $q$ sets $\stopwait[\myseq_p \% 3]$ to $\true$,
		which $p$ notices the next time it executes Line~\refln{fullcrash:try:4} and hence comes out of the loop.
		
		\item[Case 2]
		Suppose $\myseq_p = \seq$.
		Let's assume for the sake of the argument that there is no crash in the run after $\pcp = \refln{fullcrash:try:4}$ for the first time in the current passage and $\seq = \myseq_p$ forever,
		because otherwise $p$ is at a different line due to one of the above cases and the claim holds trivially.
		We note that by Condition~\ref{inv:fullcrash:cond3} $\stopwait[\seq \% 3]$ will remain $\false$ forever, which prevents $p$ from getting out of the loop at Line~\refln{fullcrash:try:4}.
		
		If $\csowner = \perp$ and $p$ notices that the next time it executes Line~\refln{fullcrash:try:5}, $p$ comes out of the loop.
		Otherwise, assume $\csowner = q$.
		By the contrapositive of the first part of Condition~\ref{inv:fullcrash:cond9} it follows that $q \neq p$.
		By Lemma~\ref{lem:fullcrash:csoperp} $\csowner$ changes to $\perp$ at a later point.
		If $p$ notices that $\csowner = \perp$ at Line~\refln{fullcrash:try:5}, we have the claim.
		Otherwise, we argue as follows that $\csowner$ could assume the value $r \neq \perp$ at most once after changing to $\perp$ as above,
		and then would become $\perp$ again when $p$ finally notices it.
		Thus, assume that $\csowner$ changes to the value $r \neq \perp$ right after becoming $\perp$ as above but before $p$ notices it at Line~\refln{fullcrash:try:5}.
		We note that this value of $r$ could only be written into $\csowner$ by $r$ itself at Lines~\refln{fullcrash:try:6}, \refln{fullcrash:try:11}, or \refln{fullcrash:try:13}.
		Thus, $r \neq p$, otherwise $p$ already left the loop and we have the claim.
		We first note that the change to $\csowner$ could not happen at Lines~\refln{fullcrash:try:11} or \refln{fullcrash:try:13}.
		This is because, by Condition~\ref{inv:fullcrash:cond7} and \ref{inv:fullcrash:cond11}, at Lines~\refln{fullcrash:try:11} and \refln{fullcrash:try:13}
		$\myseq_r = \seq$ and $r \in \mlock[\myseq_r \% 3].\csset$.
		Because $\pcp \in \{ \refln{fullcrash:try:4}\upto\refln{fullcrash:try:5} \}$ and $\myseq_p = \seq$ (which implies $\stopwait[\myseq_p \% 3] = \false$ by Condition~\ref{inv:fullcrash:cond3}) we already have $p \in \mlock[\myseq_p \% 3].\csset$ from Condition~\ref{inv:fullcrash:cond7}.
		Hence, the mutual exclusion property of $\mlock[\seq \% 3]$ implies that $r$ cannot be in $\mlock[\myseq_r \% 3].\csset$ at the same time with $\myseq_{r} = \seq$.
		It follows that the value $r$ is written into $\csowner$ at Line~\refln{fullcrash:try:6} only.
		By Condition~\ref{inv:fullcrash:cond11} and the same argument as above, it follows that $\myseq_r = \seq -1$, and by Condition~\ref{inv:fullcrash:cond7}, $r \in \mlock[\myseq_r \% 3].\csset$
		when $r$ writes its own name into $\csowner$ at Line~\refln{fullcrash:try:6}.
		Once $r$ CASes the value $r$ into $\csowner$ at Line~\refln{fullcrash:try:6}, it moves to the CS, completes the CS operation and invokes $\exitproc{r}()$.
		At Line~\refln{fullcrash:exit:1} $r$ finds that $\myseq_r = \seq - 1 \neq \seq$ 
		and hence it doesn't invoke $\mlock[\myseq_r \% 3].\exitproc{r}()$ at Line~\refln{fullcrash:exit:2} ever.
		Thereby, $r$ remains in $\mlock[\myseq_r \% 3].\csset$ forever and by the mutual exclusion property of the base lock $\mlock[\myseq_r \% 3]$, no other process comes to execute Line~\refln{fullcrash:try:6} later.
		Thus, once $r$ writes $\perp$ into $\csowner$ at Line~\refln{fullcrash:exit:3}, there is no other process to write its own name into $\csowner$ at Lines~\refln{fullcrash:try:6}, \refln{fullcrash:try:11}, or \refln{fullcrash:try:13}.
		Thus, $p$ notices at Line~\refln{fullcrash:try:5} that $\csowner = \perp$ and comes out of the loop.
	\end{itemize}

We note that from the above two cases, the claim holds.
\end{proof}

\begin{lemma}  \label{lem:fullcrash:nowait2}
If $p$ is at Line~\refln{fullcrash:try:10} at time $t$,
$p$ is at a different line at some time $t' > t$.
\end{lemma}
\begin{proof}
	We first note that $\myseq_p = \seq$ when $\pcp = \refln{fullcrash:try:10}$ by Condition~\ref{inv:fullcrash:cond11}.
	The claim holds by the same argument as given in Case~2 of the proof of Lemma~\ref{lem:fullcrash:nowait1} above.
\end{proof}

\begin{lemma}  \label{lem:fullcrash:nowait3}
	If $p$ is at Line~\refln{fullcrash:try:12} at time $t$,
	$p$ is at a different line at some time $t' > t$.
\end{lemma}
\begin{proof}
	Suppose $p$ is at Line~\refln{fullcrash:try:12} at time $t$.
	We note that $\myseq_p = \seq$ when $\pcp = \refln{fullcrash:try:12}$ by Condition~\ref{inv:fullcrash:cond11}.
	If $\csowner = q \neq \perp$, by Condition~\ref{inv:fullcrash:cond9}, $q \neq p$.
	By Lemma~\ref{lem:fullcrash:csoperp} we have that $\csowner = \perp$ at a later time $t'$.
	Thus, assume $\csowner = \perp$, by Condition~\ref{inv:fullcrash:cond14}, 
	we get $\forall r, (\myseq_r \neq \seq - 1 \vee \pc{r} \in \{ \refln{fullcrash:rem:1} \upto \refln{fullcrash:try:3}, \refln{fullcrash:try:8}, \refln{fullcrash:exit:4}, \refln{fullcrash:rec:1} \upto \refln{fullcrash:rec:8} \}$
	$\vee$ $( \pc{r} = \refln{fullcrash:try:4} \limplies \stopwait[\myseq_r  \% 3] = \true))$.
	This implies that there is no process $r$ with $\myseq_r = \seq - 1$ and $\pc{r} = \refln{fullcrash:try:6}$ once $\csowner = \perp$.
	Since $\pcp = \refln{fullcrash:try:12}$, by Condition~\ref{inv:fullcrash:cond7},
	$p \in \mlock[\myseq_p \% 3].\csset$.
	Since $\myseq_p = \seq$, by Condition~\ref{inv:fullcrash:cond7} again, there is no process $q$ with $\myseq_q = \seq$ and $\pc{q}$ to be \refln{fullcrash:try:6}, \refln{fullcrash:try:11}, or \refln{fullcrash:try:13}.
	Thus $\csowner$ doesn't change from $\perp$ to some other value once $p$ is at Line~\refln{fullcrash:try:12} and so long as $p$ doesn't notice $\csowner$ to be $\perp$.
	It follows that $p$ comes out of the loop at Line~\refln{fullcrash:try:12} to move to Line~\refln{fullcrash:try:13}.
\end{proof}

\begin{lemma} \label{lem:fullcrash:4rcathm1}
For each lock $\ell$ and each process $q$, 
if, for a process $p$, $\pcp$ points to a step of ${\ell}.\initializep()$ at time $t$, 
then $\pc{q}$ does not point to a step in ${\ell}.\tryproc{q}()$ or ${\ell}.\exitproc{q}()$ at $t$.
\end{lemma}
\begin{proof}
Suppose $\pcp = \refln{fullcrash:rec:2}$ with $\myseq_p = \seq$ at time $t$.
Assume for a contradiction that at $t$ there is a process $q$ such that $\pc{q} = \refln{fullcrash:try:3}$ with $\myseq_{q} = \seq -1$.
By Condition~\ref{inv:fullcrash:cond12} it follows that for each process $r$, if $\pc{r} \in \{ \refln{fullcrash:rec:2} \upto \refln{fullcrash:rec:5} \}$ then $\myseq_{r} = \seq - 1$,
which is a contradiction to our original assumption that $\pcp = \refln{fullcrash:rec:2}$ with $\myseq_p = \seq$.
Thus there is no process $q$ such that $\pc{q} = \refln{fullcrash:try:3}$ with $\myseq_{q} = \seq -1$ at $t$.
By the same argument we conclude that there is no process $q$ such that $\pc{q} = \refln{fullcrash:exit:2}$ with $\myseq_{q} = \seq -1$. 
By Condition~\ref{inv:fullcrash:cond11} we know that if for a process $q$ if $\pc{q} = \refln{fullcrash:try:9}$, $\myseq_{q} = \seq$.
Thus, when $p$ is pointing to a step of $\mlock[(\seq - 1) \% 3].\initializep()$ at time $t$, 
there is no process $q$ that points to a step in $\mlock[(\seq - 1) \% 3].\tryproc{q}()$ or $\mlock[(\seq - 1) \% 3].\exitproc{q}()$ at $t$.

Suppose $\pcp = \refln{fullcrash:rec:2}$ with $\myseq_p = \seq - 1$ at time $t$.
By Condition~\ref{inv:fullcrash:cond11},
for any process $q$, if $\pc{q} \in \{ \refln{fullcrash:try:3}, \refln{fullcrash:try:9}, \refln{fullcrash:exit:2} \}$,
then $\myseq_{q} \in \{ \seq -1, \seq \}$.
It follows that when $p$ is pointing to a step of $\mlock[(\seq - 2) \% 3].\initializep()$ (or equivalently $\mlock[(\seq + 1) \% 3].\initializep()$) at time $t$, 
there is no process $q$ that points to a step in $\mlock[(\seq - 2) \% 3].\tryproc{q}()$ or $\mlock[(\seq - 2) \% 3].\exitproc{q}()$ at $t$.

The claim thus follows from the above.
\end{proof}

\subsection{Proof of properties} \label{sec:correcproof}

\begin{lemma}[{\bf Mutual Exclusion}] \label{lem:fullcrash:mutex}
	At most one process is in the CS in any configuration of any run.
\end{lemma}
\begin{proof}
	Assume for a contradiction that there are two processes $p$ and $q$ in the CS in the same configuration.
	Therefore, $\pcp = \refln{fullcrash:cs:1}$ and $\pc{q} = \refln{fullcrash:cs:1}$ in the same configuration.
	By Condition~\ref{inv:fullcrash:cond9} we have $\csowner = p$ as well as $\csowner = q$ in the same configuration.
	Thus $\csowner$ has two different values in the same configuration, a contradiction.
\end{proof}

\begin{lemma}[{\bf Bounded Exit}] \label{lem:fullcrash:boundedexit}
	There is an integer $b$ such that if in any run any process $p$ invokes
	and executes $\exitprocp()$ without crashing, the method completes in at most $b$ steps of $p$.
\end{lemma}
\begin{proof}
	We note that the base algorithm satisfies the bounded exit property.
	From an inspection of the algorithm we note that the $\exitprocp()$ procedure from Figure~\ref{algo:fullcrash} has a bounded number of steps and no loops.
	The claim thus holds.
\end{proof}

\begin{lemma}[{\bf Starvation Freedom}] \label{lem:fullcrash:starvfreedom}
In every fair run, if a process $p$ executes the try method and no crash occurs during this execution of the try method, then the method will eventually complete, leading to $p$ entering the CS.
\end{lemma}
\begin{proof}
Since there are only finitely many crash steps in the run,
for the purpose of the argument take a run and pick the earliest configuration from the run such that all the crashes have already occurred.
We need to show that if a process $p$ is in $\tryprocp()$ at a time $t$,
it is in the CS at a different time $t' > t$.
From an inspection of the algorithm we note that when the method $\tryprocp()$ returns, 
it puts $p$ in the CS because every return statement returns the value $\gotocs$.
Thus, we need to show that process $p$ doesn't forever get stuck at Lines~\refln{fullcrash:try:3}, \refln{fullcrash:try:4}\upto\refln{fullcrash:try:5},
\refln{fullcrash:try:9}, \refln{fullcrash:try:10}, or \refln{fullcrash:try:12}, which will ensure that $p$ does return from $\tryprocp()$.
By Lemmas~\ref{lem:fullcrash:nowait1}, \ref{lem:fullcrash:nowait2}, and \ref{lem:fullcrash:nowait3} we know that $p$ eventually gets past the 
Lines~\refln{fullcrash:try:4}\upto\refln{fullcrash:try:5}, \refln{fullcrash:try:10}, and \refln{fullcrash:try:12} respectively.
When $\pcp = \refln{fullcrash:try:3}$, by Condition~\ref{inv:fullcrash:cond11}, $\myseq_p \in \{ \seq - 1, \seq \}$.
If $\myseq_p = \seq - 1$, by Condition~\ref{inv:fullcrash:cond15}, $\stopwait[\myseq_p \% 3] = \true \vee (\exists q, \pc{q} = \refln{fullcrash:rec:5} \wedge \myseq_{q} = \myseq_p)$.
In either case, $\stopwait[\myseq_p \% 3] = \true$ eventually, which $p$ notices at the wait loop of Line~\refln{fullcrash:try:3} and goes past the line.
Hence, we assume that whenever $p$ executes Lines~\refln{fullcrash:try:3} or \refln{fullcrash:try:9}, $\myseq_p = \seq$, 
because that is the only other possibility by Condition~\ref{inv:fullcrash:cond11}.
Therefore, we will argue next that the starvation freedom property of $\lock[\seq \% 3]$ is satisfied,
which will imply that $p$ gets past Lines~\refln{fullcrash:try:3} or \refln{fullcrash:try:9}.
We know from an inspection of the algorithm that any process that were to execute the $\initialize{}()$ at Line~\refln{fullcrash:rec:2},
would do so for $\mlock[\seq - 1]$ or $\mlock[\seq - 2]$, because by Condition~\ref{inv:fullcrash:cond11} $\myseq_p \in \{ \seq - 1, \seq \}$ for any process $p$ with $\pcp = \refln{fullcrash:rec:2}$.
It follows that from the last crash onwards, no process will execute $\mlock[\seq \% 3].\initialize{}()$, meeting the first condition for starvation freedom on the base lock.
It is straightforward that the second condition for starvation freedom on the base lock is met.
Lemmas~\ref{lem:fullcrash:nowait1}, \ref{lem:fullcrash:nowait2}, and \ref{lem:fullcrash:nowait3} ensure that the last condition for starvation freedom on the base lock is also met.
Hence we know that $\lock[\seq \% 3]$ satisfies starvation freedom.
Therefore, a process $p$ gets past Lines~\refln{fullcrash:try:3} or \refln{fullcrash:try:9} by the starvation freedom property of the base lock.
It follows that the claim holds.
\end{proof}

\begin{lemma}[{\bf Critical Section Reentry}] \label{lem:fullcrash:csr}
	In any run, if a process $p$ crashes while in the CS, 
	no other process enters the CS until $p$ subsequently reenters the CS.
\end{lemma}
\begin{proof}
The claim is immediate from Lemmas~\ref{lem:fullcrash:wfcsr} and \ref{lem:fullcrash:mutex}.
\end{proof}

\begin{lemma}[{\bf Bounded Recovery to CS/Exit}] \label{lem:fullcrash:boundedrec2csexit}
	There is an integer $b$ such that if in any run any process $p$ 
	executes $\recoverprocp()$ without crashing and with $\status_p \in \{\reccs, \recexit\}$, 
	the method completes in at most $b$ steps of $p$.
\end{lemma}
\begin{proof}
 An inspection of the $\recoverprocp()$ method reveals that the method completes within a constant number of steps,
 when a process $p$ executes $\recoverprocp()$ without crashing.
\end{proof}

\begin{lemma}[{\bf Fast Recovery to Remainder}] \label{lem:fullcrash:fastrec2rem}
	There is a constant $b$ (independent of $|\procset|$) such that if in any run any process $p$ 
	executes $\recoverprocp()$ without crashing and with $\status_p \in \{\good, \recrem\}$, 
	the method completes in at most $b$ steps of $p$.
\end{lemma}
\begin{proof}
 An inspection of the $\recoverprocp()$ method reveals that the method completes within a constant number of steps,
when a process $p$ executes $\recoverprocp()$ without crashing.
\end{proof}

\begin{lemma}[{\bf Bounded Recovery to Remainder}] \label{lem:fullcrash:boundedrec2rem}
	There is an integer $b$ such that if in any run
	$\recoverprocp()$, executed by a process $p$ with $\status_p = \rectry$, returns $\gotorem$,
	$p$ must have completed that execution of $\recoverprocp()$ in at most $b$ of its steps.
\end{lemma}
\begin{proof}
 An inspection of the $\recoverprocp()$ method reveals that the method completes within a constant number of steps,
when a process $p$ executes $\recoverprocp()$ without crashing.
\end{proof}

\subsection{RMR Complexity}
We know that the base lock incurs $O(1)$ RMR complexity to execute the procedures $\tryprocp()$, $\exitprocp()$, and $\initializep()$.
Hence, Lines~\refln{fullcrash:rec:2}, \refln{fullcrash:try:3}, \refln{fullcrash:try:9}, and \refln{fullcrash:exit:2} take $O(1)$ RMR to execute.
From an inspection of the algorithm it is clear that the rest of the lines from the algorithm in Figure~\ref{algo:fullcrash} incur $O(1)$ RMR on CC machines.
It follows that a process incurs $O(1)$ RMR per passage when executing the algorithm in Figure~\ref{algo:fullcrash}. 

\subsection{Proof of Invariant}
The proof of the invariant is in Appendix~\ref{sec:invariant-proof}

\subsection{CC Algorithm Main theorem}

The theorem below summarizes the result of the CC algorithm.

\begin{theorem}
	The algorithm in Figure~\ref{algo:fullcrash} solves the RME problem on CC machines for an arbitrary and unknown number of processes.
	It satisfies mutual exclusion, bounded exit, bounded critical section reentry, and starvation freedom.
	The RMR complexity of the algorithm on CC machines is $O(1)$ per passage.
\end{theorem}

\section{Invariant Proof}
\label{sec:invariant-proof}

\newcommand{\invcond}[1]{\ref{inv:fullcrash:cond#1}}
\newcommand{\ih}{IH}
\newcommand{\indhyp}[1]{\mbox{\ih:\invcond{#1}}}
\newcommand{\casnormal}{CAS}

\begin{lemma}
	The algorithm in Figure~\ref{algo:fullcrash} satisfies the invariant (i.e., the conjunction of all the conditions)
	stated in Figure~\ref{inv:fullcrash}, i.e., the invariant holds in every configuration of every run of the algorithm.
\end{lemma}
\begin{proof}
	We prove the lemma by induction.
	Specifically, we show 
	(i) {\em base case: } the invariant holds in the initial configuration, and 
	(ii) {\em induction step: } if the invariant holds in a configuration $C$ and 
	a step of a process takes the configuration $C$ to $C'$, then the invariant holds in $C'$.

	In the initial configuration,
	we have $\seq = 1$, 
	each of the $\mlock[0 \dots 2]$ are in their initial state, 
	$\stopwait[0 \dots 2]$ are all $\false$,
	$\csowner = \perp$, 
	$\forall p, \inuse_p = \false$, and
	$\myseq_p = 1$.
	Note, since the locks are in their initial state, 
	$\forall i \in [0, 2], \mlock[i].\status = (\phi, \phi, \phi)$.
	Since all processes are in the Remainder section, $\forall p, \pcp = \refln{fullcrash:rec:1}$.
	It follows that all the conditions hold trivially the initial configuration.
	Hence, we have the base case.
	
	For the induction step, let $C$ be an arbitrary configuration of an arbitrary run.
	The induction hypothesis states that Conditions~\invcond{1} through \invcond{14} hold in $C$.
	Let $(C, \sigma, C')$ be an arbitrary (normal or crash) step of a process in $\procset$.
	We now establish the induction step by arguing that each of Conditions~\invcond{1} through \invcond{14} holds in $C'$.
	
	We use the following notation in the proof: (i) \ih\ denotes the induction hypothesis and, for all $i \in [\invcond{1}, \invcond{14}]$, 
	\ih:$i$ is the part of \ih\ that states that Condition $i$ of the invariant holds in $C$, and
	(ii) If $D$ is any configuration and $x$ is any variable, $D.x$ is the value of $x$ in $D$.
	
	\begin{enumerate}
		\item
		\underline{Proof that Condition \invcond{1} holds in $C'$} \\
		$\seq$ changes only when some process executes Line~\refln{fullcrash:rec:4}.
		When a process $p$ executes that line, by \indhyp{11}, $\myseq_p \in \{ \seq - 1, \seq \}$.
		If $\myseq_p = \seq$, then $\seq$ increases by 1; otherwise, it remains unchanged.
		Thus, $C'.\seq$ is always greater than or equal to 1.
		
		Similarly, $\myseq_p$ changes at Line~\refln{fullcrash:try:2} or \refln{fullcrash:try:8}.
		When $\myseq_p$ changes at Line~\refln{fullcrash:try:2}, it assumes the value of $\seq$,
		hence we have $C'.\myseq_p \leq C'.\seq$.
		When $\myseq_p$ changes at Line~\refln{fullcrash:try:8}, 
		by \indhyp{11}, $\myseq_p = \seq -1$.
		Therefore, by the step $\myseq_p = \seq$ and hence $C'.\myseq_p \leq C'.\seq$.
		
		\item
		\underline{Proof that Condition \invcond{2} holds in $C'$} \\
		We establish each of the conjuncts of Condition~\invcond{2} separately as below.
		
		\begin{enumerate}
			\item[(a). ]
			\underline{$C'.\mlock[(\seq + 1) \% 3].\lockstate = (\phi, \phi, \phi)$}. \\
			Whenever the $\tryprocp()$ or $\exitprocp()$ of the base lock is called, it is called at Lines~\refln{fullcrash:try:3}, \refln{fullcrash:try:9}, or \refln{fullcrash:exit:2}.
			By \indhyp{11}, $\myseq_p \in \{ \seq - 1, \seq \}$ at these lines, which implies that these lines do not affect either of $\mlock[(\seq + 1) \% 3].\tryset$, $\mlock[(\seq + 1) \% 3].\csset$, or $\mlock[(\seq + 1) \% 3].\exitset$.
			The only place $\seq$ is changed is at Line~\refln{fullcrash:rec:4} when $\myseq_p = \seq$.
			We know from \indhyp{2} that when $\pcp = \refln{fullcrash:rec:4}$, $\mlock[(\myseq_p - 1) \% 3].\lockstate = (\phi, \phi, \phi)$.
			Since the step will increment $\seq$ by 1,
			it follows that $(C'.\myseq_p - 1) \% 3 = ((C'.\seq - 1) - 1) \% 3 = (C'.\seq - 2) \% 3 = (C'.\seq + 1) \% 3$.
			Therefore, by the step of $p$ at Line~\refln{fullcrash:rec:4} when $\myseq_p = \seq$, $\mlock[(C'.\seq  + 1) \% 3].\lockstate = (\phi, \phi, \phi)$.
			
			\item[(b). ]
			\underline{$C'.\pcp \in \{ \refln{fullcrash:rec:3}, \refln{fullcrash:rec:4} \} \limplies \mlock[(C'.\myseq_p - 1) \% 3].\lockstate = (\phi, \phi, \phi)$}. \\
				To prove this implication, assume that $C'.\pcp \in \{ \refln{fullcrash:rec:3}, \refln{fullcrash:rec:4} \}$. \\
				\\
				In case $C.\pcp = \refln{fullcrash:rec:2}$, $p$ executes $\mlock[(\myseq_p - 1) \% 3].\initializep()$, which sets $ \mlock[(\myseq_p - 1) \% 3].\lockstate$ to $(\phi, \phi, \phi)$. \\
				\\
				In case $C.\pcp \in \{ \refln{fullcrash:rec:3}, \refln{fullcrash:rec:4} \}$, by \indhyp{2} $\mlock[(C.\myseq_p - 1) \% 3].\lockstate = (\phi, \phi, \phi)$ and the step from $C$ to $C'$ is by some $q \neq p$.
				$q$ can modify the $\tryset$, $\csset$, or $\exitset$ of a lock only by executing either of Lines~\refln{fullcrash:try:3}, \refln{fullcrash:try:9}, or \refln{fullcrash:exit:2}.
				By \indhyp{11}, $\myseq_p \in \{ \seq -1, \seq \}$ when $\pc{q} \in \{ \refln{fullcrash:try:3}, \refln{fullcrash:exit:2} \}$,
				and $\myseq_p = \seq$ when $\pc{q} = \refln{fullcrash:try:9}$.
				If $q$ executed a step with $\pc{q} \in \{ \refln{fullcrash:try:3}, \refln{fullcrash:exit:2} \}$ and $\myseq_{q} = \seq - 1$,
				then by \indhyp{12}, $C.\myseq_p = \seq - 1$.
				We note that the step by $q$ doesn't affect either $\seq$ or $\myseq_p$.
				Thus the step by $q$ doesn't affect $\mlock[(C.\myseq_p - 1) \% 3].\tryset$, $\mlock[(C.\myseq_p - 1) \% 3].\csset$, or $\mlock[(C.\myseq_p - 1) \% 3].\exitset$.
				Hence, by \indhyp{2}, $\mlock[(C'.\myseq_p - 1) \% 3].\lockstate = (\phi, \phi, \phi)$. \\
				\\
				In case $q$ executed a step with $\pc{q} \in \{ \refln{fullcrash:try:3}, \refln{fullcrash:try:9}, \refln{fullcrash:exit:2} \}$ and $\myseq_{q} = \seq$,
				we know from \indhyp{11} that $C.\myseq_p \in \{ \seq - 1, \seq \}$.
				It follows from \indhyp{2} that $C.\mlock[(\myseq_p - 1) \% 3].\lockstate = (\phi, \phi, \phi)$
				and since the step by $q$ doesn't affect the $\tryset$, $\csset$, or $\exitset$ for either of $\mlock[(\seq - 1) \% 3]$ or $\mlock[(\seq - 2) \% 3]$,
				we have $C'.\mlock[(\myseq_p - 1) \% 3].\lockstate = (\phi, \phi, \phi)$.
		\end{enumerate}

		\item
		\underline{Proof that Condition \invcond{3} holds in $C'$} \\
		We establish each of the conjuncts of Condition~\invcond{3} separately as below.
		
		\begin{enumerate}
			\item[(a). ]
			\underline{$C'.\stopwait[\seq \% 3] = \false$}. \\
			A cell from $\stopwait$ is changed only at Line~\refln{fullcrash:rec:3} and \refln{fullcrash:rec:5}.
			Line~\refln{fullcrash:rec:3} sets $\stopwait[(\myseq_p - 1) \% 3]$ to $\false$, therefore it is not of concern to argue this case.
			If $\pcp = \refln{fullcrash:rec:5}$, by \indhyp{11}, $\myseq_p  = \seq -1$.
			It follows that an execution of Line~\refln{fullcrash:rec:5} doesn't modify $\stopwait[\seq \% 3]$.\\
			\\
			The only place $\seq$ is changed is at Line~\refln{fullcrash:rec:4} when $\myseq_p = \seq$.
			We know from \indhyp{3} that $\stopwait[(\seq + 1) \% 3] = \false$.
			Since the step will increment $\seq$ by 1,
			it follows that $(C'.\seq) \% 3 = (C.\seq + 1) \% 3$.
			Therefore, by the step of $p$ at Line~\refln{fullcrash:rec:4} when $\myseq_p = \seq$, $\stopwait[C'.\seq \% 3] = \false$.
			
			\item[(b). ]
			\underline{$C'.\stopwait[(\seq + 1) \% 3] = \false$}. \\
			A cell from $\stopwait$ is changed only at Line~\refln{fullcrash:rec:3} and \refln{fullcrash:rec:5}.
			Line~\refln{fullcrash:rec:3} sets $\stopwait[(\myseq_p - 1) \% 3]$ to $\false$, therefore it is not of concern to argue this case.
			If $\pcp = \refln{fullcrash:rec:5}$, by \indhyp{11}, $\myseq_p  = \seq -1$.
			It follows that an execution of Line~\refln{fullcrash:rec:5} doesn't modify $\stopwait[(\seq + 1) \% 3]$.\\
			\\
			The only place $\seq$ is changed is at Line~\refln{fullcrash:rec:4} when $\myseq_p = \seq$.
			We know from \indhyp{3} that when $\pcp = \refln{fullcrash:rec:4}$, $\stopwait[(\myseq_p - 1) \% 3] = \false$.
			Since the step will increment $\seq$ by 1,
			it follows that $(C'.\myseq_p - 1) \% 3 = ((C'.\seq - 1) - 1) \% 3 = (C'.\seq - 2) \% 3 = (C'.\seq + 1) \% 3$.
			Therefore, by the step of $p$ at Line~\refln{fullcrash:rec:4} when $\myseq_p = \seq$, $\stopwait[(C'.\seq  + 1) \% 3] = \false$.
			
			\item[(c). ]
			\underline{$C'.\pcp =\refln{fullcrash:rec:4} \limplies \stopwait[(C'.\myseq_p - 1) \% 3] = false$}. \\
			To prove this implication, assume that $C'.\pcp = \refln{fullcrash:rec:4}$. \\
			\\
			In case $C.\pcp = \refln{fullcrash:rec:3}$, $p$ sets $\stopwait[(\myseq_p - 1) \% 3]$ to $\false$ during the step. \\
			\\
			In case $C.\pcp = \refln{fullcrash:rec:4}$, by \indhyp{3} $\stopwait[(C.\myseq_p - 1) \% 3] = \false$ and the step from $C$ to $C'$ is by some $q \neq p$.
			$q$ can set a cell from $\stopwait$ to $\true$ only at Line~\refln{fullcrash:rec:5}, specifically, $q$ sets $\stopwait[\myseq_{q} \% 3]$ to $\false$ at Line~\refln{fullcrash:rec:5}.
			By \indhyp{11} when $\pc{q} = \refln{fullcrash:rec:5}$, $\myseq_{q} = \seq - 1$.
			It follows from \indhyp{12} that $\myseq_p = \seq -1$.\\
			\\
			Thus $\stopwait[(C.\myseq_p - 1) \% 3] = \stopwait[(C'.\myseq_p - 1) \% 3] = \false$.
		\end{enumerate}

		\item
		\underline{Proof that Condition \invcond{4} holds in $C'$} \\
		We establish each of the conjuncts of Condition~\invcond{4} separately as below.
		
		\begin{enumerate}
			\item[(a). ]
			\underline{$(C'.\inuse_p = \false \vee C'.\myseqp < C'.\seq \vee C'.\pcp \in \{ \refln{fullcrash:try:2},  \refln{fullcrash:exit:3}, \refln{fullcrash:exit:4}, \refln{fullcrash:rec:6}, \refln{fullcrash:rec:7} \})$}\\
			\hspace*{5mm} \underline{$\limplies$ $p \notin \mlock[C'.\seq \% 3].\set$}. \\
			We divide the implication into sub-parts as below and argue the correctness of each part.
			\begin{enumerate}
				\item[i. ]
				\underline{$C'.\inuse_p = \false$ $\limplies$ $p \notin \mlock[C'.\seq \% 3].\set$}. \\
				To prove this implication, assume that $C'.\inuse_p = \false$. \\
				\\
				In case $C.\inuse_p = \false$, by \indhyp{4} $p \notin \mlock[C.\seq \% 3].\set$ and the step from $C$ to $C'$ is by some $q \neq p$.
				The only step that can affect the condition is $q$'s changing of the value of $\seq$, which could happen at Line~\refln{fullcrash:rec:4} and that line only increments $\seq$ by 1.
				We note that \indhyp{2} implies that $\mlock[(C.\seq + 1) \% 3].\lockstate = (\phi, \phi, \phi)$ and $q$'s increment of $\seq$ means that
				$\mlock[(C'.\seq) \% 3].\lockstate = (\phi, \phi, \phi)$, which means $p \notin \mlock[C'.\seq \% 3].\set$. \\
				\\
				In case $C.\inuse_p = \true$, only $p$'s execution of one of Lines~\refln{fullcrash:exit:4} or \refln{fullcrash:rec:7} could turn $\inuse_p$ to $\false$.
				Hence, $C.\pcp \in \{ \refln{fullcrash:exit:4}, \refln{fullcrash:rec:7} \}$.
				By \indhyp{4} we have that $p \notin \mlock[C.\seq \% 3].\set$, which implies $p \notin \mlock[C'.\seq \% 3].\set$.
				\item[ii. ]
				\underline{$C'.\myseqp < C'.\seq$ $\limplies$ $p \notin \mlock[C'.\seq \% 3].\set$}. \\
				To prove this implication, assume that $C'.\myseqp < C'.\seq$. \\
				\\
				In case $C.\myseqp < C.\seq$, by \indhyp{4} $p \notin \mlock[C.\seq \% 3].\set$ and the step from $C$ to $C'$ is by some $q \neq p$.
				This case is similar to the one as argued for Case~i argued above. \\
				\\
				We know from \indhyp{1} that $C.\myseq_p \leq C.\seq$.
				Thus, the other case is that $C.\myseq_p = C.\seq$.
				It follows from an inspection of the algorithm that $C'.\seq = C.\seq + 1$, because $\myseq_p$ is never decremented.
				Thus, the step is an execution of Line~\refln{fullcrash:rec:4} by a process $q$ (possibly same as $p$).
				By \indhyp{2} we have $C.\mlock[(\seq + 1) \% 3].\lockstate = (\phi, \phi, \phi)$, which implies that $p \notin \mlock[C'.\seq \% 3].\set$.
				
				\item[iii. ] 
				\underline{$C'.\pcp \in \{ \refln{fullcrash:try:2},  \refln{fullcrash:exit:3}, \refln{fullcrash:exit:4}, \refln{fullcrash:rec:6}, \refln{fullcrash:rec:7} \}$ $\limplies$ $p \notin \mlock[C'.\seq \% 3].\set$}. \\
				To prove this implication, assume that $C'.\pcp \in \{ \refln{fullcrash:try:2},  \refln{fullcrash:exit:3}, \refln{fullcrash:exit:4}, \refln{fullcrash:rec:6}, \refln{fullcrash:rec:7} \}$. \\
				\\
				In case $C.\pcp \in \{ \refln{fullcrash:try:2},  \refln{fullcrash:exit:3}, \refln{fullcrash:exit:4}, \refln{fullcrash:rec:6}, \refln{fullcrash:rec:7} \}$,
				by \indhyp{4} $p \notin \mlock[C.\seq \% 3].\set$.
				If $p$ took a step at one of Lines~\refln{fullcrash:exit:3} or \refln{fullcrash:rec:6} to move to Lines~\refln{fullcrash:exit:4} or \refln{fullcrash:rec:7}, respectively,
				the step doesn't affect $\mlock[C.\seq \% 3].\set$.
				Which implies $p \notin \mlock[C'.\seq \% 3].\set$.
				In case the step from $C$ to $C'$ is by some $q \neq p$, 
				the case is similar to the one as argued for Case~i argued above. \\
				\\
				Assume that $C.\pcp \in \{ \refln{fullcrash:try:1},  \refln{fullcrash:exit:1}, \refln{fullcrash:exit:2}, \refln{fullcrash:rec:1}, \refln{fullcrash:rec:5} \}$ so that due to a step by $p$,
				$C'.\pcp \in \{ \refln{fullcrash:try:2},  \refln{fullcrash:exit:3}, \refln{fullcrash:exit:4}, \refln{fullcrash:rec:6}, \refln{fullcrash:rec:7} \}$.
				If $C.\pcp = \refln{fullcrash:try:1}$, by \indhyp{10} $\activep = \false$ and by \indhyp{4}, $p \notin \mlock[C.\seq \% 3].\set$, which implies that $p \notin \mlock[C'.\seq \% 3].\set$.
				If $C.\pcp = \refln{fullcrash:exit:1}$, it follows from the step that $C.\myseq_p \neq C.\seq$ and by \indhyp{1} $C.\myseq_p < C.\seq$.
				By \indhyp{4} it follows that $p \notin \mlock[C.\seq \% 3].\set$, which implies that $p \notin \mlock[C'.\seq \% 3].\set$.
				If $C.\pcp = \refln{fullcrash:exit:2}$, the step completes $p$'s execution of $\mlock[C.\myseq_p \% 3].\exitprocp()$ and hence removes $p$ from $\mlock[C.\myseq_p \% 3].\exitset$.
				By \indhyp{5} $p$ is in at most one of $\mlock[C.\myseq_p \% 3].\tryset$, $\mlock[C.\myseq_p \% 3].\csset$, or $\mlock[C.\myseq_p \% 3].\exitset$,
				which implies that $p \notin \mlock[C'.\myseq_p \% 3].\set$.
				If $C.\myseq_p = C.\seq$, we have $p \notin \mlock[C'.\seq \% 3].\set$;
				otherwise $C.\myseq_p < C.\seq$ and by \indhyp{4}, $p \notin \mlock[C'.\seq \% 3].\set$.
				If $C.\pcp = \refln{fullcrash:rec:1}$, it follows that either $C.\inuse_p = \false$ or $C.\myseq_p < C.\seq$,
				in either case $p \notin \mlock[C'.\seq \% 3].\set$ as already argued above.
				If $C.\pcp = \refln{fullcrash:rec:5}$, by \indhyp{11} $C.\myseq_p < C.\seq$, which implies that $p \notin \mlock[C'.\seq \% 3].\set$ as already argued above.
			\end{enumerate}
			
			\item[(b). ]
			\underline{$C'.\myseqp < C'.\seq - 1 \limplies p \notin \mlock[C'.\myseqp \% 3].\set$}. \\
			To prove this implication, assume that $C'.\myseqp < C'.\seq - 1$. \\
			\\
			If $C.\myseq_p < C.\seq - 1$, then by \indhyp{4} $p \notin \mlock[C.\myseq_p \% 3].\set$ and the step from $C$ to $C'$ is by some $q \neq p$.
			This step by $q$ doesn't affect $\myseq_p$ and doesn't put $p$ in any sets associated with any base lock.
			Therefore it follows that $p \notin \mlock[C'.\myseq_p \% 3].\set$. \\
			\\
			Since $C'.\myseq_p < C'.\seq - 1$, the only other case possible is that $C.\myseq_p = C.\seq - 1$.
			In this case $\seq$ was incremented due to an execution of Line~\refln{fullcrash:rec:4} by a process $q$.
			Since the step increments $\seq$, it follows that $C.\myseq_{q} = C.\seq$ and hence $p \neq q$.
			By \indhyp{4}, since $\myseq_p < \seq$, $p \notin \mlock[C.\seq \% 3].\set$ and hence $p \notin \mlock[(C.\seq - 1) \% 3].\set$.
			By \indhyp{2}, $p \notin \mlock[(C.\seq + 1) \% 3].\set$ which implies that $p \notin \mlock[C'.\seq \% 3].\set$.
			Since we have $C.\pc{q} = \refln{fullcrash:rec:4}$, by \indhyp{2}, $p \notin \mlock[(C.\myseq_{q} - 1) \% 3].\set$, which implies $p \notin \mlock[(C'.\seq + 1) \% 3].\set$ (because the step by $q$ increments $\seq$ by 1 and $C.\myseq_{q} = C.\seq$).
			It follows that $p$ is not in any of the lock sets and we have the implication.
		\end{enumerate}
		\item
		\underline{Proof that Condition \invcond{5} holds in $C'$} \\
		$p$ is moved to different sets only due to the invocation/execution of $\ell.\tryprocp()$ and $\ell.\exitprocp()$ on the base lock $\ell$. \\
		\\
		When $p$ executes Line~\refln{fullcrash:try:2} to invoke $\lock[C'.\myseq_p \% 3].\tryprocp()$ at Line~\refln{fullcrash:try:3},
		by \indhyp{4} $p \notin \lock[C.\seq \% 3].\set$.
		Similarly when it executes Line~\refln{fullcrash:try:8}, by \indhyp{11} $C.\myseq_p = C.\seq -1$ and thus by \indhyp{4} $p \notin \lock[C.\seq \% 3].\set$.
		Since $C'.\myseq_p = C.\seq$ due to the step,
		it follows that the condition holds in $C'$. \\
		\\
		When $p$ successfully completes $\lock[C.\myseq_p \% 3].\tryprocp()$ at Lines~\refln{fullcrash:try:3} or \refln{fullcrash:try:9}, $p$ is moved to $\lock[C.\myseq_p \% 3].\csset$ and removed from $\lock[C.\myseq_p \% 3].\tryset$. 
		Thus the condition holds in $C'$.\\
		\\
		When $p$ invokes $\lock[C.\myseq_p \% 3].\exitprocp()$ at Line~\refln{fullcrash:exit:2}, $p$ is moved from $\lock[C.\myseq_p \% 3].\csset$ to $\lock[C.\myseq_p \% 3].\exitset$.  
		Similarly when $p$ completes $\lock[C.\myseq_p \% 3].\exitprocp()$ successfully, $p$ is removed from $\lock[C.\myseq_p \% 3].\set$.
		Thus the condition holds in $C'$. \\
		\\
		When a process $q$ executes $\lock[C.\myseq_q \% 3].\initialize{q}()$ at Line~\refln{fullcrash:rec:2}, it only empties the three sets.
		Thereby $p$ is in none of the sets in $C'$ and thus satisfying the condition.
		
		\item
		\underline{Proof that Condition \invcond{6} holds in $C'$} \\
		When a process $p$ takes a step from Line~\refln{fullcrash:try:2} in $C$ to go to Line~\refln{fullcrash:try:3} in $C'$,
		by \indhyp{4} $p \notin \mlock[C.\seq \% 3].\set$.
		The implication then holds since invoking $\lock[\myseq_p \% 3].\tryprocp()$ at Line~\refln{fullcrash:try:3} puts $p$ into $\lock[\myseq_p \% 3].\tryset$.
		Similarly, when a process $p$ takes a step from Line~\refln{fullcrash:try:8} to go to Line~\refln{fullcrash:try:9},
		we have $C.\myseq_p = C.\seq - 1 < C.\seq$ by \indhyp{11}, and hence by \indhyp{4} $p \notin \mlock[C.\seq \% 3].\set$.
		The implication again holds since invoking $\lock[\myseq_p \% 3].\tryprocp()$ at Line~\refln{fullcrash:try:9} puts $p$ into $\lock[\myseq_p \% 3].\tryset$. \\
		\\
		If a process $q \neq p$ takes a step at Line~\refln{fullcrash:rec:2} to execute $\lock[(\myseq_{q} - 1) \% 3].\initialize{q}()$ and thus set $\lock[(\myseq_{q} - 1) \% 3].\lockstate$ to $(\phi, \phi, \phi)$,
		we first note that by \indhyp{11} $\myseq_{q} \in \{ \seq - 1, \seq \}$.
		If $C.\pcp = \refln{fullcrash:try:3}$ and $C.\myseq_p = C.\seq - 1$, then by \indhyp{12} $C.\myseq_{q} = C.\seq - 1$, which implies that $q$'s step doesn't affect $\lock[\myseq_p \% 3].\lockstate$.
		If $C.\myseq_p = C.\seq$, then since $\myseq_{q} \in \{ \seq - 1, \seq \}$, $q$'s step only affects $\lock[(C.\seq - 1)\% 3].\lockstate$ or $\lock[(C.\seq - 2) \% 3].\lockstate = \lock[(C.\seq + 1) \% 3].\lockstate$,
		and not $\lock[C.\seq \% 3].\lockstate$.
		
		\item
		\underline{Proof that Condition \invcond{7} holds in $C'$} \\
		We divide the implication into sub-parts as below and argue the correctness of each part.
		
		\begin{enumerate}
			\item[(a). ]
			\underline{$(C'.\pcp = \refln{fullcrash:try:4} \wedge C'.\stopwait[\myseq_p \% 3] = \false) \limplies p \in \mlock[C'.\myseqp \% 3].\csset$}. \\
			Here we first argue the correctness of the condition when $p$ takes a step so that $C'.\pcp = \refln{fullcrash:try:4}$.
			There are two possibilities: $C.\pcp = \refln{fullcrash:try:3}$ or $C.\pcp = \refln{fullcrash:try:5}$.
			In case $C.\pcp = \refln{fullcrash:try:3}$, we know that $C'.\stopwait[\myseq_p \% 3] = \false$ from the hypothesis of the implication.
			Since the step doesn't change $C.\stopwait[\myseq_p \% 3]$, it follows that $p$ completed the execution of $\lock[\myseq_p \% 3].\tryprocp()$ successfully.
			As a result of that $p$ is removed from $\lock[C.\myseq_p \% 3].\tryset$ and put into $\lock[C.\myseq_p \% 3].\csset$, and thereby $p \in \lock[C'.\myseq_p \% 3].\csset$.
			In case $C.\pcp = \refln{fullcrash:try:5}$, by \indhyp{7} $p \in \lock[C.\myseq_p \% 3].\csset$ and hence $p \in \lock[C'.\myseq_p \% 3].\csset$.
			In either case $p \in \lock[C'.\myseq_p \% 3].\csset$. \\
			\\
			If the step from $C$ to $C'$ is by a process $q \neq p$, only a step at Line~\refln{fullcrash:rec:2} could possibly affect $\lock[C.\myseq_p \% 3].\lockstate$.
			The argument for this situation is same as that argued for Condition~\invcond{6} above, which shows that $\lock[C.\myseq_p \% 3].\lockstate$ is unaffected by $q$'s step.
			
			\item[(b). ]
			\underline{$C'.\pcp \in \{ \refln{fullcrash:try:5} \upto \refln{fullcrash:try:7}, \refln{fullcrash:try:10} \upto \refln{fullcrash:try:14} \}$ $\limplies$ $p \in \mlock[C'.\myseqp \% 3].\csset$}. \\
			Here we need to show that when $p$ takes either a step from Line~\refln{fullcrash:try:4} to go to Line~\refln{fullcrash:try:5} or from Line~\refln{fullcrash:try:9} to go to Line~\refln{fullcrash:try:10},
			the implication holds.
			Consider $C.\pcp = \refln{fullcrash:try:4}$.
			Since $C'.\pcp = \refln{fullcrash:try:5}$, we note that $C.\stopwait[\myseq_p \% 3] = \false$, and hence by \indhyp{7} $p \in \lock[C.\myseq_p \% 3].\csset$ and thereby $p \in \lock[C'.\myseq_p \% 3].\csset$.
			If $C.\pcp = \refln{fullcrash:try:9}$, $p$ completed $\lock[\myseq_p \% 3].\tryprocp()$ to go to Line~\refln{fullcrash:try:10}, which removes $p$ from $\lock[C.\myseq_p \% 3].\tryset$ and puts in $\lock[C.\myseq_p \% 3].\csset$.
			In either case $p \in \lock[C'.\myseq_p \% 3].\csset$. \\
			\\
			If the step from $C$ to $C'$ is by a process $q \neq p$, only a step at Line~\refln{fullcrash:rec:2} could possibly affect $\lock[C.\myseq_p \% 3].\lockstate$.
			We note that $C.\pcp \in \{ \refln{fullcrash:try:5} \upto \refln{fullcrash:try:7}, \refln{fullcrash:try:10} \upto \refln{fullcrash:try:14} \}$.
			If $C.\pcp \in \{ \refln{fullcrash:try:5} \upto \refln{fullcrash:try:6} \}$ and $C.\myseq_p = C.\seq - 1$, we use the same argument as for Condition~\invcond{6} above and note that $\lock[C.\myseq_p \% 3].\lockstate$ is unaffected by $q$'s step.
			Similarly, for $C.\pcp \in \{ \refln{fullcrash:try:7}, \refln{fullcrash:try:10} \upto \refln{fullcrash:try:14} \}$,
			\indhyp{12} implies that $\myseq_{q} = \seq - 1$ and again we observe that $\lock[C.\myseq_p \% 3].\lockstate$ would be unaffected by $q$'s step.
			Otherwise, by Condition~\invcond{11} $C.\myseq_p = C.\seq$ and this case is also argued the same way as the argument for Condition~\invcond{6} above.
			In either case $\lock[C.\myseq_p \% 3].\lockstate$ is unaffected by the step.

			\item[(c). ]
			\underline{$(C'.\pcp \in \{ \refln{fullcrash:cs:1}, \refln{fullcrash:exit:1} \} \wedge C'.\myseq_p = C'.\seq)$ $\limplies$ $p \in \mlock[C'.\myseqp \% 3].\csset$}. \\
			Here we need to show that when $p$ takes a step from $C.\pcp \in \{ \refln{fullcrash:try:6}, \refln{fullcrash:try:11}, \refln{fullcrash:try:14} \}$ to $C'.\pcp = \refln{fullcrash:cs:1}$,
			the implication holds.
			When $C.\pcp = \refln{fullcrash:try:6}$, it is straightforward to see that the implication holds because the step only modifies $\csowner$.
			Similarly it follows for a step from $C.\pcp = \refln{fullcrash:try:11}$.
			For a step from $C.\pcp = \refln{fullcrash:try:14}$, the implication holds trivially by \indhyp{7}. \\
			\\
			If the step from $C$ to $C'$ is by a process $q \neq p$, only a step at Line~\refln{fullcrash:rec:2} could possibly affect $\lock[C.\myseq_p \% 3].\lockstate$.
			We know from the hypothesis of the implication that $C.\myseq_p = C.\seq$.
			By \indhyp{11} $C.\myseq_{q} \in \{ C.\seq - 1, C.\seq \}$, and in either case $q$ only affects $\lock[(C.\seq - 1) \% 3].\lockstate$ or $\lock[(C.\seq - 2) \% 3].\lockstate = \lock[(C.\seq + 1) \% 3].\lockstate$,
			not $\lock[C.\seq \% 3].\lockstate$.
			Therefore we have $p \in \mlock[C'.\myseqp \% 3].\csset$.
			
		\end{enumerate}
	
		\item
		\underline{Proof that Condition \invcond{8} holds in $C'$} \\
		When a process $p$ takes a step from Line~\refln{fullcrash:exit:1} in $C$ with $C.\myseq_p = C.\seq$ to go to Line~\refln{fullcrash:exit:1} in $C'$,
		by \indhyp{7} $p \in \mlock[C.\seq \% 3].\csset$.
		The implication then holds since invoking $\lock[\myseq_p \% 3].\exitprocp()$ at Line~\refln{fullcrash:exit:2} puts $p$ into $\lock[\myseq_p \% 3].\exitset$. \\
		\\
		If a process $q \neq p$ takes a step at Line~\refln{fullcrash:rec:2} to execute $\lock[(\myseq_{q} - 1) \% 3].\initialize{q}()$ and thus set $\lock[(\myseq_{q} - 1) \% 3].\lockstate$ to $(\phi, \phi, \phi)$,
		we first note that by \indhyp{11} $\myseq_{q} \in \{ \seq - 1, \seq \}$.
		By \indhyp{11}, since $C.\pcp = \refln{fullcrash:exit:2}$, $C.\myseq_p \in \{ C.\seq - 1, C.\seq \}$.
		If $C.\myseq_p = C.\seq - 1$, then by \indhyp{12} $C.\myseq_{q} = C.\seq - 1$, which implies that $q$'s step doesn't affect $\lock[\myseq_p \% 3].\lockstate$.
		If $C.\myseq_p = C.\seq$, then since $\myseq_{q} \in \{ \seq - 1, \seq \}$, $q$'s step only affects $\lock[(C.\seq - 1)\% 3].\lockstate$ or $\lock[(C.\seq - 2) \% 3].\lockstate = \lock[(C.\seq + 1) \% 3].\lockstate$,
		and not $\lock[C.\seq \% 3].\lockstate$.
		
		\item
		\underline{Proof that Condition \invcond{9} holds in $C'$} \\
		We establish each of the conjuncts of Condition~\invcond{9} separately as below.
		
		\begin{enumerate}
			\item[(a). ]
			\underline{$(C'.\inuse_p = \false \vee C'.\pcp \in \{ \refln{fullcrash:try:1} \upto \refln{fullcrash:try:12}, \refln{fullcrash:exit:4}, \refln{fullcrash:rec:7} \}) \limplies C'.\csowner \neq p$} \\
			We divide the implication into sub-parts as below and argue the correctness of each part.
			\begin{enumerate}
				\item[i. ]
				\underline{$C'.\inuse_p = \false$ $\limplies$ $C'.\csowner \neq p$}. \\
				Assume $C.\inuse_p = \true$.
				Here we need to argue that when $p$ takes a step to set $C.\inuse_p$ to $\false$, it does so either at Line~\refln{fullcrash:exit:4} or \refln{fullcrash:rec:7}.
				Hence, $C.\pcp \in \{ \refln{fullcrash:exit:4}, \refln{fullcrash:rec:7} \}$.
				It follows from \indhyp{9} that $C.\csowner \neq p$ and since the step doesn't affect $\csowner$, $C'.\csowner \neq p$. \\
				\\
				Assume $C.\inuse_p = \false$, we have two cases: 
				(a) $p$ takes a step where it doesn't modify $\inuse_p$, or
				(b) the step from $C$ to $C'$ is by a process $q \neq p$. 
				In the first case we have from the contrapositive of first part of \indhyp{10} that $C.\pcp \notin \{ \refln{fullcrash:try:6}, \refln{fullcrash:try:11}, \refln{fullcrash:try:13}, \refln{fullcrash:exit:3} \}$.
				Thus $p$ doesn't modify $\csowner$ due to the step, hence we have the implication.
				In the second case if $q$ at all modifies $\csowner$, it does so at one of Lines~\refln{fullcrash:try:6}, \refln{fullcrash:try:11}, \refln{fullcrash:try:13}, or \refln{fullcrash:exit:3},
				where it doesn't write $p$'s name in $C.\csowner$.
				Thus in either case the implication holds in $C'$.
				
				\item[ii. ]
				\underline{$C'.\pcp \in \{ \refln{fullcrash:try:1} \upto \refln{fullcrash:try:12}, \refln{fullcrash:exit:4}, \refln{fullcrash:rec:7} \}$ $\limplies$ $C'.\csowner \neq p$}. \\
				Suppose $C.\pcp = \refln{fullcrash:rem:1}$ such that $p$ takes a step to go to Line~\refln{fullcrash:try:1}.
				It follows that $C.\status_p$ must be $\good$, otherwise $p$ would invoke $\recoverprocp()$ instead of $\tryprocp()$.
				By \indhyp{10} $C.\inuse_p = \false$ and thus by \indhyp{9} $C.\csowner \neq p$.
				It follows that $C'.\csowner \neq p$. \\
				\\
				Suppose $C.\pcp \in \{ \refln{fullcrash:try:1} \upto \refln{fullcrash:try:11}\}$.
				In that case the step from $C$ to $C'$ is by $p$ and doesn't affect $\csowner$.
				This include the CAS operations at Lines~\refln{fullcrash:try:6} and \refln{fullcrash:try:11}, which must have failed for $p$ to go to the next line instead of the CS.
				It follows from \indhyp{9} that $C.\csowner \neq p$ and thus $C'.\csowner \neq p$. \\
				\\
				Suppose $C.\pcp = \refln{fullcrash:exit:3}$.
				In that case $p$'s execution of Line~\refln{fullcrash:exit:3} sets $\csowner$ to $\perp$ and hence the implication holds in $C'$. \\
				\\
				Suppose $C.\pcp = \refln{fullcrash:rec:6}$.
				In that case $p$ went to Line~\refln{fullcrash:rec:7} because the \ifcode condition at Line~\refln{fullcrash:rec:6} failed.
				Thus the implication holds in $C'$. \\
				\\
				Suppose $C.\pcp = C'.\pcp$.
				In that case the step from $C$ to $C'$ is by some process $q$.
				If $q$ at all modifies $\csowner$, it does so at one of Lines~\refln{fullcrash:try:6}, \refln{fullcrash:try:11}, \refln{fullcrash:try:13}, or \refln{fullcrash:exit:3},
				where it doesn't write $p$'s name in $C.\csowner$.
				Thus, the implication holds in $C'$.

			\end{enumerate}
		
			\item[(b). ]
			\underline{$C'.\pcp = \refln{fullcrash:try:13} \limplies C'.\csowner = \perp$} \\
			Suppose $C.\pcp = \refln{fullcrash:try:12}$.
			Since $C'.\pcp = \refln{fullcrash:try:13}$, $p$ left the \waittill loop at Line~\refln{fullcrash:try:12} to move to Line~\refln{fullcrash:try:13}.
			It could happne only when $\csowner = \perp$.
			Thus we have the implication. \\
			\\
			Suppose $C.\pcp = \refln{fullcrash:try:13}$.
			In that case the step from $C$ to $C'$ is by some process $q$.
			Note that by \indhyp{9} $C.\csowner = \perp$.
			We argue that $q$ cannot take a step at Line~\refln{fullcrash:try:6}, \refln{fullcrash:try:11}, \refln{fullcrash:try:13}, or \refln{fullcrash:exit:3} to modify $\csowner$.
			First, $C.\pc{q} \neq \refln{fullcrash:exit:3}$, otherwise, by \indhyp{9} $\csowner = q \neq \perp$, a contradiction.
			Second, it can't be the case that $C.\pc{q} \in \{ \refln{fullcrash:try:6}, \refln{fullcrash:try:11}, \refln{fullcrash:try:13} \}$ and $C.\myseq_{q} = \seq$.
			This is because by \indhyp{7} and \indhyp{11} we already have $p \in \lock[C.\seq \% 3]$ and hence the mutual exclusion condition of $\lock[C.\seq \% 3]$ prevents the case from occurring.
			Lastly, if $C.\myseq_{q} = C.\seq - 1$, by \indhyp{14} $\pc{q} \in \{ \refln{fullcrash:rem:1} \upto \refln{fullcrash:try:3}, \refln{fullcrash:try:8}, \refln{fullcrash:exit:4}, \refln{fullcrash:rec:1} \upto \refln{fullcrash:rec:8} \}$,
			and $q$'s step at any of these lines doesn't modify $\csowner$.
			
			\item[(c). ]
			\underline{$(C'.\pcp \in \{ \refln{fullcrash:try:14} \upto \refln{fullcrash:exit:3} \} \vee C'.\status_p = \reccs) \limplies C'.\csowner = p$} \\
			We divide the implication into sub-parts as below and argue the correctness of each part.
			\begin{enumerate}
				\item[i. ]
				\underline{$C'.\pcp \in \{ \refln{fullcrash:try:14} \upto \refln{fullcrash:exit:3} \}$ $\limplies$ $C'.\csowner = p$}. \\
				In order to argue the correctness we first assume that $C'.\pcp \in \{ \refln{fullcrash:try:14} \upto \refln{fullcrash:exit:3} \}$.
				$p$ could reach to the given line in $C'$ due to its own step from different locations of the program, or $p$ might not have changed $\pcp$ but the step was due to a process $q \neq p$.
				In the following we argue each of these cases and show the correctness of the condition. \\
				\\
				Suppose $C.\pcp = \refln{fullcrash:try:6}$. 
				Since $C'.\pcp \in \{ \refln{fullcrash:try:14} \upto \refln{fullcrash:exit:3} \}$, we note that specifically Line~\refln{fullcrash:try:6} puts $p$ into the CS (at Line~\refln{fullcrash:cs:1}) due to a successful CAS.
				It follows that the CAS writes $p$'s name into $\csowner$ and hence we have $C'.\csowner = p$.\\
				\\
				Suppose $C.\pcp = \refln{fullcrash:try:11}$. 
				Since $C'.\pcp \in \{ \refln{fullcrash:try:14} \upto \refln{fullcrash:exit:3} \}$, we note that specifically Line~\refln{fullcrash:try:11} puts $p$ into the CS (at Line~\refln{fullcrash:cs:1}) due to a successful CAS.
				It follows that the CAS writes $p$'s name into $\csowner$ and hence we have $C'.\csowner = p$.\\
				\\
				Suppose $C.\pcp = \refln{fullcrash:try:13}$. 
				Since $C'.\pcp \in \{ \refln{fullcrash:try:14} \upto \refln{fullcrash:exit:3} \}$, we note that specifically Line~\refln{fullcrash:try:13} writes $p$ into $\csowner$ and increments $\pcp$.
				Thus we have $C'.\pcp = \refln{fullcrash:try:14}$ and $C'.\csowner = p$.\\
				\\
				Suppose $C.\pcp = \refln{fullcrash:rec:6}$. 
				Since $C'.\pcp \in \{ \refln{fullcrash:try:14} \upto \refln{fullcrash:exit:3} \}$, we note that specifically Line~\refln{fullcrash:rec:6} puts $p$ back into the CS (at Line~\refln{fullcrash:cs:1}) because $p$ finds $C.\csowner = p$ and the \ifcode condition at Line~\refln{fullcrash:rec:6} is met.
				It follows that we have $C.\csowner = p$ and hence we have $C'.\csowner = p$.\\
				\\
				Suppose $C.\pcp = C'.\pcp$. 
				From \indhyp{9} we note that $C.\csowner = p$.
				It follows that the step from $C$ to $C'$ is by a process $q \neq p$.
				We argue as below that $q$'s step doesn't modify $C.\csowner$,
				and by \indhyp{9} $C.\csowner = p$, which would imply $C'.\csowner = p$.
				First, if $q$ were to take a step at Lines~\refln{fullcrash:try:6} or \refln{fullcrash:try:11}, $q$'s CAS would fail since $C.\csowner = p \neq \perp$.
				Second, $C.\pc{q} \neq \refln{fullcrash:try:13}$ because we know $C.\csowner = p$ and had $C.\pc{q}$ been \refln{fullcrash:try:13}, by \indhyp{9} $C.\csowner = \perp \neq p$, a contradiction.
				Lastly, $C.\pc{q} \neq \refln{fullcrash:exit:3}$, again because we know $C.\csowner = p$ and had $C.\pc{q}$ been \refln{fullcrash:exit:3}, by \indhyp{9} $C.\csowner = q \neq p$, a contradiction.
				Thus, from the above we conclude that $q$'s step didn't modify $C.\csowner$.
				Hence we have $C'.\csowner = p$.
				
				\item[ii. ]
				\underline{$C'.\status_p = \reccs$ $\limplies$ $C'.\csowner = p$}. \\
				Suppose $C.\status_p \neq \reccs$.
				It follows that the step is a crash step such that $p$ was in the CS in configuration $C$ (i.e., $C.\pcp = \refln{fullcrash:cs:1}$), hence it caused $C.\status_p$ to change to $\reccs$.
				Since the crash step doesn't affect $C.\csowner$ and because $C.\pcp = \refln{fullcrash:cs:1}$, by \indhyp{9} $C.\csowner = p$, which implies that $C'.\csowner = p$.\\
				\\
				Suppose $C.\status_p = \reccs$.
				We note that by \indhyp{9} $C.\csowner = p$.
				It follows that the step from $C$ to $C'$ is either a crash step or a normal step by a process $q \neq p$.
				If it was a crash step then the condition holds in $C'$ as it held in $C$.
				If it was a step by a process $q \neq p$, we argue as below that $q$'s step doesn't modify $C.\csowner$,
				and by \indhyp{9} $C.\csowner = p$, which would imply $C'.\csowner = p$.
				First, if $q$ were to take a step at Lines~\refln{fullcrash:try:6} or \refln{fullcrash:try:11}, $q$'s CAS would fail since $C.\csowner = p \neq \perp$.
				Second, $C.\pc{q} \neq \refln{fullcrash:try:13}$ because we know $C.\csowner = p$ and had $C.\pc{q}$ been \refln{fullcrash:try:13}, by \indhyp{9} $C.\csowner = \perp \neq p$, a contradiction.
				Lastly, $C.\pc{q} \neq \refln{fullcrash:exit:3}$, again because we know $C.\csowner = p$ and had $C.\pc{q}$ been \refln{fullcrash:exit:3}, by \indhyp{9} $C.\csowner = q \neq p$, a contradiction.
				Thus, from the above we conclude that $q$'s step didn't modify $C.\csowner$.
				Hence we have $C'.\csowner = p$.
			\end{enumerate}
		\end{enumerate}
	
		\item
		\underline{Proof that Condition \invcond{10} holds in $C'$} \\
		We establish each of the conjuncts of Condition~\invcond{10} separately as below. 
		\begin{enumerate}
			\item[(a). ]
			\underline{$C'.\pcp \in \{ \refln{fullcrash:try:2} \upto \refln{fullcrash:exit:4}, \refln{fullcrash:rec:2} \upto \refln{fullcrash:rec:5} \} \limplies C'.\inuse_p = \true$} \\
			In order to argue the correctness we first assume that $C'.\pcp \in \{ \refln{fullcrash:try:2} \upto \refln{fullcrash:exit:4}, \refln{fullcrash:rec:2} \upto \refln{fullcrash:rec:5} \}$.
			$p$ could reach to the given line in $C'$ due to its own step from different locations of the program, or $p$ might not have changed $\pcp$ but the step was due to a process $q \neq p$.
			In the following we argue each of these cases and show the correctness of the condition. \\
			\\
			Suppose $C.\pcp = \refln{fullcrash:try:1}$. 
			By the step $p$ sets $\inuse_p$ to $\true$ and thereby we get $C'.\inuse_p = \true$, satisfying the implication.\\
			\\
			Suppose $C.\pcp = \refln{fullcrash:rec:1}$. 
			By the step $p$ could only go to Line~\refln{fullcrash:rec:2} since $C'.\pcp \in \{ \refln{fullcrash:try:2} \upto \refln{fullcrash:exit:4}, \refln{fullcrash:rec:2} \upto \refln{fullcrash:rec:5} \}$.
			It follows that the \ifcode condition at Line~\refln{fullcrash:rec:1} was met and hence $C.\inuse_p = \true$.
			Since the step doesn't affect $\inuse_p$, we have $C'.\inuse_p = \true$.\\
			\\
			Suppose $C.\pcp = C'.\pcp$.
			By \indhyp{10} we have $C.\inuse_p = \true$.
			It follows that the step from $C$ to $C'$ is a normal step by a process $q \neq p$.
			Since $q$ will not change $\inuse_p$ anywhere in the code we have $C.\inuse_p = C'.\inuse_p$.
			Hence, $C'.\inuse_p = \true$.
			
			\item[(b). ]
			\underline{$((C'.\pcp \in \{ \refln{fullcrash:rem:1}, \refln{fullcrash:rec:1} \upto \refln{fullcrash:rec:7} \} \wedge C'.\status_p \in \{ \good, \recrem \} ) \vee C'.\pcp \in \{ \refln{fullcrash:try:1}, \refln{fullcrash:rec:8} \} )$}\\
				 \hspace*{5mm} \underline{$\limplies C'.\inuse_p = \false$} \\
				 We divide the implication into sub-parts as below and argue the correctness of each part.
				 \begin{enumerate}
				 	\item[i. ]
				 	\underline{$C'.\pcp \in \{ \refln{fullcrash:rem:1}, \refln{fullcrash:rec:1} \upto \refln{fullcrash:rec:7} \} \wedge C'.\status_p \in \{ \good, \recrem \} \limplies C'.\inuse_p = \false$}. \\
				 	In order to argue the correctness we first assume that $C'.\pcp \in \{ \refln{fullcrash:rem:1}, \refln{fullcrash:rec:1} \upto \refln{fullcrash:rec:7} \} \wedge C'.\status_p \in \{ \good, \recrem \}$.
				 	$p$ could reach to the given state in $C'$ due to its own step from different locations of the program, or $p$ might not have changed $\pcp$ but the step was due to a process $q \neq p$.
				 	In the following we argue each of these cases and show the correctness of the condition. \\
				 	\\
				 	Suppose $C.\pcp =\refln{fullcrash:exit:4}$. 
				 	In that case $p$ sets $\inuse_p$ to $\false$ and hence we have $C'.\inuse_p = \false$\\
				 	\\
				 	Suppose $C.\pcp = \refln{fullcrash:rec:8}$. 
				 	By \indhyp{10} $C.\inuse_p = \false$ and it follows by the step that $C'.\inuse_p = \false$.\\
				 	\\
				 	Suppose $C.\pcp \in \{ \refln{fullcrash:rem:1}, \refln{fullcrash:rec:1} \upto \refln{fullcrash:rec:7} \} \wedge C.\status_p \in \{ \good, \recrem \}$.
				 	By \indhyp{10} we have $C.\inuse_p = \false$.
				 	It follows that the step from $C$ to $C'$ is a normal step by $p$ such that it doesn't change $\inuse_p$ to $\true$ (which happens only at Line~\refln{fullcrash:try:1}),
				 	or a crash step (which doesn't change $\inuse_p$), or a normal step by process $q \neq p$, which again doesn't change $\inuse_p$.
				 	Thus, we have $C'.\inuse_p = \false$. \\
				 	
				 	\item[ii. ]
				 	\underline{$C'.\pcp \in \{ \refln{fullcrash:try:1}, \refln{fullcrash:rec:8} \} \limplies C'.\inuse_p = \false$}. \\
				 	Similar to the above, we consider different scenarios as below. \\
				 	\\
				 	Suppose $C.\pcp = \refln{fullcrash:rem:1}$ and $C.\status_p = \good$.
				 	In that case by \indhyp{10} we have $C.\inuse_p = \false$ which remains unchanged in $C'$. \\
				 	\\
				 	Suppose $C.\pcp = \refln{fullcrash:rec:7}$.
				 	In that case the step writes $\false$ to $\inuse_p$.
				 	Thus, we have $C'.\inuse_p = \false$.\\
				 	\\
				 	Suppose $C.\pcp = C'.\pcp$. 
				 	It follows that the step from $C$ to $C'$ is a step by some process $q \neq p$ which doesn't change the value of $\inuse_p$.
				 	Since $C.\inuse_p = \false$ by \indhyp{10}, we have $C'.\inuse_p = \false$.
				 \end{enumerate}
		\end{enumerate}
	
		\item
		\underline{Proof that Condition \invcond{11} holds in $C'$} \\
		We establish each of the conjuncts of Condition~\invcond{11} separately as below. 
		
		\begin{enumerate}
			\item[(a). ]
			\underline{$C'.\pcp \in \{ \refln{fullcrash:try:3} \upto \refln{fullcrash:try:7}, \refln{fullcrash:exit:2}, \refln{fullcrash:rec:2} \upto \refln{fullcrash:rec:4} \} \limplies C'.\myseqp \in \{ C'.\seq - 1, C'.\seq \}$} \\
			Similar to the arguments above, we consider different scenarios as below.
			Therefore assume that $C'.\pcp \in \{ \refln{fullcrash:try:3} \upto \refln{fullcrash:try:7}, \refln{fullcrash:exit:2}, \refln{fullcrash:rec:2} \upto \refln{fullcrash:rec:4} \}$. \\
			\\
			Suppose $C.\pcp = \refln{fullcrash:try:2}$. 
			By the step $\myseq_p$ takes the value of $C.\seq$.
			Thus we have $C'.\myseq_p = C.\seq = C'.\seq$.\\
			\\
			Suppose $C.\pcp = \refln{fullcrash:exit:1}$. 
			We note that of the possibilities for $C'.\pcp$ described above, only $C'.\pcp = \refln{fullcrash:exit:2}$ is the real possibility.
			It follows that when $p$ executed Line~\refln{fullcrash:exit:1} from $C$,
			the \ifcode condition is met.
			Therefore, $C.\myseq_p = C.\seq$ and hence $C'.\myseq_p = C'.\seq$.\\
			\\
			Suppose $C.\pcp = \refln{fullcrash:rec:1}$. 
			We note that of the possibilities for $C'.\pcp$ described above, only $C'.\pcp = \refln{fullcrash:rec:1}$ is the real possibility.
			It follows that when $p$ executed Line~\refln{fullcrash:rec:1} from $C$,
			the \ifcode condition is met.
			Therefore, $C.\myseq_p = C.\seq$ and hence $C'.\myseq_p = C'.\seq$.\\
			\\
			Suppose $C.\pcp \in \{ \refln{fullcrash:try:3} \upto \refln{fullcrash:try:7}, \refln{fullcrash:exit:2}, \refln{fullcrash:rec:2} \upto \refln{fullcrash:rec:4} \}$.
			Since $C'.\pcp \in \{ \refln{fullcrash:try:3} \upto \refln{fullcrash:try:7}, \refln{fullcrash:exit:2}, \refln{fullcrash:rec:2} \upto \refln{fullcrash:rec:4} \}$,
			if the step from $C$ to $C'$ was by $p$, neither the value of $\seq$ nor that of $\myseq_p$ is changed by $p$'s step.
			Therefore, $C'.\myseq_p \in \{ C'.\seq - 1, C'.\seq \}$. 
			Otherwise the step from $C$ to $C'$ is by a process $q \neq p$ and $C.\pcp = C'.\pcp$.
			If $q$ didn't change $\seq$ due to the step, then the condition continues to hold.
			Otherwise $q$ executed Line~\refln{fullcrash:rec:4} to increment $\seq$.
			Hence $C.\myseq_{q} = C.\seq$.
			We have from \indhyp{11} that $C.\myseq_p \in \{ C.\seq - 1, C.\seq \}$.
			If $C.\myseq_p = C.\seq - 1$, by \indhyp{12} $C.\myseq_q = C.\seq - 1$, a contradiction.
			Therefore, $C.\myseq_p = C.\seq$.
			It follows that after $q$ increments $\seq$, $C'.\myseq_p = C'.\seq - 1$.
			Hence, the condition holds.
			
			\item[(b). ]
			\underline{$C'.\pcp \in \{ \refln{fullcrash:try:8}, \refln{fullcrash:rec:5} \} \limplies C'.\myseqp = C'.\seq - 1$} \\
			Similar to the arguments above, we consider different scenarios as below.
			Therefore assume that $C'.\pcp \in \{ \refln{fullcrash:try:8}, \refln{fullcrash:rec:5} \}$. \\
			\\
			Suppose $C.\pcp = \refln{fullcrash:try:7}$.
			By \indhyp{11} $C.\myseq_p \in \{ C.\seq - 1, C.\seq \}$.
			Since the \ifcode condition was met for $p$ to move to Line~\refln{fullcrash:try:8},
			$C.\myseq_p = C.\seq - 1$.
			Which implies that $C'.\myseq_p = C'.\seq - 1$. \\ 
			\\
			Suppose $C.\pcp = \refln{fullcrash:try:4}$.
			Since $C'.\pcp = \refln{fullcrash:try:8}$, $\stopwait[C.\myseq_p \% 3] = \true$.
			By \indhyp{3} $\stopwait[C.\seq \% 3] = \false$.
			Hence, by \indhyp{11} $C.\myseq_p = C.\seq - 1$, which implies that $C'.\myseq_p = C'.\seq - 1$. \\
			\\
			Suppose $C.\pcp = \refln{fullcrash:rec:4}$.
			If $C.\myseq_p = C.\seq - 1$, the step doesn't change anything and hence $C'.\myseq_p = C'.\seq - 1$.
			Otherwise $C.\myseq_p = C.\seq$ and the step increments $\seq$.
			It follows that $C'.\myseq_p = C'.\seq - 1$ in either case. \\
			\\
			Suppose $C.\pcp = C'.\pcp$.
			It follows from \indhyp{11} that $C.\myseq_p = C.\seq - 1$ and that some process $q$ took a step from $C$ to $C'$.
			However, note that from \indhyp{12} $C.\pc{q} \in \{ \refln{fullcrash:rec:2} \upto \refln{fullcrash:rec:5} \}$ implies $C.\myseq_{q} = C.\seq - 1$.
			Hence, $q$ doesn't change the value of $\seq$ due to the step.
			
			\item[(c). ]
			\underline{$C'.\pcp \in [\refln{fullcrash:try:9}, \refln{fullcrash:try:14}] \limplies C'.\myseqp = C'.\seq$} \\
			Similar to the arguments above, we consider different scenarios as below.
			Therefore assume that $C'.\pcp \in [\refln{fullcrash:try:9}, \refln{fullcrash:try:14}]$. \\
			\\
			Suppose $C.\pcp = \refln{fullcrash:try:8}$.
			It follows from \indhyp{11} that $C.\myseqp = C.\seq - 1$.
			Since the step increments $\myseqp$, we have $C'.\myseqp = C.\seq$. \\
			\\
			Suppose $C.\pcp = \refln{fullcrash:try:7}$.
			We note that $C'.\pcp = \refln{fullcrash:try:12}$ and that happened only because $C.\myseqp = C.\seq$, which follows from \indhyp{11}. \\
			\\
			Suppose $C.\pcp = C'.\pcp$. 
			It follows from \indhyp{11} that $C.\myseq_p = C.\seq$ and that some process $q$ took a step from $C$ to $C'$.
			However, note that from \indhyp{12} $C.\pc{q} \in \{ \refln{fullcrash:rec:2} \upto \refln{fullcrash:rec:5} \}$ implies $C.\myseq_{q} = C.\seq - 1$.
			Hence, $q$ doesn't change the value of $\seq$ due to the step.
			
			\item[(d). ]
			\underline{$(C'.\pcp = \refln{fullcrash:rec:6} \wedge C'.\csowner = p) \limplies C'.\myseq_p < C'.\seq$} \\
			Similar to the arguments above, we consider different scenarios as below.
			Therefore assume that $C'.\pcp = \refln{fullcrash:rec:6} \wedge C'.\csowner = p$. \\
			\\
			Suppose $C.\pcp = \refln{fullcrash:rec:5}$.
			By \indhyp{11} $C.\myseqp = C.\seq - 1 < C.\seq$.
			Since the step doesn't modify $\seq$, we have $C'.\myseqp < C'.\seq$. \\
			\\
			Suppose $C.\pcp = \refln{fullcrash:rec:1}$.
			It follows that the \ifcode condition of Line~\refln{fullcrash:rec:1} is not met.
			Since $C'.\csowner = p$, the step couldn't modify $\csowner$ and we have $C.\csowner = p$.
			By \indhyp{9} $C.\inuse_p = \true$.
			Therefore, $C.\myseq_p \neq C.\seq$ and by \indhyp{1} $C.\myseq_p < C.\seq$.
			It follows that $C'.\myseqp < C'.\seq$. \\
			\\
			Suppose $C.\pcp = C'.\pcp$. 
			We first observe that $C.\csowner = p$ because we have $C'.\csowner = p$ and no process $q \neq p$ can set $\csowner$ to $p$ in the step from $C$ to $C'$ as it can only write its own name at Lines~\refln{fullcrash:try:6}, \refln{fullcrash:try:11}, or \refln{fullcrash:try:13} or the value $\perp$ at Line~\refln{fullcrash:exit:3}.
			It follows from \indhyp{11} that $C.\myseq_p < C.\seq$ and that some process $q$ took a step from $C$ to $C'$.
			However, note that if $q$ took a step, it could only increment $\seq$ by 1.
			Which means $C'.\myseq_p < C'.\seq$.
		\end{enumerate}
	
		\item
		\underline{Proof that Condition \invcond{15} holds in $C'$} \\
		\underline{$(C'.\pcp \in \{ \refln{fullcrash:try:3} \upto \refln{fullcrash:try:5} \} \wedge C'.\myseq_p = C'.\seq -1)$} \\
		\hspace*{5mm} \underline{$\limplies$ $(\stopwait[C'.\myseq_p \% 3] = \true \vee (\exists q, C'.\pc{q} = \refln{fullcrash:rec:5} \wedge C'.\myseq_{q} = C'.\myseq_p))$} \\
		\\
		Assume $(C'.\pcp \in \{ \refln{fullcrash:try:3} \upto \refln{fullcrash:try:5} \} \wedge C'.\myseq_p = C'.\seq -1)$. 
		We analyze the condition by cases as below.
			
		\begin{enumerate}
			\item[(i). ] \underline{$C.\pcp = \refln{fullcrash:try:2}$} \\
			Since Line~\refln{fullcrash:try:2} sets only $\myseq_p = \seq$ and doesn't modify $\seq$, 
			we will have a step that also modifies $\seq$ by decrementing given that we assumed $(C'.\pcp \in \{ \refln{fullcrash:try:3} \upto \refln{fullcrash:try:5} \} \wedge C'.\myseq_p = C'.\seq -1)$. 
			We conclude that this case doesn't arise.
			
			\item[(ii). ] \underline{$C.\pcp \in \{ \refln{fullcrash:try:3} \upto \refln{fullcrash:try:5} \} \wedge C.\myseq_p \neq C.\seq -1$} \\
			By our assumption about $C'$ and by \indhyp{11} $C.\myseq_p = C.\seq$.
			It follows that some process $q \neq p$ incremented $\seq$ at Line~\refln{fullcrash:rec:4} in the step from $C$ to $C'$.
			Since $q$ incremented $\seq$ at Line~\refln{fullcrash:rec:4}, by \indhyp{11} we have $C.\myseq_{q} = C.\seq = C.\myseq_p$.
			Therefore we have $\exists q, C'.\pc{q} = \refln{fullcrash:rec:5} \wedge C'.\myseq_{q} = C'.\myseq_p$.
			
			\item[(iii). ] \underline{$(C.\pcp \in \{ \refln{fullcrash:try:3} \upto \refln{fullcrash:try:5} \} \wedge C.\myseq_p = C.\seq -1)$} \\
			By \indhyp{15} we have $\stopwait[C.\myseq_p \% 3] = \true \vee (\exists q, C.\pc{q} = \refln{fullcrash:rec:5} \wedge C.\myseq_{q} = C.\myseq_p)$.\\
			If $q$ takes a step, the condition continues to hold since $q$ only writes $\true$ into $\stopwait[C.\myseq_q \% 3]$ and $C.\myseq_{q} = C.\myseq_p = C'.\myseq_p$.
			If $p$ takes a step from $C$, it doesn't modify $\myseq_p$ or $\seq$ and hence the condition continues to hold.
			If some process $r \neq q$, where $q$ is as described above, takes a step, then $r$ doesn't modify $\seq$, $\myseq_p$ or $\myseq_q$.
			Because we have $(C.\pcp \in \{ \refln{fullcrash:try:3} \upto \refln{fullcrash:try:5} \} \wedge C.\myseq_p = C.\seq -1)$, by \indhyp{12} such $r$ with $C.\pc{r} = \refln{fullcrash:rec:5}$ has $C.\myseq_{r} = C.\seq - 1$.
			Hence, the step by $r$ will only set $\stopwait[C.\myseq_p \% 3]$ to $\true$ satisfying the condition.
		\end{enumerate}
		
		\item
		\underline{Proof that Condition \invcond{12} holds in $C'$} \\
		We divide the implication into sub-parts as below and argue the correctness of each part.
		\begin{enumerate}
			\item[(a). ]
			\underline{$(C'.\pcp \in \{ \refln{fullcrash:try:3} \upto \refln{fullcrash:try:6}, \refln{fullcrash:exit:2}, \refln{fullcrash:rec:2}\upto\refln{fullcrash:rec:5} \} \wedge C'.\myseqp = C'.\seq - 1)$} \\
			\hspace*{5mm} \underline{$\limplies$ $\forall q, (\neg ( C'.\pc{q} \in \{ \refln{fullcrash:rem:1}, \refln{fullcrash:rec:1} \}$ $\wedge$ $C'.\inuse_{q} = \true$ $\wedge$ $C'.\myseq_{q} = C'.\seq)$} \\
			\hspace*{15mm} \underline{$\wedge$ $(C'.\pc{q} \in \{ \refln{fullcrash:rec:2} \upto \refln{fullcrash:rec:5} \} \limplies C'.\myseq_{q} = C'.\seq - 1))$}
			
			Assume $(C'.\pcp \in \{ \refln{fullcrash:try:3} \upto \refln{fullcrash:try:6}, \refln{fullcrash:exit:2}, \refln{fullcrash:rec:2}\upto\refln{fullcrash:rec:5} \} \wedge C'.\myseqp = C'.\seq - 1)$. 
			We analyze the condition by cases as below.
			
			\begin{enumerate}
				\item[(i). ] \underline{$C.\pcp = \refln{fullcrash:try:2}$} \\
				Since Line~\refln{fullcrash:try:2} sets only $\myseq_p = \seq$ and doesn't modify $\seq$, 
				we will have a step that also modifies $\seq$ by decrementing given that we assumed $(C'.\pcp \in \{ \refln{fullcrash:try:3} \upto \refln{fullcrash:try:6}, \refln{fullcrash:exit:2}, \refln{fullcrash:rec:2}\upto\refln{fullcrash:rec:5} \} \wedge C'.\myseqp = C'.\seq - 1)$. 
				We conclude that this case doesn't arise.
				A similar argument holds for the case if we consider $C.\pcp = \refln{fullcrash:exit:1}$ or $C.\pcp = \refln{fullcrash:rec:1}$, hence we don't argue these cases.
				
				\item[(ii).] \underline{$C.\pcp \in \{ \refln{fullcrash:try:3} \upto \refln{fullcrash:try:6}, \refln{fullcrash:exit:2}, \refln{fullcrash:rec:2}\upto\refln{fullcrash:rec:4} \} \wedge C.\myseqp \neq C.\seq - 1$} \\
				We first note that by \indhyp{11} when $\pcp = \refln{fullcrash:rec:5}$, $\myseq_p = \seq - 1$.
				Thus the case $C.\pcp = \refln{fullcrash:rec:5} \wedge C.\myseqp \neq C.\seq - 1$ cannot arise.
				We therefore consider the remaining subcases here.
				By our assumption about $C'$ and by \indhyp{11} $C.\myseq_p = C.\seq$.
				It follows that some process $q$ (possibly same as $p$) incremented $\seq$ at Line~\refln{fullcrash:rec:4} in the step from $C$ to $C'$.
				By \indhyp{1} we know that for each process $r$, $C.\myseq_{r} \leq C.\seq$.
				Thus we have, $\forall r, \neg (C.\pc{r} \in \{ \refln{fullcrash:rem:1}, \refln{fullcrash:rec:1} \} \wedge C.\inuse_{r} = \true \wedge C.\myseq_{r} = C.\seq+1)$ trivially.
				It follows that $\forall r, \neg (C'.\pc{r} \in \{ \refln{fullcrash:rem:1}, \refln{fullcrash:rec:1} \} \wedge C'.\inuse_{r} = \true \wedge C'.\myseq_{r} = C'.\seq)$ because the step from 
				$C$ to $C'$ only affected the value of $\seq$.
				We observe that for any $r$ if $C.\pc{r} \in \{ \refln{fullcrash:rec:2}\upto\refln{fullcrash:rec:4} \}$, then $C.\myseq_{r} = C.\seq$.
				Because otherwise, by \indhyp{11} $C.\myseq_{r} = C.\seq - 1$ but \indhyp{12} implies that for process $q$ that changed $\seq$, $C.\myseq_{q} = C.\seq - 1$, which cannot be the case.
				We know by \indhyp{11} that the case $C.\pcp = \refln{fullcrash:rec:5} \wedge C.\myseqp = C.\seq$ cannot arise.
				Hence we have that $C'.\pc{r} \in \{ \refln{fullcrash:rec:2}\upto\refln{fullcrash:rec:4} \}$, then $C'.\myseq_{r} = C'.\seq - 1$.
				And for the process $q$ that incremented $\seq$ we have $C'.\pc{q} = \refln{fullcrash:rec:5}$ and $C'.\myseq_{q} = C'.\seq - 1$.
				It follows that the implication holds in $C'$.
				
				\item[(iii).] \underline{$C.\pcp \in \{ \refln{fullcrash:try:3} \upto \refln{fullcrash:try:6}, \refln{fullcrash:exit:2}, \refln{fullcrash:rec:2}\upto\refln{fullcrash:rec:4} \} \wedge C.\myseqp = C.\seq - 1$} \\
				By \indhyp{12} we have $\forall q, (\neg ( C.\pc{q} \in \{ \refln{fullcrash:rem:1}, \refln{fullcrash:rec:1} \}$ $\wedge$ $C.\inuse_{q} = \true$ $\wedge$ $C.\myseq_{q} = C.\seq) \wedge (C.\pc{q} \in \{ \refln{fullcrash:rec:2} \upto \refln{fullcrash:rec:5} \} \limplies C.\myseq_{q} = C.\seq - 1))$.
				If a $q$ with $C.\pc{q} \in \{ \refln{fullcrash:rec:2} \upto \refln{fullcrash:rec:5} \}$ takes a step, 
				the condition continues to hold since $q$ doesn't change the value of $\seq$.
				If $p$ takes a step from $C$, it doesn't modify $\myseq_p$ or $\seq$ and hence the condition continues to hold.
				If some process $r \neq q$, where $q$ is as described above, takes a step, then $r$ doesn't modify $\seq$, $\myseq_p$ or $\myseq_q$.
				Therefore we have the implication in $C'$.
			\end{enumerate}
				
			\item[(b). ]
			\underline{$(C'.\pcp = \refln{fullcrash:try:6} \wedge C'.\csowner \neq \perp)$} \\
			\hspace*{5mm} \underline{$\limplies$ $\forall q, (\neg ( C'.\pc{q} \in \{ \refln{fullcrash:rem:1}, \refln{fullcrash:rec:1} \}$ $\wedge$ $C'.\inuse_{q} = \true$ $\wedge$ $C'.\myseq_{q} = C'.\seq)$} \\
			\hspace*{15mm} \underline{$\wedge$ $(C'.\pc{q} \in \{ \refln{fullcrash:rec:2} \upto \refln{fullcrash:rec:5} \} \limplies C'.\myseq_{q} = C'.\seq - 1))$}
			
			Assume $(C'.\pcp = \refln{fullcrash:try:6} \wedge C'.\csowner \neq \perp)$. 
			We analyze the condition by cases as below.
			
			\begin{enumerate}
				\item[(i). ] \underline{$C.\pcp = \refln{fullcrash:try:5}$} \\
				We note that $p$ finds $\csowner = \perp$ due to the step if $C'.\pcp = \refln{fullcrash:try:6}$ as we assume, otherwise $p$ moves to Line~\refln{fullcrash:try:4} in the loop.
				In either case, we conclude that the case doesn't arise.
				
				\item[(ii). ] \underline{$C.\pcp = \refln{fullcrash:try:6} \wedge C.\csowner = \perp$} \\
				We note that in the step some process $q$ wrote its name into $\csowner$.
				$q \neq p$ because $C.\pcp = C'.\pcp = \refln{fullcrash:try:6}$. \\
				Suppose $C.\pc{q} = \refln{fullcrash:try:13}$.
				By \indhyp{14} if $C.\myseq_p = C.\seq - 1$, then $C.\pcp \neq \refln{fullcrash:try:6}$.
				If $C.\myseq_p = C.\seq$, by \indhyp{11} $C.\myseq_{q} = C.\seq$,
				by \indhyp{7} $q \in \lock[C.\seq \% 3].\csset$,
				by \indhyp{7} again $p \in \lock[C.\seq \% 3].\csset$, a contradiction to the mutual exclusion property of $\lock[C.\seq \% 3]$.
				Thus $C.\pc{q} = \refln{fullcrash:try:13}$ cannot arise. \\
				Suppose $C.\pc{q} = \refln{fullcrash:try:11}$.
				By \indhyp{11} $C.\myseq_{q} = C.\seq$ and by mutual exclusion property of the base lock, $C.\myseq_p = C.\seq - 1$.
				Thus, by \indhyp{12} we have the implication in $C'$.
				Suppose $C.\pc{q} = \refln{fullcrash:try:6}$.
				Again we have one of $C.\myseq_{q}$ or $C.\myseq_p$ as $C.\seq - 1$.
				Therefore, again by \indhyp{12} we have the implication in $C'$.
				
				\item[(iii). ] \underline{$C.\pcp = \refln{fullcrash:try:6} \wedge C.\csowner \neq \perp$} \\
				In this case the step is by some process $q \neq p$.
				By \indhyp{12} we have 
				$\forall q, (\neg ( C.\pc{q} \in \{ \refln{fullcrash:rem:1}, \refln{fullcrash:rec:1} \}$ $\wedge$ $C.\inuse_{q} = \true$ $\wedge$ $C.\myseq_{q} = C.\seq)$
				$\wedge$ $(C.\pc{q} \in \{ \refln{fullcrash:rec:2} \upto \refln{fullcrash:rec:5} \} \limplies C.\myseq_{q} = C.\seq - 1))$.
				We note from the above that any $q$ cannot change the value of $\seq$, therefore we have the implication in $C'$.
			\end{enumerate}
			
			\item[(c). ]
			\underline{$C'.\pcp \in \{ \refln{fullcrash:try:7} \upto \refln{fullcrash:try:14} \}$} \\
			\hspace*{5mm} \underline{$\limplies$ $\forall q, (\neg ( C'.\pc{q} \in \{ \refln{fullcrash:rem:1}, \refln{fullcrash:rec:1} \}$ $\wedge$ $C'.\inuse_{q} = \true$ $\wedge$ $C'.\myseq_{q} = C'.\seq)$} \\
			\hspace*{15mm} \underline{$\wedge$ $(C'.\pc{q} \in \{ \refln{fullcrash:rec:2} \upto \refln{fullcrash:rec:5} \} \limplies C'.\myseq_{q} = C'.\seq - 1))$}
			
			Assume $C'.\pcp \in \{ \refln{fullcrash:try:7} \upto \refln{fullcrash:try:14} \}$. 
			We analyze the condition by cases as below.
			
			\begin{enumerate}
				\item[(i). ] \underline{$C.\pcp = \refln{fullcrash:try:6} \wedge C.\csowner \neq \perp$} \\
				As a result of the step $C'.\pcp = \refln{fullcrash:try:7}$.
				By \indhyp{12} we have 
				$\forall q, (\neg ( C.\pc{q} \in \{ \refln{fullcrash:rem:1}, \refln{fullcrash:rec:1} \}$ $\wedge$ $C.\inuse_{q} = \true$ $\wedge$ $C.\myseq_{q} = C.\seq)$
				$\wedge$ $(C.\pc{q} \in \{ \refln{fullcrash:rec:2} \upto \refln{fullcrash:rec:5} \} \limplies C.\myseq_{q} = C.\seq - 1))$.
				We note from the above that $p$ doesn't change the value of $\seq$, therefore we have the implication in $C'$.
				
				\item[(ii). ] \underline{$C.\pcp = \refln{fullcrash:try:4} \wedge C.\stopwait[\myseq_p \% 3] = \true$} \\
				As a result of the step $C'.\pcp = \refln{fullcrash:try:8}$.
				By \indhyp{12} we have 
				$\forall q, (\neg ( C.\pc{q} \in \{ \refln{fullcrash:rem:1}, \refln{fullcrash:rec:1} \}$ $\wedge$ $C.\inuse_{q} = \true$ $\wedge$ $C.\myseq_{q} = C.\seq)$
				$\wedge$ $(C.\pc{q} \in \{ \refln{fullcrash:rec:2} \upto \refln{fullcrash:rec:5} \} \limplies C.\myseq_{q} = C.\seq - 1))$.
				We note from the above that $p$ doesn't change the value of $\seq$, therefore we have the implication in $C'$.
				
				\item[(iii). ] \underline{$C.\pcp \in \{ \refln{fullcrash:try:7} \upto \refln{fullcrash:try:14} \}$} \\
				If $C.\pcp \neq C'.\pcp$, we note that $p$ doesn't change the value of $\seq$ or the variables $\myseq_{q}$ and $\inuse_{q}$ of some process $q \neq p$, 
				therefore we have the implication in $C'$ following from \indhyp{12}.
				Thus assume $C.\pcp = C'.\pcp$.
				In this case the step is by some process $q \neq p$.
				By \indhyp{12} we have 
				$\forall q, (\neg ( C.\pc{q} \in \{ \refln{fullcrash:rem:1}, \refln{fullcrash:rec:1} \}$ $\wedge$ $C.\inuse_{q} = \true$ $\wedge$ $C.\myseq_{q} = C.\seq)$
				$\wedge$ $(C.\pc{q} \in \{ \refln{fullcrash:rec:2} \upto \refln{fullcrash:rec:5} \} \limplies C.\myseq_{q} = C.\seq - 1))$.
				We note from the above that any $q$ cannot change the value of $\seq$, therefore we have the implication in $C'$.
				
			\end{enumerate}
		\end{enumerate}
	
		\item
		\underline{Proof that Condition \invcond{14} holds in $C'$} \\
		We establish each of the conjuncts of Condition~\invcond{14} separately as below.
		
		\begin{enumerate}
			\item[(a). ] \underline{$(\forall q, (C'.\pcp \in \{\refln{fullcrash:try:6}, \refln{fullcrash:try:7}, \refln{fullcrash:try:12} \} \wedge C'.\myseq_p = C'.\seq \wedge C'.\csowner = q)$} \\
				\hspace*{15mm} \underline{$\limplies$ $(C'.\myseq_{q} = C'.\seq - 1 \wedge q \in \lock[C'.\myseq_{q} \% 3].\csset$} \\
				\hspace*{25mm} \underline{$\wedge$ $\forall r, (q \neq r \wedge C'.\myseq_{r} = C'.\myseq_{q})$} \\
				\hspace*{35mm} \underline{$\limplies$ $(C'.\pc{r} \in \{ \refln{fullcrash:rem:1} \upto \refln{fullcrash:try:3}, \refln{fullcrash:try:8}, \refln{fullcrash:exit:4}, \refln{fullcrash:rec:1} \upto \refln{fullcrash:rec:8} \}$} \\
				\hspace*{45mm} \underline{$\vee$ $( C'.\pc{r} = \refln{fullcrash:try:4} \limplies \stopwait[C'.\myseq_r  \% 3] = \true)))$} \\
			
			Assume that $C'.\pcp \in \{\refln{fullcrash:try:6}, \refln{fullcrash:try:7}, \refln{fullcrash:try:12} \}$ $\wedge$ $C'.\myseq_p = C'.\seq$ $\wedge$ $C'.\csowner = q$.
			We divide the argument into different cases as below and argue the correctness of each case.
			
			\begin{enumerate}
				\item[(i). ] Suppose $C.\pcp = \refln{fullcrash:try:5}$. \\
				$p$ could reach Line~\refln{fullcrash:try:6} from Line~\refln{fullcrash:try:5} only after noticing that $\csowner = \perp$.
				Since we have $C'.\pcp \in \{\refln{fullcrash:try:6}, \refln{fullcrash:try:7}, \refln{fullcrash:try:12} \}$ $\wedge$ $C'.\myseq_p = C'.\seq$ $\wedge$ $C'.\csowner = q$,
				it follows that $C.\csowner = \perp$ and $C'.\csowner = q$, which is not possible.
				We conclude that the case cannot arise.
				
				\item[(ii). ] Suppose $C.\pcp \in \{\refln{fullcrash:try:6}, \refln{fullcrash:try:7}, \refln{fullcrash:try:12} \}$ $\wedge$ $C.\myseq_p \neq C.\seq$. \\
				By \indhyp{11} we have $C.\myseq_p = C.\seq - 1$, but by our assumption $C'.\myseq_p = C'.\seq$.
				We note from the algorithm that $\seq$ is never decremented and no process other than $p$ can change $\myseq_p$.
				Since $C.\pcp \in \{\refln{fullcrash:try:6}, \refln{fullcrash:try:7}, \refln{fullcrash:try:12} \}$ and $C'.\pcp \in \{\refln{fullcrash:try:6}, \refln{fullcrash:try:7}, \refln{fullcrash:try:12} \}$,
				if at all $p$ took a step from $C$ to $C'$, it was either at Line~\refln{fullcrash:try:6} to go to Line~\refln{fullcrash:try:7}, or at Line~\refln{fullcrash:try:7} to go to Line~\refln{fullcrash:try:12}.
				We note that at either line $p$ doesn't modify $\myseq_p$, and hence it follows that the case cannot arise.
				
				\item[(iii). ] Suppose $C.\pcp \in \{\refln{fullcrash:try:6}, \refln{fullcrash:try:7}, \refln{fullcrash:try:12} \}$ $\wedge$ $C.\myseq_p = C.\seq$ $\wedge$ $C.\csowner \neq q$ for any $q$. \\
				First we note that $C.\csowner = \perp$.
				Since $C'.\csowner = q$, it follows that $q$ wrote its own name into $\csowner$ at one of Lines~\refln{fullcrash:try:6}, \refln{fullcrash:try:11}, or \refln{fullcrash:try:13}.
				We also note that $q \neq p$ because if $q$ were $p$, writing its own name into $\csowner$ at Lines~\refln{fullcrash:try:6} or \refln{fullcrash:try:11}
				would mean $C'.\pcp = \refln{fullcrash:cs:1}$ and writing it at Line~\refln{fullcrash:try:13} would mean $C'.\pcp = \refln{fullcrash:try:14}$,
				which contradicts our assumption.
				If $q$ were to write its name at Line~\refln{fullcrash:try:13}, by \indhyp{11} $C.\myseq_{q} = C.\seq$.
				By \indhyp{7} $q \in \lock[C.\seq \% 3].\csset$.
				Since $C.\pcp \in \{\refln{fullcrash:try:6}, \refln{fullcrash:try:7}, \refln{fullcrash:try:12} \}$ and $C.\myseq_p = C.\seq$,
				by \indhyp{7} $p \in \lock[C.\seq \% 3].\csset$.
				But because $p \neq q$, it follows from the mutual exclusion property of $\lock[C.\seq \% 3]$, that $q$ didn't write its name at Line~\refln{fullcrash:try:13}.
				The same argument holds for Line~\refln{fullcrash:try:11} and the case for Line~\refln{fullcrash:try:6} and $C.\myseq_{q} = C.\seq$.
				Hence, by \indhyp{11} we note that $q$ wrote its name in $\csowner$ at Line~\refln{fullcrash:try:6} with $C.\myseq_{q} = C.\seq - 1$.
				It follows that $C'.\myseq_{q} = C'.\seq - 1$.
				By \indhyp{7} it follows that $q \in \lock[C'.\myseq_{q} \% 3].\csset$.
				For any $r$ with $C.\myseq_{r} = C.\myseq_{q}$ and $q \neq r$, we have the following.
				By \indhyp{7} $C.\pc{r} \notin \{ \refln{fullcrash:try:5} \upto \refln{fullcrash:try:7}, \refln{fullcrash:try:10} \upto \refln{fullcrash:try:14} \}$
				and if $C.\pc{r} = \refln{fullcrash:try:4}$, $C.\stopwait[C.\myseq_{r} \% 3] = \true$, all due to the mutual exclusion property of $\lock[C.\myseq_{q} \% 3]$.
				By \indhyp{11} $C.\pc{r} \neq \refln{fullcrash:try:9}$, because $C.\myseq_{r} = C.\myseq_{q} = C.\seq - 1$.
				By \indhyp{9} $C.\pc{r} \notin \{ \refln{fullcrash:try:14} \upto \refln{fullcrash:exit:3} \}$ because $C.\csowner = \perp$.
				It follows that for each $r$ with $q \neq r$ and $C.\myseq_{r} = C.\myseq_{q}$,
				$C.\pc{r} \in \{ \refln{fullcrash:rem:1} \upto \refln{fullcrash:try:3}, \refln{fullcrash:try:8}, \refln{fullcrash:exit:4}, \refln{fullcrash:rec:1} \upto \refln{fullcrash:rec:8} \}$, or
				if $C.\pc{r} = \refln{fullcrash:try:4}$, then $\stopwait[C.\myseq_r  \% 3] = \true$.
				Which means the implication holds in $C'$.
				
				\item[(iv). ] Suppose $C.\pcp \in \{\refln{fullcrash:try:6}, \refln{fullcrash:try:7}, \refln{fullcrash:try:12} \}$ $\wedge$ $C.\myseq_p = C.\seq$ $\wedge$ $C.\csowner = q$. \\
				In this case we only need to argue how the inner implication would be affected by a step of a process $r$ such that $q \neq r$ and $C.\myseq_{q} = C.\myseq_{r}$.
				By \indhyp{14} we know that for such an $r$
				$C.\pc{r} \in \{ \refln{fullcrash:rem:1} \upto \refln{fullcrash:try:3}, \refln{fullcrash:try:8}, \refln{fullcrash:exit:4}, \refln{fullcrash:rec:1} \upto \refln{fullcrash:rec:8} \}$,
				or $( \pc{r} = \refln{fullcrash:try:4} \limplies \stopwait[C.\myseq_r  \% 3] = \true)$.\\
				If $r$ takes a step at Line~\refln{fullcrash:try:3} and moves to Line~\refln{fullcrash:try:4},
				by \indhyp{14} we have $q \in \lock[C.\myseq_{q} \% 3].\csset$.
				It follows that $r$'s step was not due to completion of $\lock[C.\myseq_{r} \% 3].\tryproc{r}()$, but due to breaking out of the \waittill loop at Line~\refln{fullcrash:try:3}.
				This can happen only when $\stopwait[C.\myseq_{r} \% 3] = \true$, which implies that the condition holds in $C'$.\\
				If $r$ takes a step at Line~\refln{fullcrash:try:4}, we know from \indhyp{14} that $\stopwait[C.\myseq_r \% 3] = \true$.
				It follows that $r$ moves to Line~\refln{fullcrash:try:8}, satisfying the condition in $C'$. \\
				If $r$ takes a step at Line~\refln{fullcrash:try:8}, it increments its own $\myseq_{r}$ and thus satisfies the condition trivially.
				
			\end{enumerate}
			
			\item[(b). ]
			We divide the implication into sub-parts as below and argue the correctness of each part.
			
			\begin{enumerate}
				\item[(i). ]
				\underline{$((C'.\pcp \in \{ \refln{fullcrash:try:7}, \refln{fullcrash:try:12} \} \wedge C'.\myseq_p = C'.\seq \wedge C'.\csowner = \perp)$} \\
				\hspace*{5mm} \underline{$\limplies$ $(\forall r, (C'.\myseq_r \neq C'.\seq - 1 \vee C'.\pc{r} \in \{ \refln{fullcrash:rem:1} \upto \refln{fullcrash:try:3}, \refln{fullcrash:try:8}, \refln{fullcrash:exit:4}, \refln{fullcrash:rec:1} \upto \refln{fullcrash:rec:8} \}$}\\
				\hspace*{25mm} \underline{$\vee$ $( C'.\pc{r} = \refln{fullcrash:try:4} \limplies C'.\stopwait[C'.\myseq_r  \% 3] = \true))$} \\
				\hspace*{12mm} \underline{$\wedge$ $\exists q, q \in \lock[(C.\seq - 1) \% 3].\csset)$} \\
				Assume $C'.\pcp \in \{ \refln{fullcrash:try:7}, \refln{fullcrash:try:12} \} \wedge C'.\myseq_p = C'.\seq \wedge C'.\csowner = \perp$.
				We divide the argument into different cases as below and argue the correctness of each case. \\
				\\
				Suppose $C.\pcp = \refln{fullcrash:try:6}$.
				For $p$ to move to Line~\refln{fullcrash:try:7} from Line~\refln{fullcrash:try:6}, $C.\csowner \neq \perp$.
				However, we have $C'.\csowner = \perp$.
				It follows that $p$ didn't take a step from Line~\refln{fullcrash:try:6} with $C.\csowner = \perp$.
				$p$ can't take a step from Line~\refln{fullcrash:try:6} with $C.\csowner \neq \perp$ either because we have $C.\csowner = \perp$ and the step cannot modify $\csowner$.
				It follows that this case doesn't arise. \\
				\\
				Suppose $C.\pcp \in \{ \refln{fullcrash:try:7}, \refln{fullcrash:try:12} \}$ and $C.\myseq_p \neq C.\seq$. 
				By \indhyp{11} we have $C.\myseq_p = C.\seq - 1$.
				However, the step couldn't be by $p$ and any other $q \neq p$ doesn't modify $C.\myseq_p$.
				It follows that this case doesn't arise. \\
				\\
				Suppose $C.\pcp \in \{ \refln{fullcrash:try:7}, \refln{fullcrash:try:12} \}$, $C.\myseq_p = C.\seq$, and $C.\csowner \neq \perp$.
				Since $\pcp$ doesn't change by the step, the step was by some $q \neq p$.
				This $q$ must have written $\perp$ into $\csowner$ only at Line~\refln{fullcrash:exit:3}, because we assumed $C'.\csowner = \perp$.
				By \indhyp{9} $C.\csowner = q$.
				Thus, $C'.\pc{q} = \refln{fullcrash:exit:4}$, and we note from the first part of the conjunction of \indhyp{14} that the implication holds in $C'$.\\
				\\
				Suppose $C.\pcp \in \{ \refln{fullcrash:try:7}, \refln{fullcrash:try:12} \}$, $C.\myseq_p = C.\seq$, and $C.\csowner = \perp$.
				It follows that the step from $C$ to $C'$ was by a process $r \neq p$ because $\pcp$ hasn't changed from $C$ to $C'$.
				In this case we only need to argue how the inner implication would be affected by a step of a process $r$ such that $C.\myseq_{r} = C.\seq - 1$.
				By \indhyp{14} we know that for such an $r$
				$C.\pc{r} \in \{ \refln{fullcrash:rem:1} \upto \refln{fullcrash:try:3}, \refln{fullcrash:try:8}, \refln{fullcrash:exit:4}, \refln{fullcrash:rec:1} \upto \refln{fullcrash:rec:8} \}$,
				or $( \pc{r} = \refln{fullcrash:try:4} \limplies \stopwait[C.\myseq_r  \% 3] = \true)$.\\
				If $r$ takes a step at Line~\refln{fullcrash:try:3} and moves to Line~\refln{fullcrash:try:4},
				by \indhyp{14} we have $\exists q, q \in \lock[C.\myseq_{r} \% 3].\csset$.
				It follows that $r$'s step was not due to completion of $\lock[C.\myseq_{r} \% 3].\tryproc{r}()$, but due to breaking out of the \waittill loop at Line~\refln{fullcrash:try:3}.
				This can happen only when $\stopwait[C.\myseq_{r} \% 3] = \true$, which implies that the condition holds in $C'$.\\
				If $r$ takes a step at Line~\refln{fullcrash:try:4}, we know from \indhyp{14} that $\stopwait[C.\myseq_r \% 3] = \true$.
				It follows that $r$ moves to Line~\refln{fullcrash:try:8}, satisfying the condition in $C'$. \\
				If $r$ takes a step at Line~\refln{fullcrash:try:8}, it increments its own $\myseq_{r}$ and thus satisfies the condition trivially.
				
				\item[(ii). ] 
				\underline{$C'.\pcp \in \{ \refln{fullcrash:try:13}, \refln{fullcrash:try:14} \}$} \\
				\hspace*{5mm} \underline{$\limplies$ $(\forall r, (C'.\myseq_r \neq C'.\seq - 1 \vee C'.\pc{r} \in \{ \refln{fullcrash:rem:1} \upto \refln{fullcrash:try:3}, \refln{fullcrash:try:8}, \refln{fullcrash:exit:4}, \refln{fullcrash:rec:1} \upto \refln{fullcrash:rec:8} \}$}\\
				\hspace*{25mm} \underline{$\vee$ $( C'.\pc{r} = \refln{fullcrash:try:4} \limplies C'.\stopwait[C'.\myseq_r  \% 3] = \true))$} \\
				\hspace*{12mm} \underline{$\wedge$ $\exists q, q \in \lock[(C.\seq - 1) \% 3].\csset)$} \\
				Assume $C'.\pcp \in \{ \refln{fullcrash:try:13}, \refln{fullcrash:try:14} \}$.
				We divide the argument into different cases as below and argue the correctness of each case. \\
				\\
				Suppose $C.\pcp =\refln{fullcrash:try:12}$.
				Since $p$ moved from Line~\refln{fullcrash:try:12} to Line~\refln{fullcrash:try:13}, it follows that $p$ noticed $\csowner = \perp$ at Line~\refln{fullcrash:try:12}.
				Thus, $C.\csowner = \perp$.
				Therefore, by \indhyp{14} the implication holds in $C'$. \\
				\\
				Suppose $C.\pcp \in \{ \refln{fullcrash:try:13}, \refln{fullcrash:try:14} \}$.
				We note that if $p$ took a step from $C$ to $C'$, the implication continues to hold in $C'$ as it held in $C$.
				Otherwise the step from $C$ to $C'$ was by a process $r \neq p$ because $\pcp$ hasn't changed from $C$ to $C'$.
				We note from \indhyp{9} that no process $q$ could invoke $\lock[(\seq - 1) \% 3].\exitproc{q}()$ at Line~\refln{fullcrash:exit:2}, otherwise we would have $\csowner = q$.
				Thus, we have $\exists q, q \in \lock[(C'.\seq - 1) \% 3].\csset)$.
				In this case we only need to argue how the inner implication would be affected by a step of a process $r$ such that $C.\myseq_{r} = C.\seq - 1$.
				We skip the argument for this as it is similar to the argument for Case~(i) above.
			\end{enumerate}
		\end{enumerate}
	\end{enumerate}
	From the above arguments for the individual inductive steps it follows that the invariant holds in every configuration of every run of the algorithm.
\end{proof}

\end{document}